\documentclass[namedreferences,hyperref,optionalrh]{spr-sola}
\usepackage{graphicx}        % For eps figures, newer & more powerfull
\usepackage{amssymb}        % useful mathematical symbols
\usepackage{color}           % For color text: \color command
\newcommand{\spc}{\hbox{S$^3$PC}}
\newcommand{\sunrise}{{\sc Sunrise}\ }
\newcommand{\sunriset}{{\sc Sunrise iii}\ }

\newcommand{\arcsec}{^{\prime\prime}}
\newcommand{\fdeg}{.\!\!^{\circ}}
\newcommand{\farcsec}{.\!\!^{\prime\prime}}

\renewcommand{\vec}[1]{{\mathbfit #1}}

\chardef\us=`\_

\begin{document}

\begin{frontmatter}
\title{TuMag: the tunable magnetograph for the \sunriset mission}

%%%%%%%%%%%%%%%%%%%%%%%%%%%%%%%%%%%%%%%%%%%%%%%%%%%
%% Authors Names
%
\author[addressref={1,2},corref,email={jti@iaa.es}]{\inits{J.C.}\fnm{J.C.}~\lnm{del Toro Iniesta}\orcid{0000-0002-3387-026X}}
\author[addressref={1,2},corref,email={orozco@iaa.es}]{\inits{D.}\fnm{D.}~\lnm{Orozco Su\'arez}\orcid{0000-0001-8829-1938}}
\author[addressref={1,3}]{\inits{A.}\fnm{A.}~\lnm{\'Alvarez-Herrero}\orcid{0000-0001-9228-3412}}
\author[addressref={1,4}]{\inits{E.}\fnm{E.}~\lnm{Sanchis Kilders}\orcid{0000-0002-4208-3575}}
\author[addressref={1,5}]{\inits{I.}\fnm{I.}~\lnm{P\'erez-Grande}\orcid{0000-0002-7145-2835}}
\author[addressref={1,6,7}]{\inits{B.}\fnm{B.}~\lnm{Ruiz Cobo}\orcid{0000-0001-9550-6749}}
\author[addressref={1,2}]{\inits{L.R.}\fnm{L.R.}~\lnm{Bellot Rubio}\orcid{0000-0001-8669-8857}}
\author[addressref={1,2}]{\inits{M.}\fnm{M.}~\lnm{Balaguer Jim\'enez}\orcid{0000-0003-4738-7727}}
\author[addressref={1,2}]{\inits{A.C.}\fnm{A.C.}~\lnm{L\'opez Jim\'enez}\orcid{0000-0002-6297-0681}}
\author[addressref={1,2}]{\inits{D.}\fnm{D.}~\lnm{\'Alvarez Garc\'{\i}a}\orcid{0000-0002-8169-8476}}
\author[addressref={1,2}]{\inits{J.L.}\fnm{J.L.}~\lnm{Ramos M\'as}\orcid{0000-0002-8445-2631}}
\author[addressref={1,2}]{\inits{J.P.}\fnm{J.P.}~\lnm{Cobos Carrascosa}\orcid{0000-0002-5847-7181}}
\author[addressref={1,2}]{\inits{P.}\fnm{P.}~\lnm{Labrousse}\orcid{0009-0003-8123-0210}}
\author[addressref={1,2}]{\inits{A.J.}\fnm{A.J.}~\lnm{Moreno Mantas}\orcid{0009-0002-3396-3359}}
\author[addressref={1,2}]{\inits{J.M.}\fnm{J.M.}~\lnm{Morales-Fern\'andez}\orcid{0000-0002-5773-0368}}
\author[addressref={1,2}]{\inits{B.}\fnm{B.}~\lnm{Aparicio del Moral}\orcid{0000-0003-2817-8719}}
\author[addressref={1,2}]{\inits{A.}\fnm{A.}~\lnm{S\'anchez G\'omez}\orcid{0009-0008-7320-5716}}
\author[addressref={1,2}]{\inits{E.}\fnm{E.}~\lnm{Bail\'on Mart\'{\i}nez}\orcid{0009-0004-3976-2528}}
\author[addressref={1,2}]{\inits{F.J.}\fnm{F.J.}~\lnm{Bail\'en}\orcid{0000-0002-7318-3536}}
\author[addressref={1,2}]{\inits{H.}\fnm{H.}~\lnm{Strecker}\orcid{0000-0003-1483-4535}}
\author[addressref={1,2}]{\inits{A.L.}\fnm{A.L.}~\lnm{Siu-Tapia}\orcid{0000-0003-0175-6232}}
\author[addressref={1,2}]{\inits{P.}\fnm{P.}~\lnm{Santamarina Guerrero}\orcid{0000-0001-7094-518X}}
\author[addressref={1,2}]{\inits{A.}\fnm{A.}~\lnm{Moreno Vacas}\orcid{0000-0002-7336-0926}}
\author[addressref={1,2}]{\inits{J.}\fnm{J.}~\lnm{Ati\'enzar Garc\'{\i}a}}
\author[addressref={1,2}]{\inits{A.J.}\fnm{A.J.}~\lnm{Dorantes Monteagudo}\orcid{0000-0003-3316-3095}}
\author[addressref={1,2}]{\inits{I.}\fnm{I.}~\lnm{Bustamante}}
\author[addressref={1,2}]{\inits{A.}\fnm{A.}~\lnm{Tobaruela}\orcid{0009-0009-4178-4554}}
%\author[addressref={1,2}]{\inits{A.T.}\fnm{A.T.}~\lnm{Gallego-Calvente}\orcid{0000-0002-6428-8045}}
\author[addressref={1,3}]{\inits{A.B.}\fnm{A.}~\lnm{Fern\'andez-Medina}\orcid{0000-0002-1232-4315}}
\author[addressref={1,3}]{\inits{A.}\fnm{A.}~\lnm{N\'u\~nez Peral}}
\author[addressref={1,3}]{\inits{M.}\fnm{M.}~\lnm{Cebollero}}
\author[addressref={1,3}]{\inits{D.}\fnm{D.}~\lnm{Garranzo-Garc\'{\i}a}\orcid{0000-0002-9819-8427}}
\author[addressref={1,3}]{\inits{P.}\fnm{P.}~\lnm{Garc\'{\i}a Parejo}\orcid{0000-0003-1556-9411}}
\author[addressref={1,3}]{\inits{A.}\fnm{A.}~\lnm{Gonzalo Melchor}\orcid{0000-0003-1600-4826}}
\author[addressref={1,3}]{\inits{A.}\fnm{A.}~\lnm{S\'anchez Rodr\'{\i}guez}}
\author[addressref={1,3}]{\inits{A.}\fnm{A.}~\lnm{Campos-Jara}\orcid{0000-0003-0084-4812}}
\author[addressref={1,3}]{\inits{H.}\fnm{H.}~\lnm{Laguna}}
\author[addressref={1,3}]{\inits{M.}\fnm{M.}~\lnm{Silva-L\'opez}\orcid{0000-0002-8384-7658}}
\author[addressref={1,4}]{\inits{J.}\fnm{J.}~\lnm{Blanco Rodr\'{\i}guez}\orcid{0000-0002-2055-441X}}
\author[addressref={1,4}]{\inits{J.L.}\fnm{J.L.}~\lnm{Gasent Blesa}\orcid{0000-0002-1225-4177}}
\author[addressref={1,4}]{\inits{P.}\fnm{P.}~\lnm{Rodr\'{\i}guez Mart\'{\i}nez}}
\author[addressref={1,4}]{\inits{A.}\fnm{A.}~\lnm{Ferreres}\orcid{0000-0003-1500-1359}}
\author[addressref={1,4}]{\inits{D.}\fnm{D.}~\lnm{Gilabert Palmer}}
\author[addressref={1,5}]{\inits{I.}\fnm{I.}~\lnm{Torralbo}\orcid{0000-0001-9272-6439}}
\author[addressref={1,5}]{\inits{J.}\fnm{J.}~\lnm{Piqueras}}
\author[addressref={1,5}]{\inits{D.}\fnm{D.}~\lnm{Gonz\'alez-Bárcena}}
\author[addressref={1,5}]{\inits{A.J.}\fnm{A.J.}~\lnm{Fern\'andez}}
\author[addressref={1,6}]{\inits{D.}\fnm{D.}~\lnm{Hern\'andez Exp\'osito}}
\author[addressref={1,6}]{\inits{E.}\fnm{E.}~\lnm{P\'aez Ma\~n\'a}}
\author[addressref={7}]{\inits{E.}\fnm{E.}~\lnm{Magdaleno Castell\'o}}
\author[addressref={7}]{\inits{M.}\fnm{M.}~\lnm{Rodr\'{\i}guez Valido}\orcid{0000-0003-0873-9857}}

\author[addressref={8,12}]{\inits{A.}\fnm{Andreas}~\lnm{Korpi-Lagg}\orcid{0000-0003-1459-7074}}
\author[addressref={8}]{\inits{A.}\fnm{Achim}~\lnm{Gandorfer}\orcid{0000-0002-9972-9840}}
\author[addressref={8}]{\inits{S.~K.}\fnm{Sami K.}~\lnm{Solanki}\orcid{0000-0002-3418-8449}}

%%%%%%%%%%%%%%%%%%%%%%%%%%%%%%
%% Lead CoIs
%%%%%%%%%%%%%%%%%%%%%%%%%%%%%%
\author[addressref={9}]{\inits{T.}\fnm{Thomas}~\lnm{Berkefeld}}
\author[addressref={10}]{\inits{P.}\fnm{Pietro}~\lnm{Bernasconi}\orcid{0000-0002-0787-8954}}
\author[addressref={8}]{\inits{A.}\fnm{Alex}~\lnm{Feller}\orcid{0009-0009-4425-599X}}
\author[addressref={11}]{\inits{Y.}\fnm{Yukio}~\lnm{Katsukawa}\orcid{0000-0002-5054-8782}}
\author[addressref={8}]{\inits{T.~L.}\fnm{Tino~L.}~\lnm{Riethm\"uller}\orcid{0000-0001-6317-4380}}

%%%%%%%%%%%%%%%%%%%%%%%%%%%%%%
%% Sunrise CoIs
%%%%%%%%%%%%%%%%%%%%%%%%%%%%%%
\author[addressref={8}]{\inits{S.}\fnm{H.N.}~\lnm{Smitha}\orcid{0000-0003-3490-6532}}
\author[addressref={11}]{\inits{M.}\fnm{Masahito}~\lnm{Kubo}\orcid{0000-0001-5616-2808}}
\author[addressref={13}]{\inits{V.}\fnm{Valent\'{\i}n}~\lnm{Mart\'{\i}nez Pillet}\orcid{0000-0001-7764-6895}}
\author[addressref={8}]{\inits{S.}\fnm{Bianca}~\lnm{Grauf}}
\author[addressref={9}]{\inits{S.}\fnm{Alexander}~\lnm{Bell}}
\author[addressref={10}]{\inits{S.}\fnm{Michael}~\lnm{Carpenter}\orcid{0000-0003-3490-6532}}

%%%%%%%%%%%%%%%%%%%%%%%%%%%%%%%%%%%%%%%%%%%%%%%%%%%
%% Runningheads
%
\runningauthor{Del Toro Iniesta, Orozco Su\'arez et al.}
\runningtitle{TuMag for \sunriset}

%%%%%%%%%%%%%%%%%%%%%%%%%%%%%%%%%%%%%%%%%%%%%%%%%%%
%% Affilations 
%% id should be the same with \author addressref value.
\address[id={1}]{Spanish Space Solar Physics Consortium (\href{https://s3pc.es}{\spc})}
\address[id={2}]{Instituto de Astrof\'{\i}sica de Andaluc\'{\i}a (IAA-CSIC), Apdo. de Correos 3004, E-18080 Granada, Spain}
\address[id={3}]{Instituto Nacional de T\'ecnica Aeroespacial (INTA), Ctra. de Ajalvir, km. 4, E-28850 Torrejón de Ardoz, Spain}
\address[id={4}]{Universitat de Val\`encia Estudi General (UVEG), Avda. de la Universitat s/n, E-46100 Burjassot, Spain}
\address[id={5}]{Instituto de Microgravedad ``Ignacio da Riva'' (IDR-UPM), Plaza Cardenal Cisneros 3, E-28040 Madrid, Spain}
\address[id={6}]{Instituto de Astrof\'{\i}sica de Canarias, V\'{\i}a L\'actea, s/n, E-38205 La Laguna, Spain}
\address[id={7}]{Universidad de La Laguna, E-38205 La Laguna, Spain}
\address[id={8}]{Max-Planck-Institut f\"ur Sonnensystemforschung, Justus-von-Liebig-Weg 3, 37077 G\"ottingen, Germany}
\address[id={12}]{Aalto University, Department of Computer Science, Konemiehentie 2, 02150 Espoo, Finland}
\address[id={9}]{Institut f\"ur Sonnenphysik, Sch\"oneckstr. 6, 79104 Freiburg, Germany}
\address[id={10}]{Johns Hopkins Applied Physics Laboratory, 11100 Johns Hopkins Road, Laurel, Maryland, USA}
\address[id={11}]{National Astronomical Observatory of Japan, 2-21-1 Osawa, Mitaka, Tokyo 181-8588, Japan}
\address[id={13}]{National Solar Observatory, 3665 Discovery Dr. 3rd, Boulder CO 80303, USA (Currently at address \#6)}

%%%%%%%%%%%%%%%%%%%%%%%%%%%%%%%%%%%%%%%%%%%%%%%%%%%
%%% Abstract 
\begin{abstract}
\sunriset is a balloon-borne solar observatory dedicated to the investigation of the processes governing the physics of the magnetic field and the plasma flows in the lower solar atmosphere. The gondola hosts a 1-m aperture telescope that feeds three post-focus instruments. 

One of these instruments, the Tunable Magnetograph (TuMag), is a tunable imaging spectropolarimeter in visible wavelengths. It is designed to probe the vector magnetic field, $\vec{B}$, and the line-of-sight (LoS) velocity, $v_{\rm LoS}$, of the photosphere and the lower chromosphere. It provides polarized images with a $63\arcsec \times\, 63\arcsec$ field of view (FoV) of the Sun in four polarization states. These images can later be processed on ground to retrieve maps of the aforementioned solar physical quantities. The quasi-simultaneous observation of two spectral lines sensitive to $\vec{B}$ and $v_{\rm LoS}$ in the photosphere and lower chromosphere provides excellent diagnostic measurements of the magnetic and dynamic coupling in these layers. When combined with the other two instruments on board, observing in the infrared and ultraviolet regions of the spectrum, TuMag's diagnostic potential is expected to be greatly enhanced.

Building upon heritage of instruments like IMaX and SO/PHI, the key technologies employed for TuMag are a liquid-crystal-variable-retarder-based polarimeter and a solid, LiNbO$_3$ Fabry--P\'erot etalon as a spectrometer. However, it also incorporates several innovative features, such as home-made, high-sensitivity scientific cameras and a double filter wheel. The latter makes TuMag the first balloon-borne instrument of its type capable of tuning between spectral lines. Specifically, it can sequentially observe any two out of the three spectral lines of Fe {\sc i} at 525.02 and 525.06 nm and of Mg {\sc i} at 517.3 nm. Time cadences range from 30 to 100 seconds, depending on the observing mode and the specific pair of spectral lines targeted. 

Laboratory measurements have demonstrated outstanding performance, including a wavefront root-mean-square error better than $W\, \sim \lambda/13$ for image quality, a full-width-at-half-maximum (FWHM) of 8.7~pm for the filtergraph transmission profile, and polarimetric efficiencies $\varepsilon_i > 0.54$, where $i=2, 3, 4$ correspond to Stokes $Q$, $U$, and $V$. Here we report on the concept, design, calibration, and integration phases of the instrument, as well as on the data reduction pipeline.
\end{abstract}

%%%%%%%%%%%%%%%%%%%%%%%%%%%%%%%%%%%%%%%%%%%%%%%%%%%
%% Keywords
%

\end{frontmatter}
%-------------------------------------------------

%%%%%%%%%%%%%%%%%%%%%%%%%%%%%%%%%%%%%%%%%%%%%%%%%%%
%% Sections
%
\section{Introduction}
\label{sec:intro}

   The increasing evidence for a key role of the magnetic field in defining the state of the Sun and its many observable features across all layers of its atmosphere has been accompanied by the evolution of instrumentation since the discovery by \cite{hale1908} of a magnetic field in sunspots. From the pioneering work of Hale by comparing more than 200 solar and laboratory spectral lines, spectrographs combined with polarimeters have been one of the main sources of information. The availability of narrow-band filtergraphs \citep[see][for a historical review]{bonaccini+etal1989} opened a new observational approach to the problem: spatial integrity is better preserved than in the case of spectrographs at the expense of a loss in spectral integrity. Both types of instruments have been improving in parallel to the advance of technology and have demonstrated to be complementary to each other. The instrument described in this publication is one of the filtergraph-based instruments. Members of this class are, for instance, the filtergraph at the heart of the Flare Genesis project  \citep{1994AdSpR..14b..89R}, the Michelson Doppler Imager \citep{1995SoPh..162..129S} aboard the ESA/NASA's {\sc Solar Heliospheric Observatory} \citep[SoHO; ][]{1995SoPh..162....1D}, the Italian Panoramic Monochromator at THEMIS \citep{1998A&AS..128..589C}, the TESOS spectrometer at the Vacuum Tower Telescope \citep{kentischer+etal1998}, the Interferometric Bidimensional Spectrometer at the Dunn Solar Telescope \citep{2001MmSAI..72..554C}, the Helioseismic and Magnetic Imager \citep{2002AGUFMSH52A0494S} on board the NASA's {\sc Solar Dynamics Observatory} \citep{2012SoPh..275....3P}, the CRisp Imaging SpectroPolarimeter at the Swedish Solar Telescope (\citeauthor{2008ApJ...689L..69S} \citeyear{2008ApJ...689L..69S}; \citeauthor{2008A&A...489..429V} \citeyear{2008A&A...489..429V}), the Narrowband Filter Imager \citep[NFI;][]{2008SoPh..249..167T} aboard the {\sc Hinode} satellite \citep{Kosugi_2007}, the Visible Imaging Polarimeter \citep{2010A&A...520A.115B}, the Imaging Magnetograph eXperiment \citep[IMaX;][]{2011SoPh...268...57M} instrument aboard \sunrise \citep{2011SoPh..268....1B}, the G\"ottingen Fabry--P\'erot Interferometer at GREGOR \citep{2013OptEn..52h1606P}, the Visible Tunable Filter \citep{2016SPIE.9908E..4NS} at the Daniel K. Inouye Solar Telescope \citep{2020SoPh..295..172R}, the Polarimetric and Helioseismic Imager \citep[SO/PHI;][]{Solanki_2020} on board the ESA/NASA's {\sc Solar Orbiter} mission \citep{Muller_2020}, and the Photospheric Magnetic field Imager \cite[PMI;][]{2020JSWSC..10...54S} for the ESA's {\sc Vigil} mission. 
   
   Together with SUSI \citep[\sunrise UV Spectropolarimeter and Imager;][]{SUSI2024} and SCIP \citep[\sunrise Chromospheric Infrared spectroPolarimeter;][]{SCIP2024}, the Tunable Magnetograph (TuMag) is part of the \sunriset suite of post-focus instruments \citep{Sunriseiii2024}. A further Correlating Wavefront Sensor \citep[CWS;][]{WCS2023} is in charge of real-time correction of image motion. TuMag is a tunable, dual-beam imaging spectropolarimeter that builds upon technologies already used in Flare Genesis, IMaX, and SO/PHI: it employs liquid crystal variable retarders as polarization modulator and a solid, LiNbO$_3$ etalon in double pass as spectrometer. The use of Fabry--P\'erot etalons in astronomy and in particular in solar physics has been extensively assessed by \cite{1998A&amp;AS..129..191B}, \cite{2000A&amp;AS..146..499V}, \cite{2006A&amp;A...447.1111S}, \cite{2010A&A...515A..85R}, and  \citeauthor{2019ApJS..241....9B} (\citeyear{2019ApJS..241....9B}, \citeyear{2019ApJS..242...21B}, \citeyear{Bail_n_2020}, \citeyear{2021ApJS..254...18B}). A thorough study of the performance of solar magnetographs and spectropolarimeters, and in particular of those using our two critical technologies was presented by \cite{2012ApJS..201...22D}. The heritage from IMaX and SO/PHI has been key for developing  TuMag. Indeed, it incorporates the polarization modulation package developed for SO/PHI (although the linear polarizer is removed) and the spare Fabry--P\'erot etalon of IMaX. Some necessary technical innovations are overviewed in the following paragraphs.
   
   TuMag is the first aerospace imaging vector spectropolarimeter able to tune among three spectral lines, namely, those of Fe {\sc i} at 525.02 and 525.06 nm and that of Mg {\sc i}{\small b}$_{\rm 2}$ at 517.3 nm. Line selection is made by rotating to high precision a filter wheel which hosts three different pre-filters, one per spectral line. Robustness and simplicity drove former designs (e.g., IMaX or SO/PHI) to limit the wavelength sampling to just one spectral line. However, the capability for switching among two or three lines offers the opportunity to probe the vector magnetic field, $\vec{B}$, and line-of-sight (LoS) velocity, $v_{\rm LoS}$, in different layers of the atmosphere. This is especially important for \sunriset as the mission is aimed at studying the magnetic coupling between the photosphere and the chromosphere.
   
   TuMag uses two home-made cameras \citep[SPGCams;][]{2023FrASS..1067540O} based on the GPIXEL back-illuminated GSENSE400BSI $2\, {\rm k} \times 2\, {\rm k}$ pixel detector. These cameras capture a solar FoV slightly larger than $63\arcsec \times\, 63\arcsec$, which is $44\,\%$ greater than that of IMaX. The cameras are in a dual-beam configuration, that is, each camera observes orthogonal polarization states of light simultaneously. The camera electronics is based on a mezzanine of three printed-circuit-boards (PCBs) for driving, hosting, and controlling the temperature of the image sensor, respectively. An especially tailored window protects the sensor from dust, humidity, and other volatile materials. The three cameras of SCIP are also SPGCams, but they have their external windows removed.
   
   Besides the pre-filters, the double TuMag filter wheel hosts several elements for calibration: an empty vane with a frame to limit the FoV and help align the two camera images; a set of pinholes that help determine the instrument point spread function (PSF) and check for any ghost images produced by the system or the external illumination; a linear polarizer; and a set of micro-polarizers\footnote{Patent pending. Patent application number: EP2282703.1 "Calibration target and method for the snapshot calibration of imaging polarimeters", Alberto Álvarez-Herrero and Pilar García Parejo, July 22, 2022.} that can help in retrieving many elements of the instrument's Mueller matrix.
   
   Thermal stability is preserved by using especially designed boxes to insulate those critical subsystems which require an accurate thermal control: the filter wheel and the etalon. These boxes are thermally controlled by means of heaters and proportional-integral-derivative (PID) control systems. SPGCams are thermally controlled by a direct conductive connection to the instrument radiator. 
   
   The optical bench is made of composite material, especially devised to preserve stiffness while reducing weight and minimizing thermal conductivity \citep{2023Senso..23.6499F}. The metallic box containing the optical unit (O-Unit) is wrapped with a mylar blanket. The top cover hosts the radiator and the connector panel that links the O-Unit with the electronics unit (E-Unit).
   
\begin{figure}
   \centerline{
   \includegraphics[width=\columnwidth]{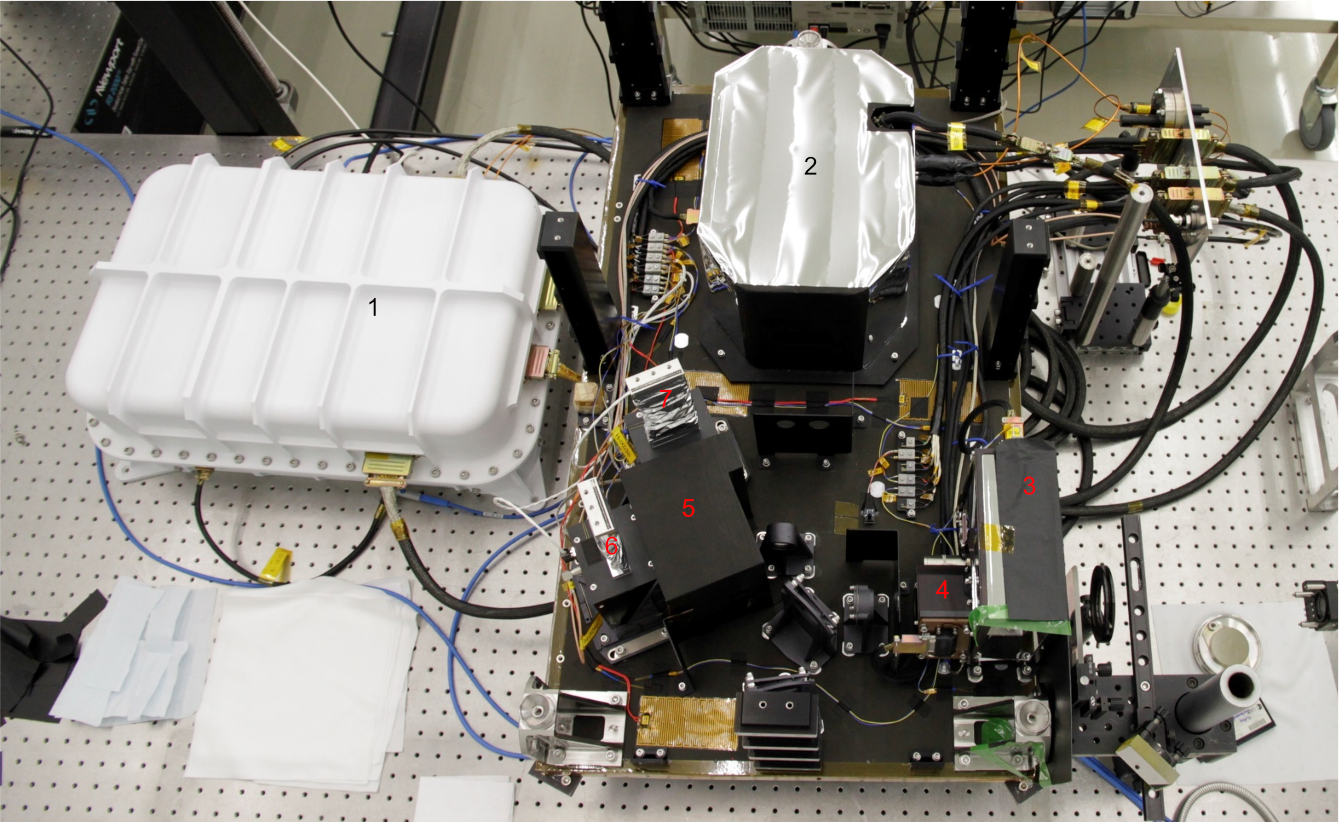}
   \caption{TuMag during assembly, integration, and verification at INTA facilities. Main units and elements are labeled with numbers. The electronics unit (1) is in its flight, closed and pressurized configuration. Within the uncovered O-Unit one can discern the enclosure of the etalon oven (2), the enclosure of the filter wheel (3), the polarization modulation package (4), the enclosure of the beam splitter (5), and the two SPGCams with their cold fingers on top (6 and 7). Some ancillary optics, the harness, the pillars for the O-Unit cover, and the isostatic mounts for lifting the instrument can also be seen.}}
   \label{TuMagAIV}%
   \end{figure}

   Likewise IMaX, the electronics unit is enclosed in a pressurized box in order to use commercial-off-the-shelf (COTS) electronic components in the very low pressure environment during the flight. Unlike IMaX, and following the SO/PHI's design recipe, the electronic components are placed in four newly designed PCBs: the data processing unit (DPU); the analog, mechanisms, and heaters driver board (AMHD); the power converter module (PCM); and the high-voltage power supply (HVPS). The E-unit is linked to the O-unit by an especially designed harness. Together with the other instruments, the TuMag O-Unit is located in the platform for instruments (PFI), on top of the telescope structure, while the E-Unit is fixed to a lateral plate attached to the \sunriset gondola. A picture of the entire system during integration is shown in Fig.~\ref{TuMagAIV}.
   
   All the control software (CSW) and firmware (FW) is completely new with respect to IMaX's. So is the ground segment equipment (GSE) of the instrument, which allows communication with the \sunriset instrument control unit for sending telecommands and receiving telemetries. All four E-Unit, CSW, FW, and GSE have been designed to maximize the commonalities between TuMag and SCIP, since our team has been in charge of those subsystems for the two instruments.
   
%-------------------------------------- 
%-------------------------------------- 

   TuMag has been designed, developed, and manufactured by the Spanish Space Solar Physics Consortium (\spc), led by the Solar Physics Group (SPG) of the Instituto de Astrof\'\i sica de Andaluc\'\i a (IAA), of the Consejo Superior de Investigaciones Cient\'\i ficas (CSIC), and including the Instituto Nacional de T\'ecnica Aeroespacial (INTA), the Universitat de Val\`encia (UV), the Instituto de Microgravedad ``Ignacio da Riva" (IDR) of the Universidad Polit\'ecnica de Madrid (UPM), and the Instituto de Astrof\'\i sica de Canarias (IAC). Besides the coordination and overall responsibility of the instrument, the SPG-IAA-CSIC has scientifically defined TuMag, designed and manufactured the SPGCams, the CSW, the FW, and the GSE. It is responsible for the overall E-Unit where it has designed and manufactured the DPU, the AMHD, and the HVPS. The harness has also been designed by the SPG and manufactured by the Max-Planck-Institut f\"ur Sonnensystemforschung (MPS), except for the high-voltage harness which has been also manufactured by SPG. INTA has been overall in charge of the O-Unit design and manufacturing, including its thermal, stray light, and radiation controls. All the activities related to assembly, integration, and verification (AIV) of the instrument as a whole have been coordinated by INTA and carried out at its facilities. UV has been in charge of the PCM and the pressurized electronics box (E-box). IDR-UPM has been responsible for the thermal design and thermal hardware of the E-Unit and the SPGCams. IAC has contributed to the firmware design for both the SPGCams and the DPU. The data reduction pipeline has been developed by SPG, UV, and IAC.
   
   This paper describes the instrument and its subsystems in detail, all the way from the scientific drivers and definition to the critical design and AIV phase (on-ground calibration). It aims at providing a useful reference for the analysis of the data gathered by TuMag, which will be available at the SPG's Web page (\url{http://spg.iaa.es}), the \spc's Web page (\url{http://s3pc.es}),  as well as in the main \sunrise data storage at MPS (\url{http://https://www.mps.mpg.de/en}).

\section{Tuning among three spectral lines}
\label{sec:drivers}

   IMaX flew twice on 5-6 day long flights of Sunrise (Sunrise I and II). Thanks to the combination of good image quality with good pointing stability, stable spectroscopic resolution, and, above all, excellent polarimetric sensitivity allowed data of very high quality to be obtained during both flights. Image quality was not only granted by the instrument but also by the telescope and ISLiD\footnote{ISLiD stands for Image Stabilization and Light Distribution.} system. Pointing stability provided by the gondola, the navigation system, and the CWS was secured for periods of up to some 25 minutes. The spectral resolution was 6.5 pm (8.5 pm Gaussian). The polarimetric sensitivity reached values above 900 for the signal-to-noise ratio. During the first flight, the two instruments on board were devoted to exploring the quiet Sun: the flight took place during a very quiet period of solar activity. The situation changed for the second flight when both SuFI \citep[{\sc Sunrise} Filter Imager;][]{2011SoPh..268...35G} and IMaX were used to study active regions as well. 

   When preparing a third edition of the stratospheric balloon mission, one would like to preserve the successful features of the first two flights while innovating in a new instrument aimed at fulfilling the goals of the new edition. {\sc Sunrise iii} is devoted to investigating the magnetic coupling of the photosphere and the chromosphere. Hence, while SUSI and SCIP cover the UV and IR regions of the spectrum, TuMag provides a view of both the photosphere and the (low) chromosphere by observing two different visible spectral lines quasi-simultaneously. Moreover, TuMag must enable better retrievals of the solar atmospheric properties by sampling the lines better than IMaX. While the latter scanned the (single) spectral line at 4 wavelengths plus the nearby continuum, the new instrument does it with at least 8 (for the photospheric line) or 10 (for the chromospheric line) wavelengths plus the nearby continuum. IMaX used to take some 33 s for one line. TuMag observes two lines in less than 66 s (see Sect.~\ref{sec:observinganddata}).\footnote{The requirement of 90\;s in Table~\ref{tab:table1} may seem too conservative. It can nevertheless be reached if extra polarimetric precision is required in specific observing modes. See Sect.~\ref{sec:observinganddata}.} This improved sampling and wider coverage should provide better diagnostics of the solar physical quantities through inversion techniques that were constrained by the limited information provided by IMaX.

   Obviously, the choice for the pair of lines depends on the instrument characteristics. Not all combinations of two spectral lines are observable with a filtergraph-based instrument. Indeed, the two lines should be located in the spectrum in such a way that compatible etalon interferometric orders can be used after only changing the pre-filter. Among the feasible combinations, we chose the Fe {\sc i} lines at 525 nm and the Mg {\sc i}{\small b}$_2$ line at 517 nm. The Fe {\sc i} line at 525.02 nm is a very well known line in solar polarimetry because of its large Land\'e factor (large Zeeman sensitivity) in the green, which produces clear polarization signatures. While strong polarization signals help diagnose weak magnetic fields, this line is very temperature dependent since it belongs to the first Fe {\sc i} multiplet, hence with a very low excitation potential. This dependence on temperature may be an obstacle for diagnostics, especially when observing large and dark sunspot umbrae. An alternative line, equally unblended, which is very close to it is the Fe {\sc i} line at 525.06 nm. With a much lower Land\'e factor and a smaller dependence on temperature, the is well suited for active regions magnetic diagnostics. A comparison of the expected behavior of both lines as observed with IMaX was carried out by \cite{2008PhDT........82O}. That comparison is applicable to TuMag. Our design decision has been to keep both lines available for the third edition of \sunrise.

   The expected polarimetric behavior of the Mg {\sc i}{\small b}$_2$ line at 517 nm has been thoroughly studied by \cite{1995A&A...299..596B} and \citeauthor{2018MNRAS.481.5675Q} (\citeyear{2018MNRAS.481.5675Q}, see references therein). Besides their finding of very good polarization signals in standard model atmospheres, the authors indicate the line as a good diagnostic for the magnetic field strength of the middle chromosphere and an excellent complement to the information provided by lines like the Ca {\sc ii} triplet at 854.2 nm, which is observed by SCIP. The diagnostic potential of the combination of these two lines has recently been demonstrated by the analysis of quiet Sun imaging spectropolarimetric observations presented by \cite{2021ApJ...911...41G}.

%--------------------------------------------------------------------

   \begin{table}
      \caption[]{TuMag scientific requirements}
         \label{tab:table1}
\begin{tabular}{ll}
            \hline
            {\rm Requirement}      &  {\rm Value} \\
%            \noalign{\smallskip}
            \hline
            \noalign{\smallskip}
            {\rm Field of view (FoV)}           & $63\arcsec \times\, 63\arcsec$    \\
            {\rm Stand-alone wavefront error} & $W \sim \lambda/14$     \\
            {\rm Phase diversity plate thickness} & $(1.5 \pm 0.2)\, \lambda$    \\
             {\rm Spatial sampling}        & $3 \times 3\, {\rm pixels}$ \\
            {\rm Plate scale}           & $0.0378\arcsec/{\rm pixel}$  \\
            {\rm Polarimetry mode}      & {\rm Dual\ beam}   \\
            {\rm Polarimetric efficiencies}  & $\varepsilon_{1,2,3} \lessapprox 1/\sqrt{3}$ \\
            {\rm Signal-to-noise ratio} & ${\rm (S/N)}_0 \gtrapprox 1700$            \\
            {\rm Vector magnetograph}    & {\rm Yes}   \\
            {\rm Longitudinal magnetograph}  & {\rm Yes}   \\
            {\rm Spectral resolution}   & $<\, 9\ {\rm pm}$      \\
            {\rm Spectral lines}    & Fe {\sc i} 525.02 nm, Fe {\sc i} 525.06 nm, and  Mg {\sc i}{\small b}$_2$ 517.27 nm \\
            {\rm Time for a two-line observation}  &  $< 90\ {\rm s}$ \\
            \noalign{\smallskip}
            \hline
\end{tabular}
   \end{table}

%--------------------------------------------------------------------

\section{Scientific definition of the instrument}
\label{sec:definition}
   TuMag is conceived to be a tunable imaging spectropolarimeter. As such, it should provide high spatial resolution images, at several quasi-monochromatic wavelengths across two or more spectral lines, and with high polarimetric sensitivity in four polarization states in order to retrieve, in a reasonable short time interval, the four Stokes parameters of light, $\vec{S} = (S_0, S_1, S_2, S_3)$\, $\equiv (I, Q, U, V)$. In this \emph{vector magnetograph} configuration the three components of the vector magnetic field, $\vec{B} \equiv (B, \gamma, \varphi)$, can be retrieved from the measurements. If instead, only two polarization states, $I \pm V (= S_0 \pm S_3)$, are measured, the configuration is called \emph{longitudinal magnetograph} and information only about the longitudinal component of the magnetic field, $B_{\rm LoS} = B \cos\gamma$, can be obtained. The standard observation time for two spectral lines is roughly estimated from the typical dynamical times of the solar atmosphere. IMaX observed just one line in four wavelengths plus one in the continuum in some 33 s. TuMag wants to improve and achieve a two-line (plus continuum) sampling, more than eight wavelength samples each, in less than 90 s. As we shall see in Sec.~\ref{sec:observinganddata}, the integrated instrument is able to reach that goal in less than 66 s, but we wanted to be conservative during the development phases. The possibility to adjust almost all instrument parameters (wavelength samples, cadence, line selection, polarimetry mode) made it mandatory to combine specific settings in so-called observing modes, optimized to achieve the scientific goals defined for \sunriset. The modes also allow for an efficient, low-risk operation of TuMag during the flight, and simplify the post-flight data reduction and analysis. The final time will depend on the choice. The main scientific requirements of the instrument are summarized in Table \ref{tab:table1}. $\lambda$ in that table stands for a reference wavelength which is taken as 525 nm. Index 0 for (S/N) refers to Stokes $I$ (see Sect.~\ref{sec:polperformance}).
  
   \subsection{Image quality}
   \label{sec:imagequality}
   
      To preserve an excellent image quality across a field of view of $63\arcsec \times\, 63\arcsec$, TuMag should achieve a wavefront error of $W\,\sim\, \lambda/14$,  include means for phase diversity (PD) data acquisition, and provide images sampled beyond the Nyquist critical value. The chosen FoV accommodates a full medium-size active region. According to the Mar\'echal criterion, an rms wavefront error of $W = \lambda/14$ is typically considered as an indicator of a diffraction-limited instrument and is equivalent to a Strehl ratio of approximately 0.8. Our various laboratory experiments, including modulation-transfer-function measurements with the slanted-slit method \citep{2013SPIE.8788E..2JH}, conducted during AIV, yielded values $W\lessapprox \lambda/13$, hence confirming the instrument's excellent optical quality in standalone conditions. Final image quality can deteriorate from additional aberrations introduced by the telescope and ISLiD system, as well as  from residual gondola jitter not fully corrected by the CWS. Nevertheless, images can always be restored if $W \gtrapprox\, \lambda/5$ \citep{2009PhDT........78V}, provided the spatial point spread function (PSF) of the instrument is known through, for instance, PD techniques. 
   
      Unlike IMaX and other instruments, TuMag is equipped with unconventional  PD capabilities. Within the filter wheel, a plane-parallel plate is located to introduce a known defocus. The plate's width, determined to be $1.45\,\lambda$ after a careful assessment made by \cite{2022ApJS..263....8B}, provides focused and defocused images sequentially rather than simultaneously. The feasibility of such a sequential procedure, considering the solar evolution between the two images, has been verified by \cite{2022ApJS..263....7B}. Further work reassures the TuMag's PD capabilities if several images with different defocus are used as coming, for instance, from a focus run of the instrument \citep{2022ApJS..263...43B}. Sampling beyond the Nyquist critical value is accomplished in TuMag by imaging the diffraction-limit size, $\lambda/D$, into $3\,\times\,3$ pixels of our custom SPGCam. This spatial sampling helps reduce the noise during the restoration process.

   \subsection{Polarimetric performance}
   \label{sec:polperformance}

      Polarimetry is crucial to detect the (often weak) signal fingerprints of magnetic fields in the spectrum of light. Magnetographic detectability, thus, relies on the instrument ability to discern between polarization signal (i.e., $\vec{S}$) and noise. The signal-to-noise ratio, S/N, is then key to quantify the polarimetric performance of instruments. We typically evaluate S/N at continuum wavelengths because they are virtually unpolarized to the levels of $10^{-5}$ or weaker. Consensus in the community establishes ${\rm S/N} = 1000$ as a good requirement for polarization detection in many of the processes we aim to study with TuMag. This value will be considered a default requirement although different values of S/N can be specified as needed for various scientific purposes. S/N is a simple parameter in regular photometry or spectroscopy but requires a further specification in polarimetry since the latter is a differential photometric technique. Following \cite{2012ApJS..201...22D}, if we call $\sigma_i$ the rms in the continuum of $S_i$, then 
      \begin{equation}
         {\rm (S/N)}_i \equiv \left( \frac{S_0}{\sigma_i} \right)_{\rm c} = \frac{\varepsilon_i}{\varepsilon_0} \left( \frac{S_0}{\sigma_0} \right)_{\rm c},
      \end{equation}
      where index c stands for continuum wavelengths. Therefore, a requirement of (S/N)$_{1,2,3} = 1000$ implies (S/N)$_0 \gtrapprox 1700$, assuming the Stokes $I$ polarimetric efficiency $\varepsilon_{0} = 1$ and all the three remaining efficiencies to be $\varepsilon_{1,2,3} \lessapprox 1/\sqrt{3}$ \citep{2000ApOpt..39.1637D}. Such signal-to-noise levels in Stokes $I$ require some $1.5\cdot10^6$ photo-electrons in each of the two cameras of our dual-beam instrument, provided that photon noise is the dominant noise source.\footnote{The sensor in our SPGCams has a quantum efficiency of 0.953 at 525~nm (see below).} This amount of electrons cannot be reached in single shots by currently available sensors, which are usually filled to 50 \% of the full well capacity. (Specifically, our GSENSEI400BSI sensor has a full well of $9.1\cdot10^4$ electrons.) Therefore, accumulation of several, $N_{\rm a}$, individual frames is needed in order to fulfill the S/N requirement. This accumulation scheme was successfully proven for the first time with ASP \citep{1992SPIE.1746...22E} and later used by instruments like the TIP I and II (\citeauthor{martinez+etal1999} \citeyear{martinez+etal1999}; \citeauthor{collados+etal2007} \citeyear{collados+etal2007}), ZIMPOL \citep[][and references therein]{2004A&A...422..703G}, {\sc Hinode}/SP \citep{2001ASPC..236...33L}, IMaX \citep{2011SoPh...268...57M}, and SO/PHI \citep{Solanki_2020}. Accumulation after applying reconstruction techniques (speckle or blind deconvolution) has also been used in other ground-based filter magnetographs (\citeauthor{2008A&A...480..265B} \citeyear{2008A&A...480..265B}; \citeauthor{2008A&A...489..429V} \citeyear{2008A&A...489..429V}; \citeauthor{2009ApJ...700L.145V} \citeyear{2009ApJ...700L.145V}; \citeauthor{2015ApJ...803...93E} \citeyear{2015ApJ...803...93E}).

      TuMag is a dual-beam polarimeter because of the advantages  such a configuration brings about for correcting the residual seeing-induced (in the case of ground-based instruments) or jitter-induced (in the case of aerospace-borne instruments) spurious polarization signals (\citeauthor{lites1987} \citeyear{lites1987}; see \citeauthor{2003isp..book.....D}, \citeyear{2003isp..book.....D} for an extended explanation). Moreover, the two cameras increase the S/N by a factor $\sqrt{2}$ as compared to one single camera. This means that (S/N)$_0 \gtrapprox 1200$ in each of the TuMag's individual cameras.

      Good S/N and dual-beam configuration are not enough for an efficient polarimeter. Since polarimetry is differential photometry, we need to modulate the polarization of incoming light as quickly as possible. Following our experience with IMaX and SO/PHI (which indeed ensure fulfillment of our requirements), polarization modulation is made in TuMag with a pair of nematic, liquid crystal variable retarders (LCVRs). 

      Two families of polarimetric observing modes (see Sect.~\ref{sec:observinganddata}) are baselined for the instrument, namely vector and longitudinal. Depending on the polarimetric observing mode, each TuMag SPGCam must then take $N_{\rm p} = 4$ (2) polarized images, which are the result of accumulating $N_{\rm a}$ single exposures for reaching (S/N)$_0 \gtrapprox 1200$, at each of the $N_{\lambda}$ wavelength samples. The $N_{\rm p}$ image acquisition cycle can be repeated a number of $N_{\rm c}$ times (see Sect.~\ref{sec:obsfirmware}) in order to enhance (S/N)$_0$ in special modes. If we call $t_{\rm exp}$ the single-shot exposure time, then any single observing series of images has an effective exposure time $t_{\rm eff} = N_{\rm c} N_{\lambda} N_{\rm p} N_{\rm a}  t_{\rm exp}$. Real implementation of modes can last a bit longer (see Sect.~\ref{sec:obsfirmware}).         

   \subsection{Spectroscopic performance}
   \label{sec:specperformance}
      The TuMag's spectral resolution is required to be better than 9 pm. The instrument is designed to work across three spectral lines. Tunability among the lines requires the use of three pre-filters which select a free spectral range of the Fabry--P\'erot etalon, hence selecting the correct interferometric order of the latter corresponding to the chosen spectral line. Those pre-filters are hosted in the double filter wheel at the entrance of the instrument. We suggest the interested reader the paper by \cite{2011SoPh...268...57M} for a rationale and an explanation on the selection of the pre-filters and the LiNbO$_3$ etalon. TuMag uses the IMaX spare etalon.

      The spectroscopic and polarimetric performance of an anisotropic etalon like ours depends to a great deal on whether it is illuminated by a collimated or a telecentric beam \citep[see][and references therein]{2019ApJS..242...21B}. Following our considerations for IMaX, TuMag's etalon is located in a collimated beam. Experience shows that neither image quality degradation nor etalon's central transmission peak shift are a significant problem for the instrument performance.

 %---------------------------------------------------
   \begin{table}
      \caption[]{Photon budget parameters}
         \label{tab:table2}
      \begin{minipage}{\textwidth}

         \begin{tabular}{llll}
            \hline
            {\rm Name}      &  {\rm Symbol}   &   {\rm Value}   &  {\rm Units}\footnote{Units are per pixel where necessary. Symbol ph stands for photons.} \\
            \hline
            {\rm Solar spectral radiance (525 nm)}  & $N_\odot$  & $7.49 \cdot 10^{19}$   &   ${\rm ph}\ {\rm s}^{-1} {\rm sr}^{-1} {\rm pm}^{-1} {\rm m}^{-2}$   \\
%            {\rm Solar spectral radiance (517 nm)}  & $N_\odot$  & $5.89 \cdot 10^{19}$   &   ${\rm ph}\ {\rm s}^{-1} {\rm sr}^{-1} {\rm pm}^{-1} {\rm m}^{-2}$   \\
            {\rm Solar spectral radiance (517 nm)}  & $N_\odot$  & $6.15 \cdot 10^{19}$   &   ${\rm ph}\ {\rm s}^{-1} {\rm sr}^{-1} {\rm pm}^{-1} {\rm m}^{-2}$   \\
            {\rm Pixel angular area}   & $\phi^2$  & $3.36 \cdot 10^{-14}$   &  sr  \\
            {\rm Pixel full well}   & $F_{\rm W}$  & $9 \cdot 10^{4}$   &  e$^-$  \\
            {\rm Spectral FWHM}   &   $\Delta \lambda$   &   8.7   & pm \\
            Effective collecting area   &   $A_{\rm D}$   &   0.7   & m$^2$   \\
            Sunrise throughput (525 nm)\footnote{A conservative estimation for both wavelengths, excluding TuMag's.}  &   $\tau_{\rm Sunrise}$   &   0.49 \\ 
            Sunrise throughput (517 nm)  &   $\tau_{\rm Sunrise}$   &   0.49 \\
            TuMag throughput (525 nm)   &   $\tau_{\rm TuMag}$   &   0.13   \\
            TuMag throughput (517 nm)   &   $\tau_{\rm TuMag}$   &   0.14   \\
            Exposure time (individual frame)   & $t_{\rm exp}$   &   0.04166   &   s   \\
            Sensor's quantum efficiency   &   $Q$   &   0.953   \\
            Analog-to-digital conversion gain   &   $G$   &   0.053   \\
            Sensor's offset   &   $\sigma_{\rm off}$   &   150   &   DN   \\
            Sensor's readout noise   &   $\sigma_{\rm ro}$   &   2   &   DN   \\
            Sensor's dark current noise\footnote{Worst case scenario for a sensor temperature of 30 degree.}   &   $\sigma_{\rm dc}$   &   50   &   e$^-$ s$^{-1}$   \\
            \hline
         \end{tabular}

      \end{minipage}

   \end{table}
% The dark current (DN/s) is the slope of the sensor output (DN) versus the exposure time (s). Then conversion factor. CHECK Camera report. 

   \subsection{Photon budget}
   \label{sec:photonbudget}

      As seen in the previous Section, polarimetric performance and signal-to-noise ratio are intimately related to each other. Therefore, an estimate of the expected (S/N)$_0$ is in order. Following \cite{2011SoPh...268...57M}, the continuum intensity received by each of the TuMag detector pixels in a single frame is given by 
      \begin{equation}
          S_{0, \rm c} = N_{\odot}\, \phi^2\, \Delta\lambda\, A_{\rm D}\, \tau\, t_{\rm exp},
      \end{equation}
      where subindex c indicates continuum wavelengths, $N_{\odot}$ is the solar spectral radiance just outside the Earth's atmosphere at a specified wavelength (525 or 517 nm), $\phi^2$ is the angular area subtended by every individual pixel on the plane of the sky, $\Delta\lambda$ is the spectral resolution of the instrument (FWHM of the etalon transmission curve multiplied by the pre-filter transmission curve), $A_{\rm D}$ stands for the effective collecting area of the telescope (difference between the areas of the main and secondary mirrors), $\tau = \tau_{\rm Sunrise}\tau_{\rm TuMag}$, stands for the throughput of the whole system (at 525 or 517 nm), and $t_{\rm exp}$ is the individual frame's exposure time. The symbols and values for these parameters are specified in Table \ref{tab:table2}. With all these values, $S_{0, \rm c} = 3.1\ (2.6) \cdot 10^4\ {\rm photons}/{\rm px}$ at 525 (517 nm). Keeping the same symbol after multiplying the intensity by the quantum efficiency of the detector, $S_{0, \rm c} = 2.93\ (2.5) \cdot 10^4\ {\rm electrons}/{\rm px}$, which is almost one third of the detector full well, $F_{\rm W} = 9 \cdot 10^{4}$ electrons. We decided not to fill the pixel well any further in order to avoid the so-called fixed pattern noise to be the largest contribution to the single exposure noise \citep[][]{2023FrASS..1067540O}. Measured in digital numbers (or counts; DN), $S_{0, \rm c} \simeq 1.7\ (1.4) \cdot 10^3\ {\rm DN}/{\rm px}$ after taking the analog/digital gain conversion of the SPGCam, $G = 0.053$ into account and an offset of $\sigma_{\rm off} = 150$ DN. If noise is added quadratically, the expected signal-to-noise ratio in the continuum is estimated in $(\rm{S/N})_{0} \simeq 170$ (160) for every single frame. 

      The different observing modes can modify the number of accumulations and of polarization states that are needed to estimate the final (S/N)$_{0, \rm c}$ for an individual data set. For instance, observing mode number 1 (see Sect. \ref{sec:observing}) uses $N_{\rm a} = 16$ frame accumulations in each of four polarization states. In this mode, the final expected signal-to-noise ratio for every data set is then $(\rm{S/N})_{0} \simeq 2000$ ($1800$) after using data from the two cameras, hence fulfilling the polarimetric sensitivity requirement.

%-------------------------------------- 
   \begin{figure}
   \centering
   \includegraphics[width=\columnwidth]{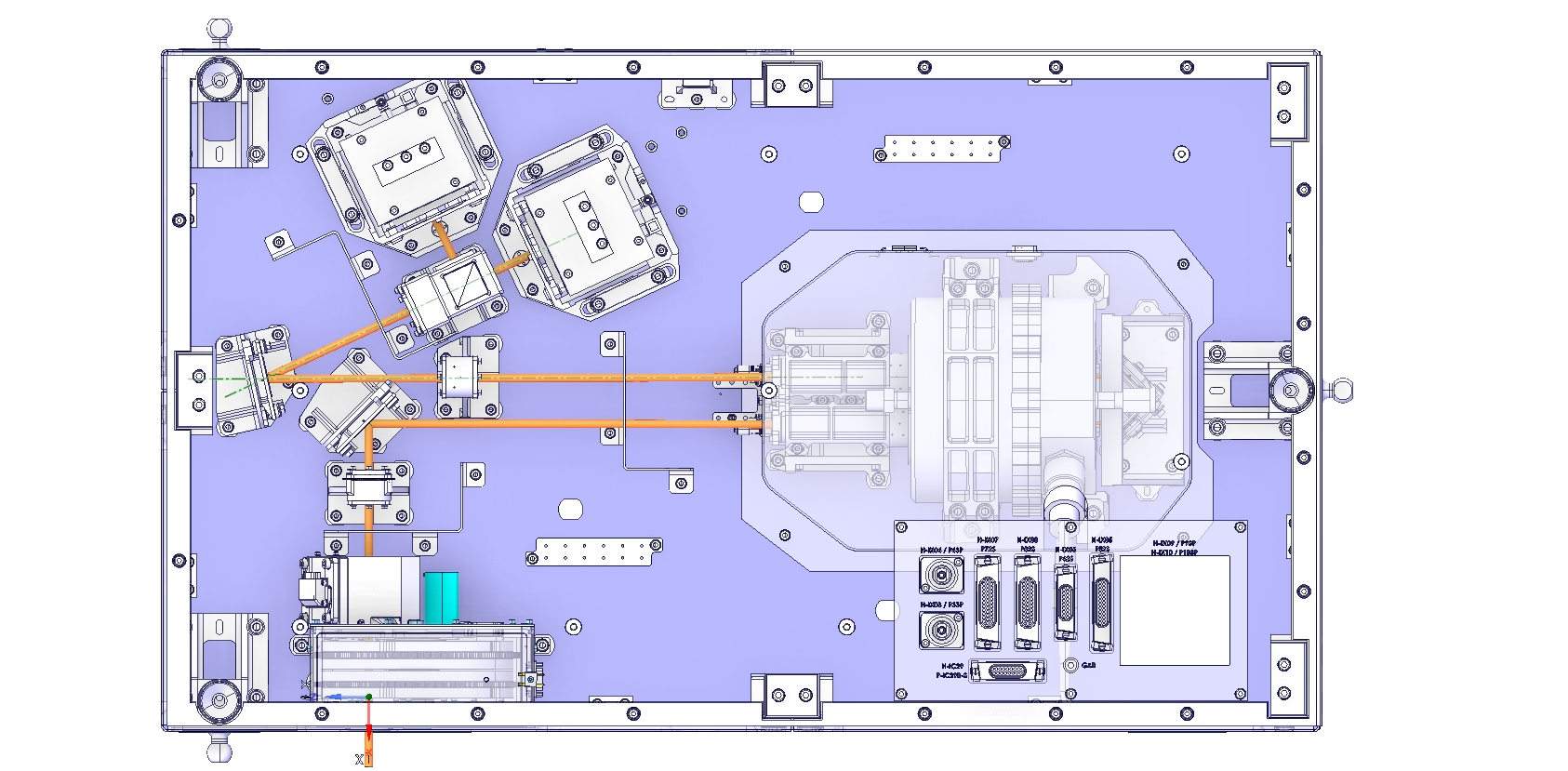}
   \caption{TuMag light path is marked in yellow (see text for details).}
   \label{fig:lightpath}
   \end{figure}

%-------------------------------------- 

   \subsection{Conceptual light and data paths}
   \label{sec:path}

      A sketch of the light path through the instrument is presented in Fig.~\ref{fig:lightpath}. The first element light encounters when reaching TuMag is a blocking filter and the filter wheel assembly, which hosts the pre-filters, among other components (see Sect.~\ref{sec:filterwheel}). In normal observing modes, light then impinges on the polarization modulation package (PMP; see Sect.~\ref{sec:pmp}). During calibration, light can go through the PD plate or other calibration devices before reaching the PMP. After it, a flat mirror drives the beam to the Fabry--P\'erot etalon. Once the etalon is traversed for the first time, a pair of flat mirrors at 45$^\circ$ redirect the light beam to the etalon once more in order to narrow down the filtergraph transmission profile (see Sect.~\ref{sec:etalon}). Another flat mirror brings the ray to the beam splitter where polarization is analyzed into its two orthogonal components, each of which are sent to the cameras (see Sect.~\ref{sec:cameras}).

      Data from the cameras are acquired by the DPU, where they are accumulated $N_{\rm a}$ times per polarization state and wavelength, accumulated, provided with headers, lossless compressed and sent to the {\sc Sunrise} instrument control system (ICS) for storage. Eventually, accumulated images can be cropped, binned, and/or lossy compressed according to the users' needs. Those operations are carried out by the DPU as well before sending the images to the ICS.

\section{Main subsystems in the optical unit}
\label{sec:subsystems}

   \subsection{Filter wheel and polarization modulation package assembly}
   \label{sec:filterwheel}

   Three different solar spectral lines can be tuned for the observation. This  can be done thanks to a double-disk filter wheel system (FW; \citeauthor{2022SPIE12188E..3AS} \citeyear{2022SPIE12188E..3AS}) designed to accommodate the three available narrow band filters (NBFs), one per spectral line, that are described in Sect.~\ref{sec:prefilters}. Each spectral line lies within the NBF spectral bandpass (FWHM $\simeq 0.1$\, nm) and is spectrally scanned by voltage tuning of the etalon (FWHM $\simeq 7$\, pm).\footnote{The final transmission profile of the filtergraph results from the product of these two filter transmissions and also broadened by other non-ideal effects like scattered light from secondary lobes of the etalon \citep[see][]{2011SoPh...268...57M}.}

   The central wavelength of the NBFs shows high thermal sensitivity ($\Delta \lambda/\Delta T \!\simeq 0.005$ nm/$^{\circ}$C) and incidence angle dependence. Consequently, the FW design has stringent requirements in terms of temperature stability and angular and position repeatability:
   \begin{itemize}
      \item The FW allows a tip/tilt mounting range for the various optical elements of up to $2\fdeg5$ with a tolerance of $\pm\, 0\fdeg 03$.
      \item The FW ensures position repeatability after switching between vanes with an accuracy better than $\pm\, 0\fdeg 01$ in tip/tilt, and an offset of 0.02 mm from the optical axis. 
      \item The FW provides a temperature stability of $\pm\, 0.5^{\circ}$C around their set-point for a continuous period of at least 12 hours.
      \item The FW minimizes the switching time (including the stabilization time) between consecutive vanes to less than 6 s (typically less than 5 s).
   \end{itemize}
    
\begin{figure}
   \centering
   \includegraphics[width=\columnwidth]{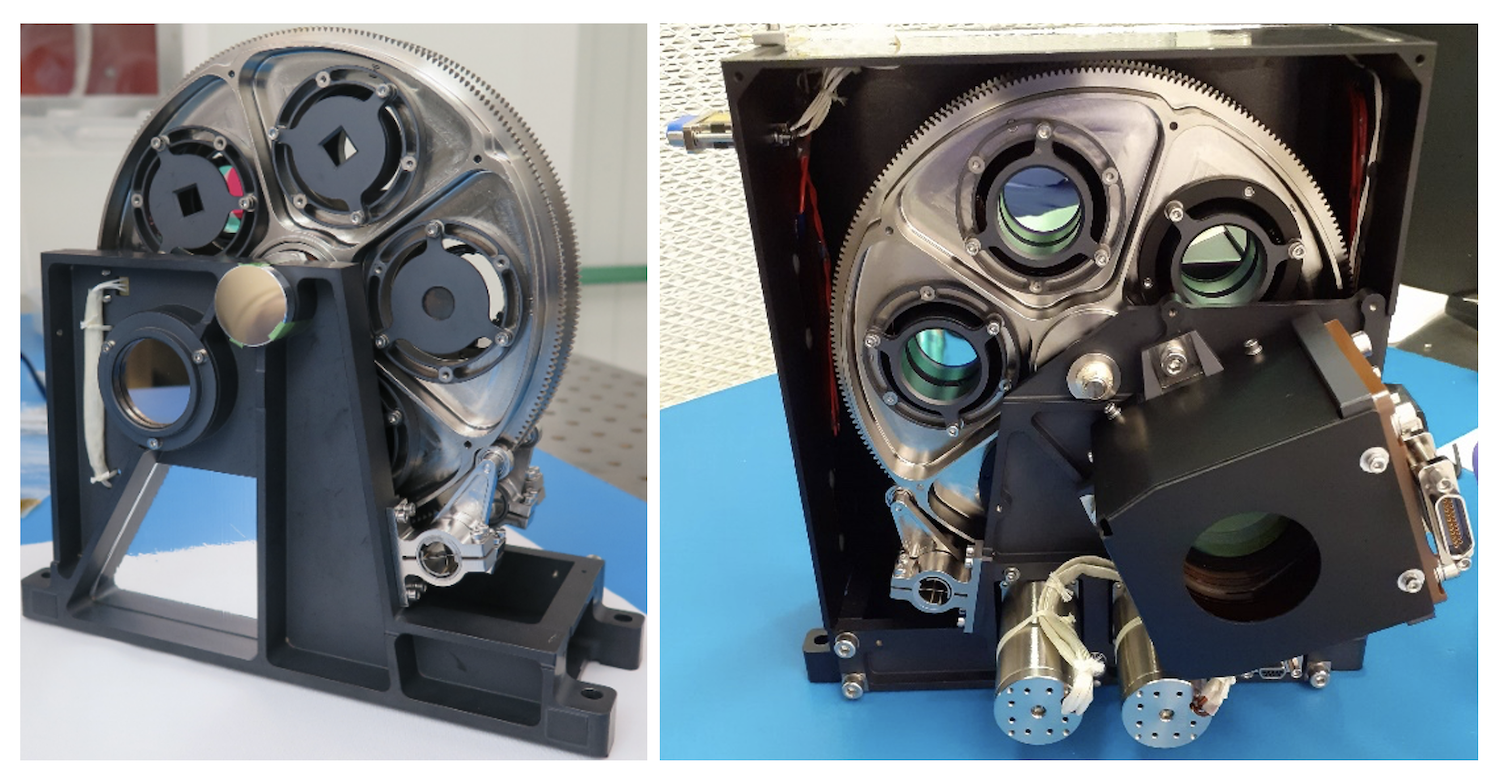}
   \caption{Left: Filter wheel assembly's front view (from disk 1) without thermal cover. The common blocking filter and an auxiliary reference mirror to be used during the unit alignment with the instrument can be seen. Right: FW rear view (from disk 2) with the PMP and NBFs.}
   \label{fig:fw}
   \end{figure}

\begin{figure}
   \centering
   \includegraphics[width=\columnwidth]{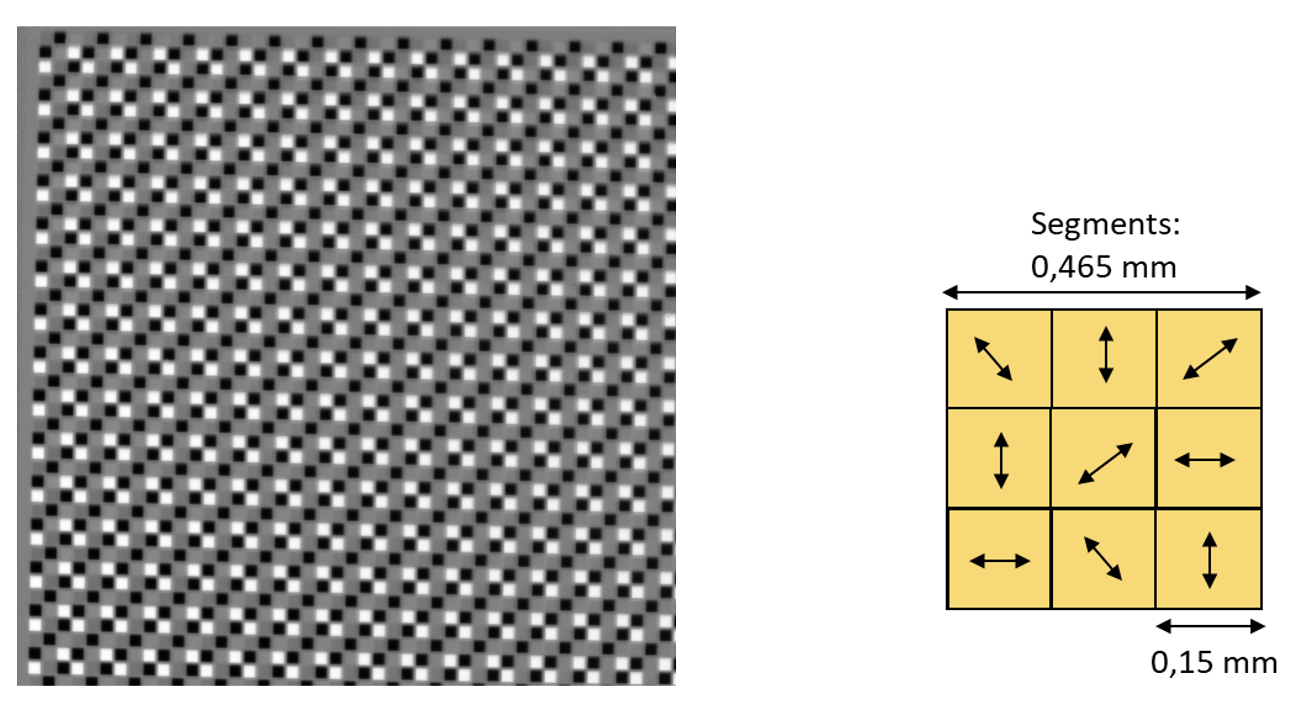}
   \caption{Left: Image of the micropolarizers calibration target using an unpolarized light source to illuminate the instrument. The 16 $\times$ 16 patches can be seen. Right: Individual patch scheme indicating the micropolarizer transmission axis.}
   \label{fig:micropol}
   \end{figure}

   Besides holding the interchangeable pre-filters, the FW provides further capabilities for image restoration, image quality diagnostic, and polarimetric calibration. It also includes elements for its alignment with the rest of the instrument and the telescope. They are all hosted in a double-disk wheel with 5 vanes per disk (see Fig.~\ref{fig:fw} and Table~\ref{tab:fw}). The F4$^{\prime}$ (square) field stop in vane 0 of disk \#1 can be used for checking the alignment in the laboratory with a dummy filter in vane -II of disk \#2, and for post-facto correcting any possible misalignment between the two camera images during data reduction.\footnote{Note that instead to the customary nomenclature F4 for the optical interface between the telescope and the instrument, here we write F4$^{\prime}$. The reason is easily understood in Sect.~\ref{sec:optics}, where the difference between the two foci is explained.} The dummy filter is just a transparent substrate similar to the NBFs that mimic the beam deviation produced by NBF$_3$. Vane 0 of disk \#1 is mainly used for regular observations (with vanes 0, I, and -I of disk \#2). The 11 mm thick plane-parallel plate (NBK7 Schott glass) for PD observations is used in vane I of disk \#1 together with vanes 0, I, and -I of disk \#2. As commented on in Sect.~\ref{sec:imagequality}, its width is equivalent to 1.5 $\lambda$. A set of pinholes are located in vane II of disk \#1 that are used for alignment in the laboratory (with vane II of disk \#2) and for PSF estimations (with vanes 0, I, and -I of disk \#2). Vanes -I and -II of disk \#1 are used with vanes 0, I, and -I of disk \#2 for polarization calibration purposes. The micropolarizers calibration target is illustrated in Fig.~\ref{fig:micropol}. It consists in 16 $\times$ 16 patches across the FoV, each made up of 3 $\times$ 3 elements with different orientations of linear polarization.

   The filter wheel assembly also accommodates a common blocking filter (see Sect. \ref{sec:optics}) and an alignment mirror on the front side (removed for flight) and the PMP (see Sect. \ref{sec:pmp} on the rear side. The whole FW assembly is enclosed in a thermal cover for thermal stability purposes (see Sect.~\ref{sec:thermal}).

%-------------------------------------- 

%-------------------------------------- 

%-------------------------------------- 

%-------------------------------------- 

 %---------------------------------------------------
   \begin{table}
      \caption[]{Filter wheel optical elements}
         \label{tab:fw}
      \begin{minipage}{\textwidth}

         \begin{tabular}{rll}
            \hline
            {\rm Vane}      &  {\rm Disk \#1}   &   {\rm Disk \#2}   \\
            \hline
            {0}  & {\rm F4}$^{\prime}$ {\rm frame (field stop)}  & {\rm NBF}$_1\!$: $\lambda_0 = 525.02$ {\rm nm}   \\
            {\rm I}  & {\rm Phase diversity plate plus F4$^{\prime}$ {\rm frame}}   &   {\rm NBF}$_2\!$: $\lambda_0 = 525.06$ {\rm nm}   \\
            {\rm II}   & {Pinhole set}   &  {\rm Empty}  \\
            {\rm -I}   & {\rm Linear polarizer plus F4$^{\prime}$ {\rm frame}}  & {\rm NBF}$_3\!$: $\lambda_0 = 517.27$ {\rm nm}   \\
            {\rm -II}   &   {\rm Micropolarizers calibration target}  &   {\rm Dummy filter}  \\
            \hline
         \end{tabular}

      \end{minipage}

   \end{table}

   \subsubsection{The TuMag pre-filters}
   \label{sec:prefilters}

\begin{figure}
   \centering
   \includegraphics[width=\columnwidth]{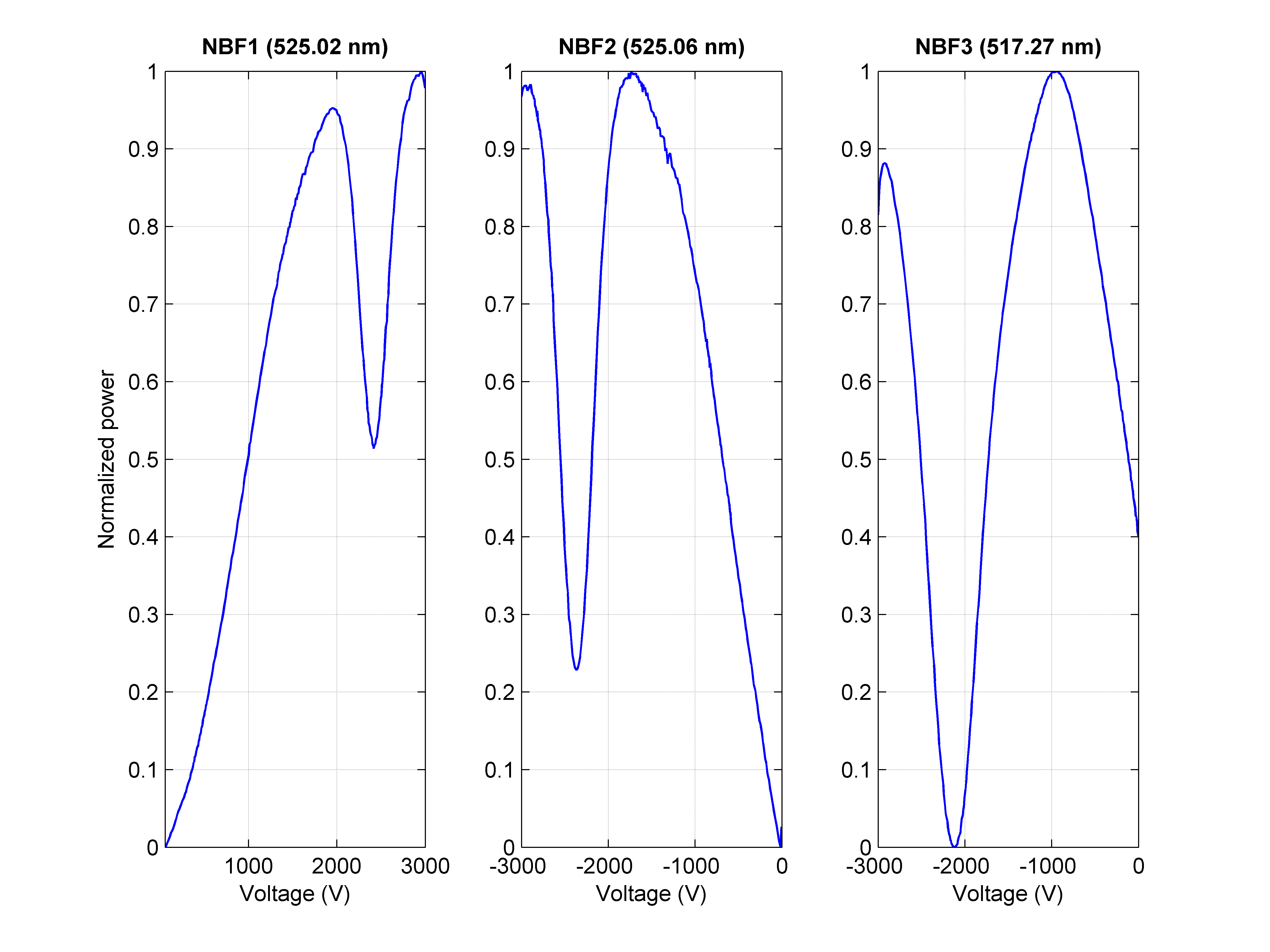}
   \caption{Voltage scan of the three NBF transmission profiles integrated in the FW at the working temperature. The three solar lines can be seen in normalized power.}
   \label{fig:prefilters}
   \end{figure}

   TuMag pre-filters were custom manufactured by Materion$^{\rm TM}$: a FWHM of $0.1 \pm 0.05$ nm, a peak transmittance better than 70\%, and a wavefront error better than $\lambda/10$ were required. They were also required to be centered at the rest wavelengths of the selected spectral lines and to have a thermal stability better than $0.005$ nm/°C, with a clear aperture of 20 mm. The final pre-filters fulfill these requirements and showed peak transmission of 88\%, 82\%, and 83\% at 525.02 nm, 525.06 nm, and 517.3 nm, respectively, and at zero-degree incidence angle. Their spectral profile was checked in the laboratory with solar light collected by a coelostat. The first tests were performed only with the FW \citep{2022SPIE12188E..3AS}; finally, the test was repeated once both the pre-filters and the blocking filter were mounted in the instrument \citep{2022SPIE12184E..2FA}. A scan of the solar line made by tuning the etalon allowed us to find the solar lines of interest. Plots of the three lines at an operating nominal temperature of 27$^{\circ}$C can be seen in Fig.~\ref{fig:prefilters}. The 525.06 nm line lies almost right at the maximum transmission, $\lambda_0$, of its pre-filter. Both the 525.02 nm line and the 517.27 nm line lie on the increasing slopes of their corresponding pre-filter transmissions. The strong dependence of $\lambda_0$ with the tip/tilt angle and temperature led us to accept these mounting positions. Similar measurements at several temperatures close to the nominal one were also made in order to have enough data for correcting the pre-filter transmission if needed. 

%-------------------------------------- 

%-------------------------------------- 

   \subsubsection{Polarization modulation package}
   \label{sec:pmp}

   Directly inherited from the SO/PHI instrument \citep{2018OExpr..2612038A,Solanki_2020}, the polarization modulation package is a very compact unit which modulates the state of polarization of the incident sunlight. It consists of two anti-parallel nematic LCVRs oriented with their fast axes at 45$^{\circ}$ with respect to each other. The PMP introduces four known modulation states allowing to deduce the Stokes vector of the incoming light \citep{2011SoPh...268...57M, Solanki_2020}. 

   The PMP includes an active thermal control system to obtain repeatable values of the optical retardances. The temperature set point is 35$^{\circ}$C to guarantee a response time shorter than 100 ms and a suitable temperature difference with the environment in order to assure the active control. That control allows a thermal stability of $\pm\, 0.5^{\circ}$C during acquisition. It is achieved by means of a PID controlled electronic driver that commands a pair of flat kapton heaters with a maximum power of 4 W, which are on the aluminum cell mounts, along with their temperature sensors (PT100). The external PMP structure is made of Vespel\textsuperscript{\textregistered} (SP1), providing high thermal insulation for the whole package and the required mechanical features for the expected loads.

   The polarimetric analysis is completed with the polarizing beam splitter, which acts as a double analyzer in front of the cameras.

   \subsection{The TuMag Fabry--P\'erot etalon}
   \label{sec:etalon}

   Thanks to the electro-optical effect, a LiNbO$_3$ etalon is used as a tunable filter to scan the solar line by applying high voltage [-4000\, V, +4000\, V] thus avoiding any mobile part. As a counterforce, precautions have to be taken in order to prevent electrical discharges since the flight pressure conditions are close to 3 mbar, where the maximum likelihood for air ionization occurs in the atmosphere. This element is direct heritage of IMaX. Indeed, the subsystem is a spare device of that instrument. In the same way as IMaX, the Fabry--P\'erot interferometer is in a collimated beam \citep[][and references therein]{2023Ap&SS.368...55B} and in double-pass configuration \citep{2006SPIE.6265E..2GA} to improve the spectral resolution. The etalon has a thermal control better than $\pm\, 0.05^{\circ}$C.

   \subsection{The TuMag cameras}
   \label{sec:cameras}
      As mentioned in Sect.~\ref{sec:intro}, specific, custom cameras have been designed and manufactured for both the TuMag and SCIP instruments of {\sc Sunrise iii} \citep[SPGCams;][]{2023FrASS..1067540O}, based on the GPIXEL SENSE GSENSE400BSI sensor. In both instrument cameras, the detector has $2\, {\rm k} \times 2\, {\rm k}$ pixels, which, in the case of TuMag, image a FoV of $63\arcsec \times\, 63\arcsec$. The sensor works in continuous, rolling shutter configuration. The scientific and system-engineering requirements for the TuMag cameras are summarized in Table~\ref{tab:tablecam}.

 %---------------------------------------------------
   \begin{table}
      \caption[]{Main camera scientific and system-engineering requirements}
         \label{tab:tablecam}
      \begin{minipage}{\textwidth}

         \begin{tabular}{llll}
            \hline
            {\rm Name}      &  {\rm Symbol}   &   {\rm Value}   &  {\rm Units (Type)}\footnote{Units are per pixel where necessary.} \\
            \hline
            Sensor's quantum efficiency   &   $Q$   &   $\geq 0.8$   \\
            {\rm Pixel full well}   &  & $\geq 6 \cdot 10^{4}$   &  e$^-$  \\
            Photon flux budget  & $F$  & $\geq 3 \cdot 10^{4}$   &   e$^-$   \\
            Photon response non-uniformity  & $f_{\rm prnu}$  & $< 0.01\, F$  &   \\
            Digital depth  &   &  12 & bit \\
            Pixel area  &    &   $11 \times 11$ & $\mu$m$^2$\\ 
            Effective collecting area   &      &   $2048 \times 2048$   & pixels   \\
    %        Sensor's built-in processing   &    &   No   \\
            Configurable readout speed  &  & $\leq 48$ & frames$\,{\rm s}^{-1}$  \\
    %        Configurable exposure time  & & Yes \\
    %        Configurable pixel readout gain & & Yes \\
            Readout noise   & $N_{\rm r}$  & $<10$   &  e$^{-}$  \\
            Fixed pattern noise\footnote{For the specified photon flux.}   &   $N_{\rm dc,fpn}$   &   $<100$   & e$^{-}$ \\
            Dark current noise at 20$^{\circ}$C   & $N_{\rm dc}$ & $< 50$   &   e$^{-}$  \\
    %        Configurable ROI area\footnote{ROI stands for region of interest.}   &      &   Yes  \\
    %        Externally triggered image capture   &   &  Yes   &  CoaXPress\footnote{Digital interface for high-speed data transmission.}   \\
            Maximum camera envelope   &     &   $100 \times 80 \times 73$ & mm$^{3}$  \\
            Optical entrance window   &    &  Yes  & Non-removable   \\
            Housing material   &    &   Black anodized   &    Al 6082-TL   \\
            Thermal interface   &     &   Yes   &   Cold finger   \\
    %        Overall camera weight &  & 1 & kg \\ 
    %        Housing electrical insulation\footnote{From camera electronics.}  &  & $> 1\, {\rm M}\Omega$ \\
    %        Sensor's surface roll$^f$ & & $<1^{\circ}$ &  \\
    %        Sensor's surface tip/tilt$^f$ & & $<1^{\circ}$ &  \\
    %        Adjustable sensor's tip/tilt & & Yes \\
    %        Camera's FPGA programming port & & Yes \\
    %        Camera's FPGA debugging port & & Yes \\
            CoaXPress coaxial connector & & Yes & DIN 1.0/2.3 jack \\
    %        Thermal control power connector & & Yes \\
    %        Near-to-vacuum operations & & Yes \\
    %        Camera operational temperature\footnote{Reference at the housing.} & & $[-20,40]^{\circ}$C \\
    %        Sensor's configurable thermal control & & Yes \\
            Default sensor's operational temperature & & 20.0 $\pm$\, 0.5 & $^{\circ}$C \\
            Power consumption & & $< 6.9$ & W  \\ 
            \hline 
         \end{tabular}

      \end{minipage}

   \end{table}

%-------------------------------------- 
   \begin{figure}
   \centering
   \includegraphics[width=\columnwidth]{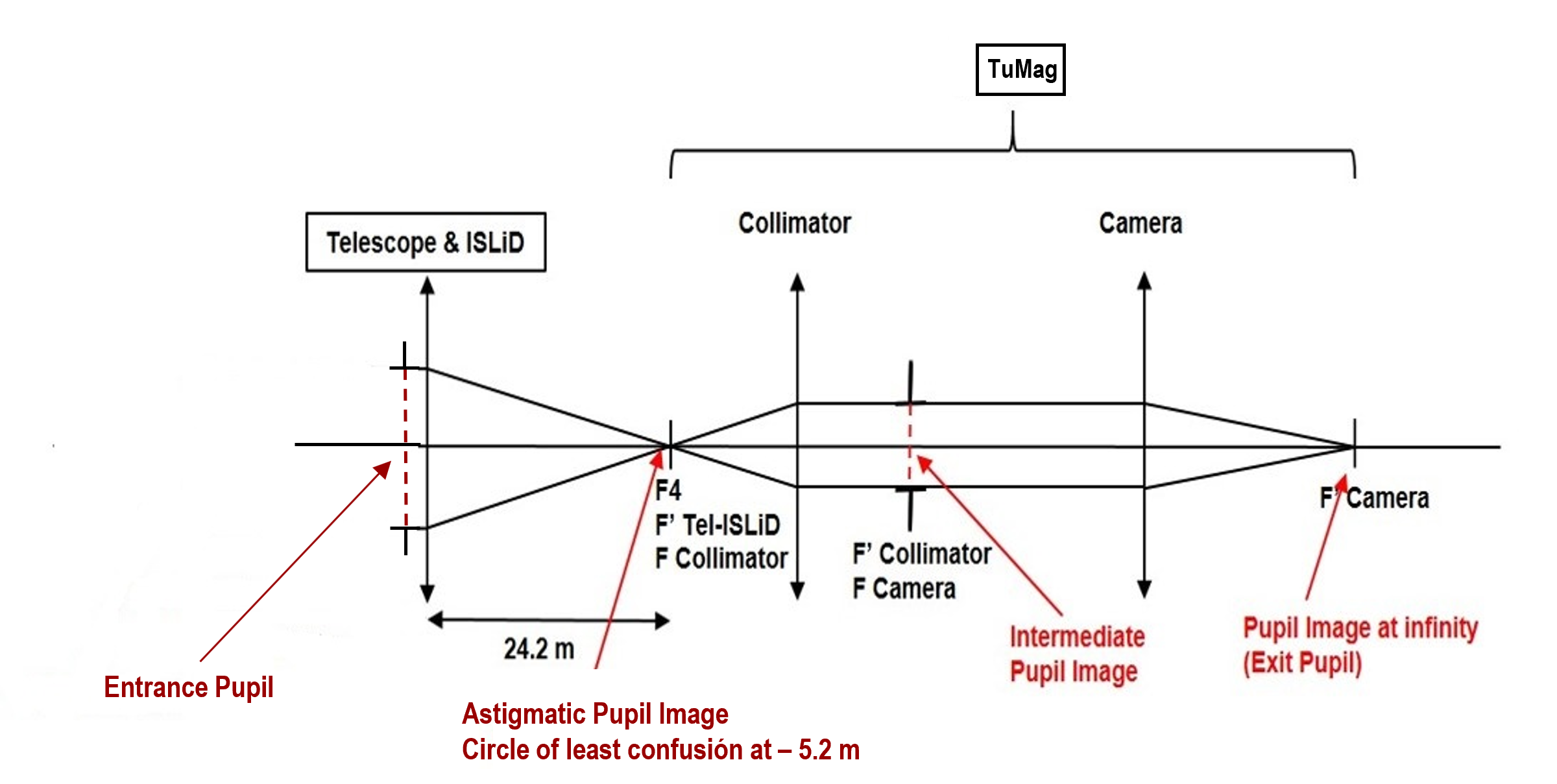}
   \caption{Optical block diagram of the whole system from the telescope entrance pupil to the final image on one of the TuMag's two cameras.}
   \label{fig:optblockdiag}
   \end{figure}

%-------------------------------------- 

\section{Optics and opto-mechanics}
\label{sec:optics}

\begin{figure}
   \centering
   \includegraphics[width=\columnwidth]{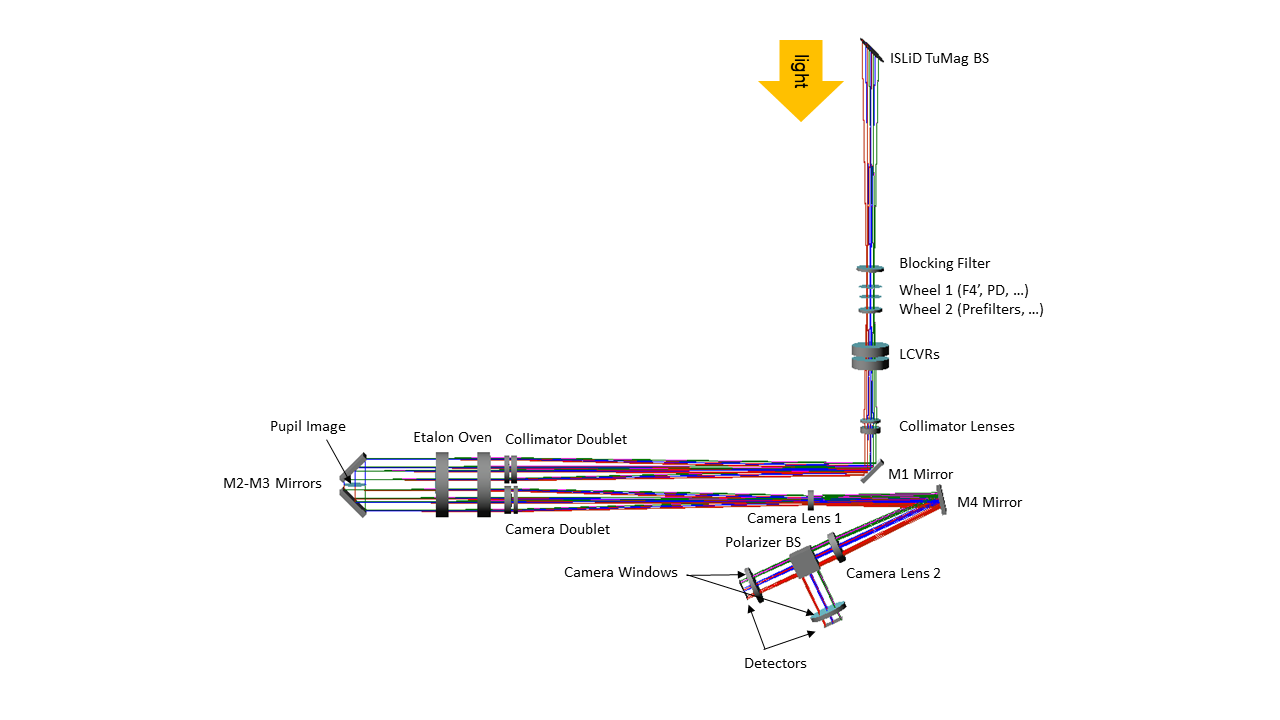}
   \caption{TuMag setup and ray tracing scheme.}
   \label{fig:optdetail}
   \end{figure}

   Following IMaX heritage, TuMag is an optical relay system with a polarization dual-beam configuration. It re-images the {\sc Sunrise} telescope plus ISLiD system image into two SPGCam cameras each with orthogonal, linearly polarized beams. A block diagram of the whole optical system, from the telescope aperture to the final image in one of the two cameras is shown in Fig.~\ref{fig:optblockdiag}. The {\sc Sunrise} telescope plus ISLiD system (focal length of 24.2 m) supplies the F4 optical interface and an astigmatic pupil image whose circle of least confusion is 5241.5 mm away from F4. The common blocking filter with a wide bandpass ($\simeq 46$ nm) and a central wavelength of 520 nm is placed in front of the F4 focus interface (see Sect. \ref{sec:prefilters}) in order to reject the unwanted spectral range. The filter produces a displacement of 0.95 mm in F4. The new focus position is called F4$^{\prime}$ (see Sect. \ref{sec:filterwheel}) in order to avoid misunderstandings. TuMag consists of a collimator system, whose F-number matches that of the telescope plus ISLiD, and a camera system that finally focuses the beam into the SPGCam cameras. An intermediate pupil image is formed between the collimator and the camera systems. Right there, an aperture stop is located, slightly oversized for AIV purposes. The optical system can be seen in more detail in Fig.~\ref{fig:optdetail}.

%-------------------------------------- 

%-------------------------------------- 

   The optical system is refractive except for four mirrors used for folding and packaging purposes. Light enters from an ISLiD beam splitter and finds the blocking filter, the FW, and the PMP. Each NBF has its own working tilt angle: $2\fdeg 06$ for 525.06 nm, $1\fdeg 71$ for 525.02 nm, and $0\fdeg 97$ for 517.27 nm. Additionally, the blocking filter is tilted by $1\fdeg 52$ but in the opposite direction to the NBFs in order to avoid retro-reflections to the ISLiD. After the LCVRs in the PMP, the collimator system consists of two lenses and a doublet with 549.91 mm focal length and 24.2 F-number, matching the \sunriset telescope and ISLiD F-number. In the collimated space, the solid, LiNbO$_3$ etalon is enclosed in an oven to ensure both thermal and pressure stability. The Fabry--P\'erot interferometer works in double pass, i.e., the light goes through the etalon, is retro-reflected by a system of two mirrors and passes back again through the etalon. Note that the already mentioned intermediate pupil image and aperture stop are located right in the middle of the M2 and M3 folding mirrors. The beam is finally focused by a camera system consisting of a doublet and two lenses. The camera system focal length is 1356.7 mm and produces an image with an F-number of 60 and with $0\farcsec 0378$/pixel image scale. The achromatization and athermalization condition for the doublets has been taken into account during the optical design phase. 
   
   A polarizing beam splitter divides the light into its two orthogonal components of linear polarization that are sent to the cameras. TuMag works with a lineal field of view (FoV) of $7.42\, \times\, 7.42$ mm$^2$ (at F4$^{\prime}$ focus) and a magnification of 2.47, covering up to $1666\,\times\, 1666$ pixel$^2$ on the detector, equivalent to a FoV of $63^{\prime\prime}\,\times \, 63^{\prime\prime}$ over the Sun.

%-------------------------------------- 
   \begin{figure}[ht]
   \centering
   \includegraphics[width=\linewidth]{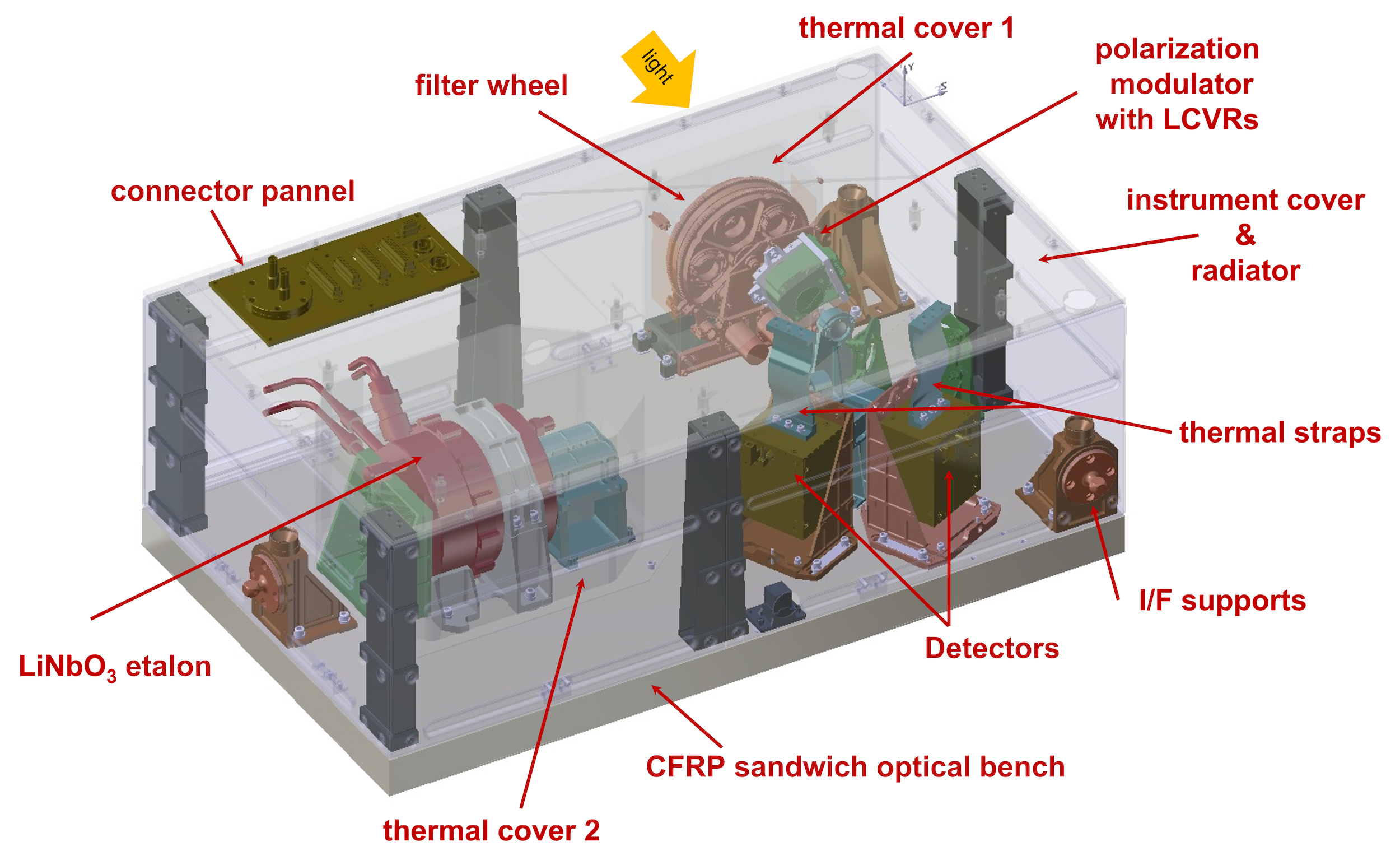}
   \caption{O-Unit's design general view.}
   \label{fig:ounit}%
   \end{figure}
%-------------------------------------- 

   The TuMag O-Unit's mechanical structure consists of a composite baseplate on which the different opto-mechanical assemblies are located. These assemblies typically consist of supports and mounts that hold the optics (lenses, mirrors, etc.) in position with respect to the optical axis. Ti6Al5V has been mostly used in these parts due to its low thermal expansion coefficient (TEC) that minimizes the thermo-elastic deformations due to temperature variations. This TEC is also close to that of the optics, hence avoiding induced stresses by dissimilar strains. The titanium parts are finished with Astroblack\textsuperscript{\textregistered}, a dark ceramic coating that avoids stray light reflections. Optical elements are bonded to their mounts using Mapsil QS-1123 adhesive. The mechanical mountings provide adjustment capabilities in the plane perpendicular to the optical axis (cameras), in the optical axis (doublets, collimator lens), and in tip/tilt (folding mirrors) either by shimming or by pre-loaded spring mechanisms.

   The main structural part of the O-Unit is the composite optical bench. It consists of a sandwich panel made up of two carbon-fiber-reinforced-polymer (CFRP) skins of 16 Prepreg Torayca M40J-MTM44-1 plies with the following stack-up: ($[+45/90/-45/0]_{\rm 2s}$), and a 40 mm aluminum (AL5056) honeycomb core. The resultant sandwich features excellent properties in terms of mass saving and thermo-elastic stability \citep{2023Senso..23.6499F}. Aluminum inserts are bonded (Hysol\textsuperscript{\textregistered} EA 9394) to the raw sandwich panel and subsequently mechanized and threaded to serve as mounting points for the opto-mechanical assemblies. They are located within a position tolerance of 0.1 mm (in radius) and feature a flatness better than 0.04 mm, which allows accurate mounting of the different assemblies. A general view of the O-Unit can be seen in Fig.\ \ref{fig:ounit} and further details of its design can be found in \cite{2022SPIE12188E..3AS} and \cite{2022SPIE12184E..2GA}.

\section{Electronics and firmware}
\label{sec:electronics}
   
Environmental conditions for a stratospheric flight are very harsh. In some regards, like those related to high voltage (needed to tune the etalon), conditions are even harsher than vacuum since the minimum of the Paschen curve for air \citep{1889AnP...273...69P} is about the ambient pressures of a few millibars. Since the budget for this type of mission is only a fraction of a space-borne one, we use a design based on COTS components. The electronics, thus, has to be enclosed in a pressurized vessel under temperature control. The reason is twofold: on the one hand, we can use commercial devices without any concern about near-vacuum behavior; on the other hand, we can use a commercial high-voltage power supply (HVPS). The development time is another important factor to take into account: the use of  general-purpose, modern components ---not qualified for space---, together with state-of-the-art software tools, sharply reduces the engineering efforts.

\begin{figure}[ht]
   \centering
   \includegraphics[width=\linewidth]{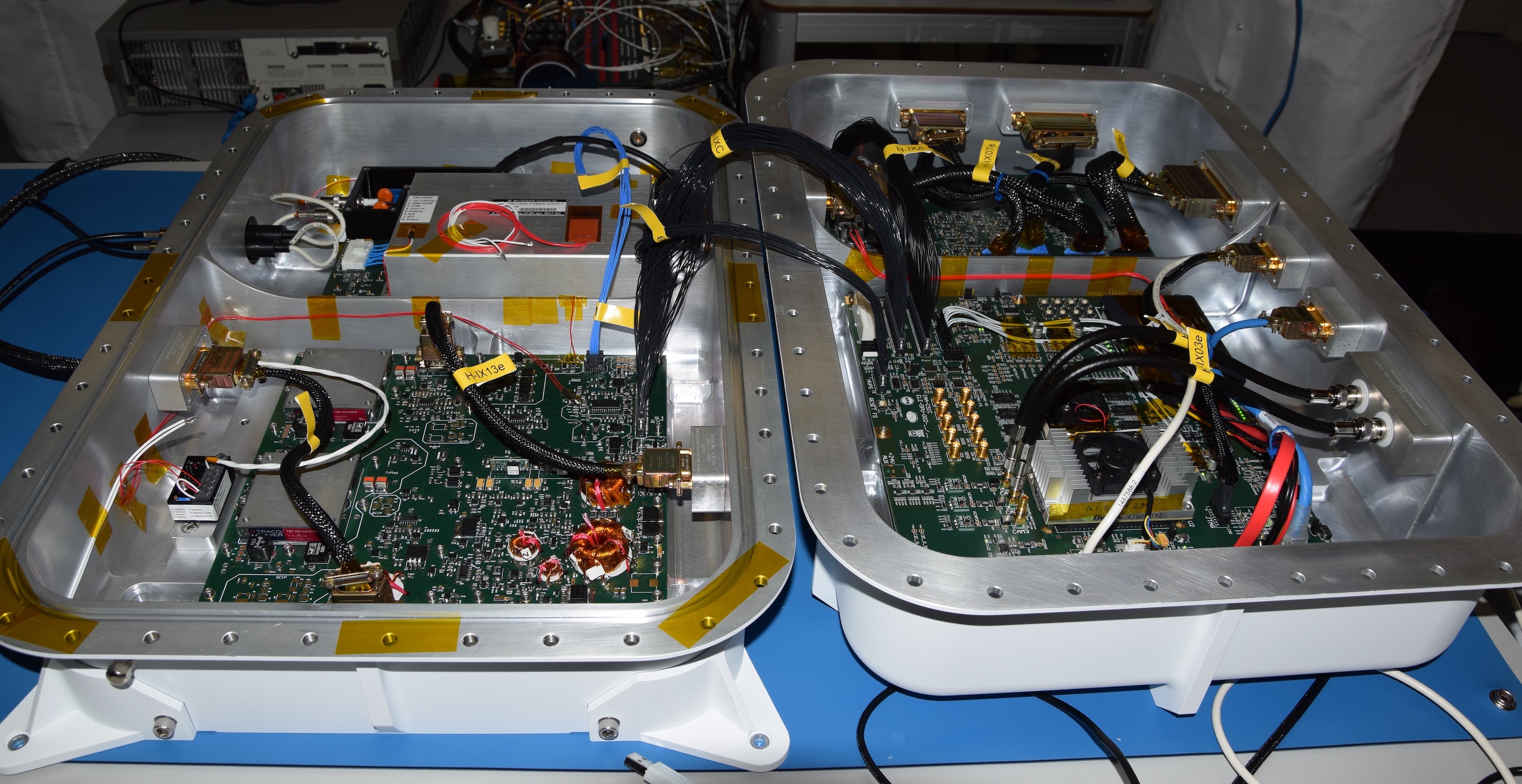}
   \caption{TuMag's flight electronic unit. The left half contains the PCM (bottom) and HVPS (top) boards. The right half contains the DPU (bottom) and the AMHD (top) boards.}
   \label{fig:E-Unit}%
   \end{figure}

Following SO/PHI's design, the electronic components are placed in four specifically tailored PCBs (see Fig.~\ref{fig:E-Unit}): the digital processing unit (DPU); the analog, mechanisms, and heaters driver board (AMHD); the power converter module (PCM); and the high-voltage power supply (HVPS).

%-------------------------------------- 
%-------------------------------------- 

   \subsection{The DPU board}
   \label{sec:dpu}
   
\begin{figure}
   \centering
   \includegraphics[width=\linewidth]{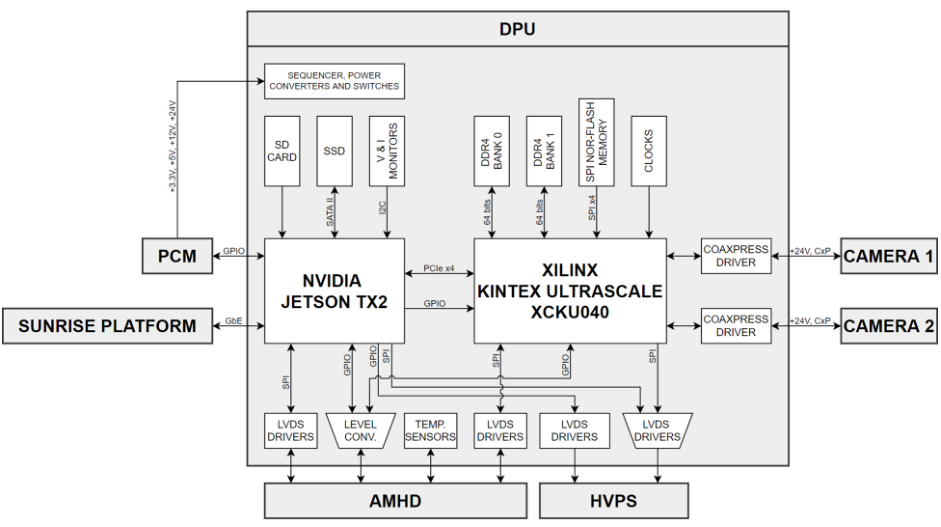}
   \caption{Functional block diagram of the DPU board. The central blocks represent the system controller (SyC; NVIDIA\textsuperscript{TM} Jetson TX2) and the frame grabber (FG; Xilinx\textsuperscript{TM} Kintex Ultraescale).}
   \label{fig:DPU}%
   \end{figure}

   The data processing unit (DPU) board includes two main blocks: a system controller (SyC) and a frame grabber (FG), which also acts as an image processor. Both devices are connected using a high-bandwidth peripheral component interconnect express (PCIe) bus. A block diagram of the DPU can be seen in Fig.~\ref{fig:DPU}. The SyC contains an ARM\textsuperscript{\textregistered} multiprocessor architecture and a tailored Linux-based operating system. It is suitable for carrying out tasks like image storage management, instrument status monitoring by housekeeping procurement, and communications with the ICS and with the ground support software. The FG has to carry out intensive image processing tasks such as acquisition, accumulation,
   %polarization demodulation \textbf{only for SCIP!}
   and lossless compression of the images. A number of other blocks surround these two main ones and receive, input, or interface them with labeled signals. Their functions are briefly described in the following paragraphs.

%-------------------------------------- 
%-------------------------------------- 

   \subsubsection{The System Controller (SyC)}
   \label{sec:sc}
   
   The system controller is in charge of booting the system, managing the instrument configuration, reading the images from the FG, generating thumbnails for selected images, collecting status telemetry data (also called housekeeping, HK) from all its internal subsystems, carrying out watch-dog tasks, launching contingencies when  necessary, and communicating with the {\sc Sunrise} instrument control unit.

   The SyC is based on an NVIDIA\textsuperscript{TM} Jetson TX2, embedded multi-media-card (eMMC) device with 32 GB that includes the operating system and all the specifically designed software (see Sect.~\ref{sec:control_software}). An external SSD of 8 TB from Samsung works as a communication buffer. This external SSD is connected to the SyC by a serial advanced technology attachment (SATA 2) interface bus. The NVIDIA\textsuperscript{TM} Jetson TX2 has the following features:

   \begin{itemize}
      \item 8 GB, 128-bit low-power double data rate (LPDDR4) memory running at 1866 MHz,
      \item eMMC internal storage of 32 GB,
      \item power consumption between 7.5 W and 15 W,
      \item one Gigabit-Ethernet interface,
      \item one 4-line PCIe (generation 2) bus at 5 Gbps,
      \item one SATA 2 bus,
      \item Quad-Core ARM\textsuperscript{\textregistered} Cortex\textsuperscript{\textregistered}-A57 MPCore.
   \end{itemize}
   
   The SyC communicates with the {\sc Sunrise} ICS through the Gigabit-Ether\-net interface, with the PCM with a general purpose input/output (GPIO) interface, with the FG with the PCIe interface, and it inputs the latter subsystem through a general purpose input/output interface. A further secure digital (SD) card was included for eventual needs, which is not finally used. Voltages and currents are sampled by ADC chips and read by the SyC through an inter-integrated circuit (I2C) serial data bus. The SyC communicates with the AMHD through a serial peripheral interface (SPI), employing low-voltage differential signaling (LVDS) drivers to convert the digital signals. Those LVDS drivers inputting signals to the HVPS receive the signals through both a GPIO and an SPI interfaces. The SyC also uses a GPIO interface in the communication with the AMHD, employing voltage level converters. The temperature sensors are sampled by the AMHD and then read by the SyC within the HK information packet. 
       
   \subsubsection{The Frame Grabber (FG)}
   \label{sec:fg}
   %The frame grabber carries out the image acquisition from the two scientific cameras while keeping them synchronized with other O-Unit subsystems. The  observations generate some raw data rates which cannot be absorbed by the system. Hence, an on-board data reduction is mandatory. The FG is also responsible for this.

\begin{figure}
   \centering
   \includegraphics[width=0.7\linewidth]{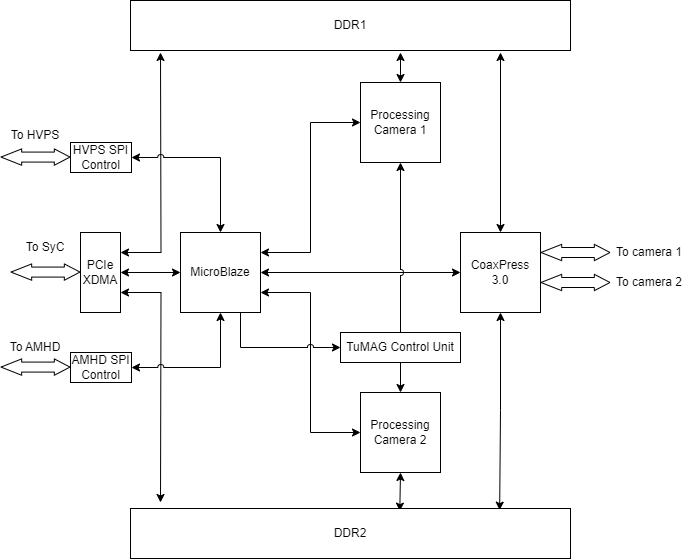}
   \caption{Functional block of the FG internal design.}
   \label{fig:fpgainter}
   \end{figure}

   The frame grabber carries out the image acquisition and processing from the two scientific cameras while keeping them synchronized with other O-Unit subsystems. Implemented on a Xilinx Kintex\textsuperscript{\textregistered} Ultrascale XCKU040-2FFVA1156I FPGA device, it includes the logic needed for interfacing the cameras, processing the images, and transferring the data stream to the SyC. A block diagram of its internal design is shown in Fig.~\ref{fig:fpgainter}
   %The FG design is implemented on a Xilinx\textsuperscript{TM} Kintex Ultrascale XCKU040-2FFVA1156I FPGA device. The design includes the logic needed for interfacing the cameras, processing the images, and transferring the data stream to the SC. 
   The brain of the system is a 32-bit MicroBlaze\textsuperscript{TM} microprocessor, responsible for configuring and monitoring all peripherals connected to the main bus, based on AMBA\textsuperscript{\textregistered} AXI4. Cameras are controlled through a CoaxPress 3.0 Host IP soft-core from EASII\textsuperscript{TM} that implements two independent lines. This configuration features two full-duplex links that perform downstream at 3.125 Gbps for image acquisition and HK, and upstream at 20 Mbps for configuring and triggering the cameras. 
   
%-------------------------------------- 
%-------------------------------------- 

   Streams of images from both cameras are processed in two processing blocks with the aim of reducing the resulting data rate. That processing pipeline is identical for both cameras and includes image accumulation and standard image compression. These processing tasks are backed by two external SDRAM DDR4 memory banks, essential for buffering intermediate and final processing results. Each bank is connected to the frame grabber through a 64-bit data bus and is controlled with the Xilinx\textsuperscript{TM} memory controller generator IP core. Each processing pipeline is connected to a different DDR bank, which guarantees total parallel processing.
   
   On the FG side, the PCIe interface with the SyC is controlled by a Xilinx\textsuperscript{TM} PCIe XDMA IP core. With this configuration, the FG is mapped as a slave in the SyC’s memory map so that the SyC can send commands to the FG and read captured images buffered in the external DDR4 memories.
   Two SPI interfaces are devised for commanding and monitoring optical subsystems. One of them is used to directly command the digital-to-analog converter (DAC) controlling the high voltage power supply (HVPS) that actuates for tuning the etalon. The other SPI interface is connected to the AMHD card, which interprets SPI commands and translates them to either the filter wheel motors or the LCVR DACs.
   
   A synchronization box, called TuMag control unit, is present in the FPGA architecture for keeping the synchronization between O-Unit subsystems and cameras during observations. It properly generates camera triggers combined with interrupt signal targeting the microprocessor, which, upon an event, generates SPI commands targeting the etalon, the filter wheels, or the LCVRs. All these logic and interfaces enable the FG to perform spectropolarimetric observations autonomously. A simple command from the SyC, containing an observing mode code, triggers a synchronized sequence of commands targeting the filter wheel, the etalon, the LCVRs and  camera triggers, as described in paragraph~\ref{sec:obsfirmware}.

   \subsection{The AMHD board}
   \label{sec:amhd}
   
%-------------------------------------- 
   \begin{figure}
   \centering
   \includegraphics[width=\linewidth]{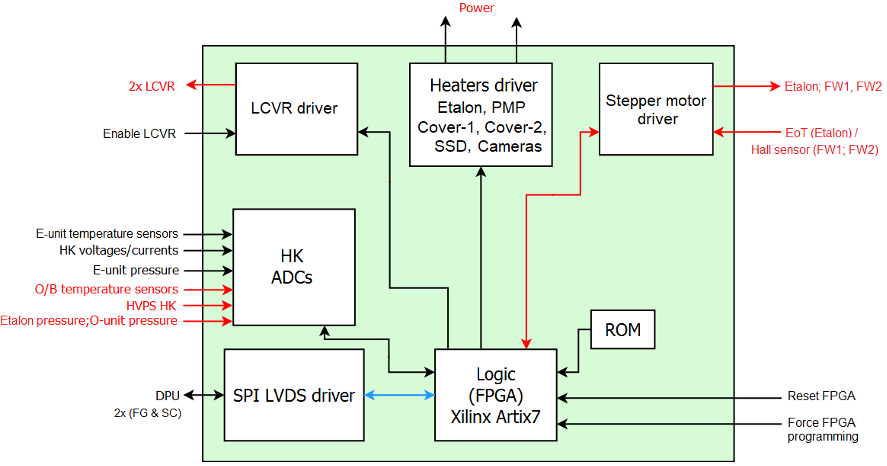}
   \caption{Functional block diagram of the AMHD board.}
   \label{fig:AMHD}%
   \end{figure}
%-------------------------------------- 

    The analog, mechanism, and heater driver (AMHD) interfaces the O-Unit as well as several boards in the E-Unit to acquire HK data, to generate the LCVRs driving voltage, to drive the mechanisms, and, finally, to stabilize the instrument temperatures by controlling the heaters. An AMHD functional diagram is shown in Fig.~\ref{fig:AMHD}. From top to bottom and from left to right, several blocks can be distinguished, namely, (i) the LCVR driver, (ii) the housekeeping analog-to-digital converters (ADCs), (iii) the driver controlling the interfaces with the DPU, (iv) the heaters' driver, (v) the logic ``brain'' of the board, (vi) the stepper motor driver, and (vii) a block of read-only memories (ROM):

    \begin{enumerate}
        \item To drive the LCVRs of the PMP, the AMHD generates two amplitude-modul\-ated, square signals of 2 kHz after receiving the `enable' signal from them. 
        
        \item The housekeeping block is in charge of driving and reading the different sensors that provide an overview of the instrument behavior. Through this functional block, the AMHD acquires voltages and current consumption of the subsystem power supplies, with a specific line for the high-voltage power supply (HVPS); a bunch of temperatures from the E-Unit and the O-Unit; and pressures of the E-box and the etalon.
        
        \item The interfaces between the DPU and the AMHD consist in two SPI interfaces and several logic lines. 
        
        \item The AMHD turns on/off the heater power supplies for the two cameras. (Their temperature is internally regulated by its own camera electronics.) The AMHD also regulates the temperature of five subsystems, four in the O-Unit (etalon, PMP, two thermal covers over the filter wheel and the etalon, and the optical bench) and one in the E-Unit (the DPU solid state drive card). Depending on the required level of stability, two types of controller algorithms are implemented. Thermostat controllers are dedicated to low temperature stability needs (optical bench and DPU solid state drive card) and PID  controllers for high stability (the remaining four). The PID output provides the duty cycle of a pulse-width modulation signal (100 Hz) that drives the heater.
     
        \item The AMHD includes an FPGA (Xilinx\textsuperscript{TM} Artix 7) to handle all the tasks. Once the DPU configures the FPGA (through either the system controller or the frame grabber, depending on the instrument functional mode) it operates autonomously or on demand. 
    
        \item As for the mechanisms, the AMHD receives information on the status of the Hall sensor of the filter wheels and the end-of-travel of the etalon mechanism. It also drives two bipolar stepper motors for moving the filter wheels and a unipolar stepper motor that controls the tilt of the etalon.
       
        \item The firmware for the FPGA is stored in a ROM memory and loaded into the FPGA once the AMHD is powered on. In case of need, this ROM can be reprogrammed in the laboratory, but not during the flight.
    \end{enumerate}

      \begin{figure}
         \centering
            \includegraphics[width=0.8\textwidth]{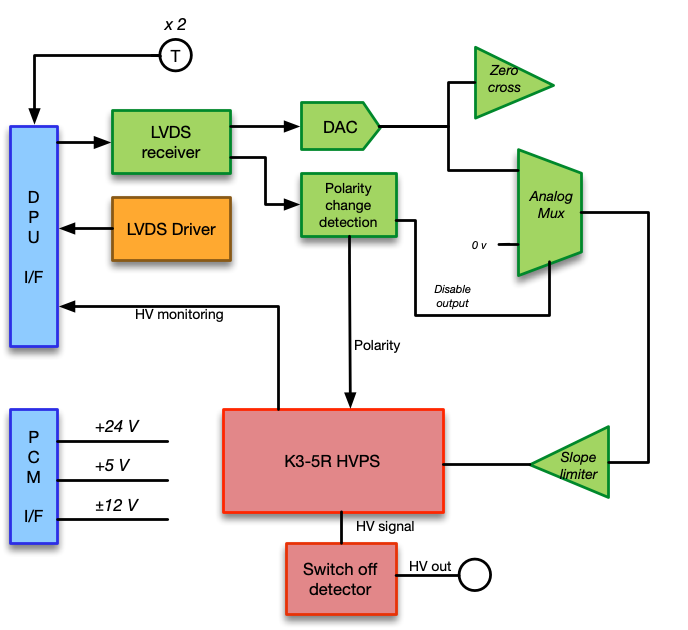}
        	\caption{HVPS block diagram.}
        	\label{fig:HVPS}
      \end{figure}
   
   \subsection{The HVPS board}
   \label{sec:hvps}
      The high voltage power supply board is in charge of supplying high voltage to the etalon in order to tune the different wavelength samples during the spectral line scan. Based on a commercial module, the K3-5R from Matsusada Precision Inc., the HVPS is able to provide either positive or negative voltages between 0 V to 5000 V, with a maximum current of 600 $\mu$A, enough to drive the etalon. Figure~\ref{fig:HVPS} shows a  block diagram of the board. The K3-5R supply is a commercial bipolar one with a good linearity in the full range (including the zero crossing). It has a polarity input transistor-transistor logic level to select positive or negative voltages. Its input low-level voltage ranges from 0 through 10 V. Its monitoring output, also with a $\pm \,10$ V range, shows a proportional low-level signal to the voltage applied to the etalon.

      Within the E-Unit, it has interfaces with the PCM and DPU (blue blocks in the figure). In the O-Unit it has an interface with the etalon (using custom designed high voltage cables; output labelled HV at the bottom). The board is powered by the PCM with + 24 V, + 5 V and ± 12 V signals. It has a typical consumption of around 5 W, with a maximum of 8 W.  The DPU is responsible of controlling the HVPS using  one unidirectional SPI interface (through LVDS signals), with an extra line to assure that the high voltage is applied to the etalon, and a polarity line. The SPI controls a DAC a full output range from 0 to 5 V. An auxiliary output LVDS driver is also included, which is not finally used. Since slopes greater than 1500 V/s are dangerous for the etalon integrity, a circuitry to avoid an uncontrolled change in the voltage is included. It consists mainly of two parts: 
	  \begin{itemize} 
	     \item The polarity change detection: this circuitry detects any voltage polarity request, allowing it only when the high voltage is close to 0 V, using the zero crossing detector. Otherwise, the high voltage drops to zero and stays at this level until the DPU changes the polarity in a controlled way through an analog multiplexer with one of its inputs set to zero. 
	     \item The slope limiter: the voltage slope is limited by hardware at 1500 V/s, hence avoiding stepper slopes. This contains also an amplifier of gain 2 to cover the full range of K3-5R input from 0 to 10 V. 
      \end{itemize}

      Although there is no direct physical interface between the AMHD board and the HVPS board, TuMag uses the cables interconnecting DPU with HVPS and DPU with AMHD, to send the HK signals: the K3-5R module monitoring output and two AD590-type temperature detectors (white circle labelled T in the figure). 

      Finally, to prevent an uncontrolled etalon's switching off with undesired voltage speed slope, a discharge circuitry is also included in the board, allowing a slower discharge from the etalon.

   \subsection{The PCM board}
   \label{sec:pcm}

        The Power Converter Module is the electronic subsystem that powers the TuMag E-Unit and provides it with the required voltages. It is also the link to the gondola interface of the instruments: from a primary input received from the platform with a variable voltage $V_{\rm in} = [20.4, 29.2]$~V, up to five secondary output lines are produced, namely, $+\, 3.3$~V, $+\, 5$~V, $+\,12.0$~V, $-\, 12.0$~V, and $+\,24$~V. Each line supplies one or more boards, as indicated in Table~\ref{tab:PCM-OutputPower} along with the approximate, maximum average power consumed in the so-called operational mode of the instrument. In addition, an unregulated $\sim\,+\, 24$~V (average of $V_{\rm in}$) output voltage is also delivered from the PCM primary side, after a \textit{common mode filter} and an \textit{in-rush current limiter circuit} through the AMHD board, to the various heaters in the instrument. A power distribution section allows switching on/off each subsystem individually under control of the DPU.
        
        \begin{table}
           \caption{TuMag operational mode maximum average power of each output.}
           \label{tab:PCM-OutputPower}
           \begin{tabular}{r r l}
        	  \hline
        	  Voltage & Power & Subsystem \\
        	  \hline
        	  +\,3.3~V & 5.0~W  & DPU, AMHD \\
        	  +\,5.0~V & 4.5~W & DPU, AMHD, HVPS \\
        	  +\,12.0~V & 40.0~W & DPU, AMHD, HVPS, Mechanisms \\
        	  $-$\,12.0~V & 0.5~W & AMHD \\
        	  +\,24.0~V & 10.0~W & Cameras + Mechanisms\\
        			    & 20.0~W & HVPS \\
        	  $V_{\rm in}$ & 75.0~W & Heaters \\
        	  \hline
        	  Total approx. & 155.0~W & \\
        	  \hline
          \end{tabular}
        \end{table}

\begin{figure}
        	\centering
        	\includegraphics[width=1\textwidth]{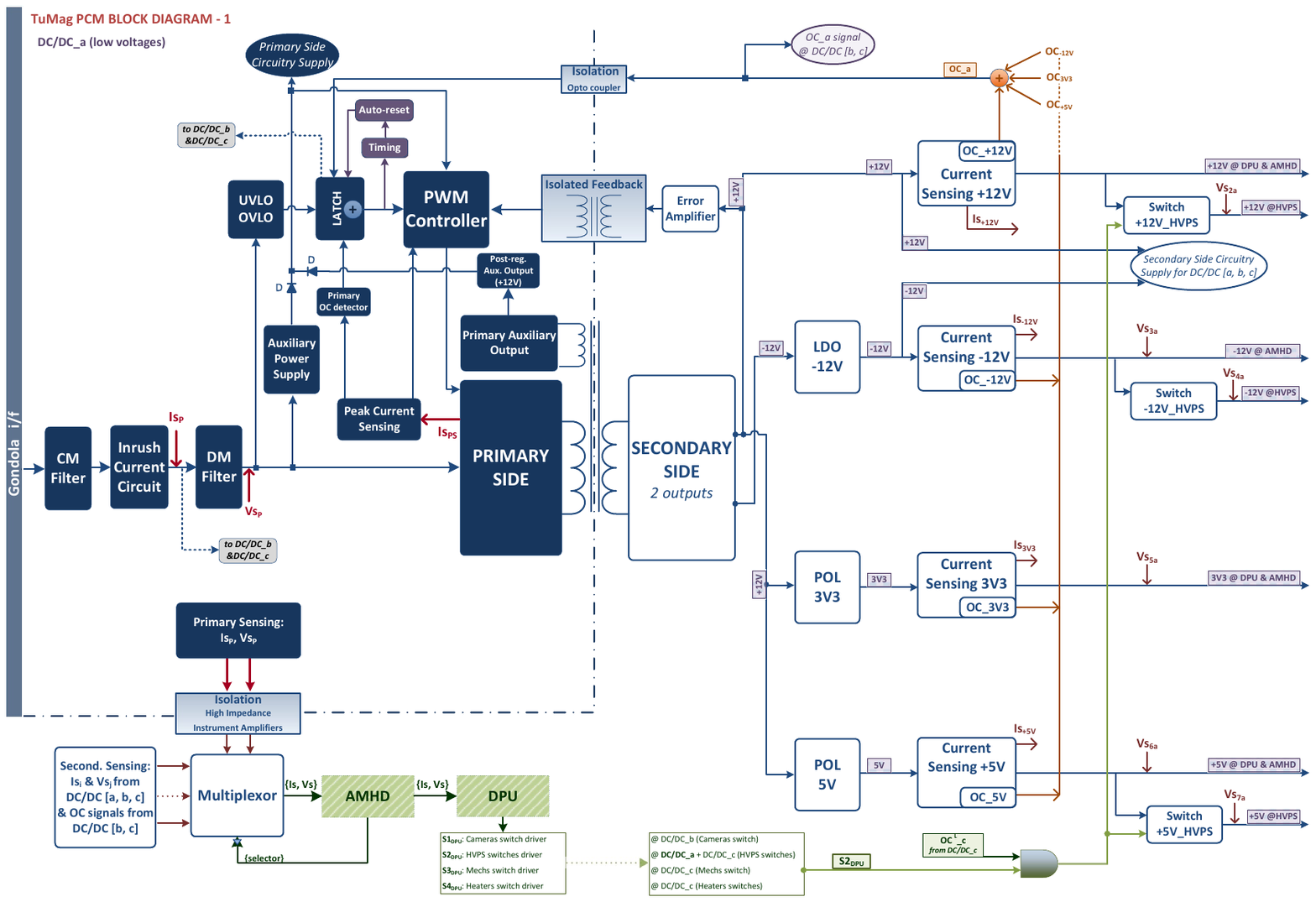}
        	\caption{Power converter module's custom DC/DC$_{\rm A}$ converter block diagram.}
        	\label{fig:BlockDiagramDC_DC_A}
        \end{figure}

        The PCM is based on one custom-designed DC/DC converter (DC/DC$_{\rm A}$) and two COTS converters (DC/DC$_{\rm B}$ and DC/DC$_{\rm C}$) using extended temperature range, commercial type electronic components placed on a double-sided PCB. The switching frequency of DC/DC$_{\rm A}$ is fixed at 125~kHz with an RC oscillator, while that of the COTS converters is 320~kHz. The PCM functionalities are:
        
        \begin{itemize}
            \item DC/DC$_{\rm A}$ is the custom, main part of the PCM. Galvanic isolation is provided and the $+\,12.0$~V is the regulated output, which feeds two \textit{points of load} that generate the $+\,3.3$~V and $+\,5.0$~V voltage lines. This circuitry partially supplies the DPU and the AMHD while providing the low voltages (below $+\, 24.0$~V) for the HVPS. Its functional block diagram is presented in Fig.~\ref{fig:BlockDiagramDC_DC_A}.
            \item DC/DC$_{\rm B}$ is based on a dedicated COTS DC/DC of $+\,24$~V, that also provides galvanic isolation and feeds the cameras.
            \item DC/DC$_{\rm C}$ is based on a second dedicated COTS DC/DC of $+\,24$~V that also provides galvanic isolation. This third converter is used to provide the HVPS and mechanisms with a voltage of $+\, 24.0$~V.
            \item The primary bus voltage ($V_{\rm in}$) supplies the instrument heaters through the PCM as well.
        \end{itemize}
        
        Since the power consumption of the instrument depends on the functional mode, so do the PCM losses that are shown in Table~\ref{tab:PCM-allPower}. While there is only one safe mode, only the maximum values are shown in the table for the operational and ascent modes. According to these losses, the PCM efficiency reaches 84~\% at maximum power processing mode, but is of course dependent on the specific operational mode.

        Like the other boards, the PCM includes stiffeners and screws, which enable its integration in the electronics pressurized box. Several protections are also present in the PCM for safety reasons: undervoltage lock-out, overvoltage lock-out, primary overcurrent detector, and secondary overcurrent protection on all the outputs.
       
        \begin{figure}
	        \centering
	        \includegraphics[width=1\textwidth]{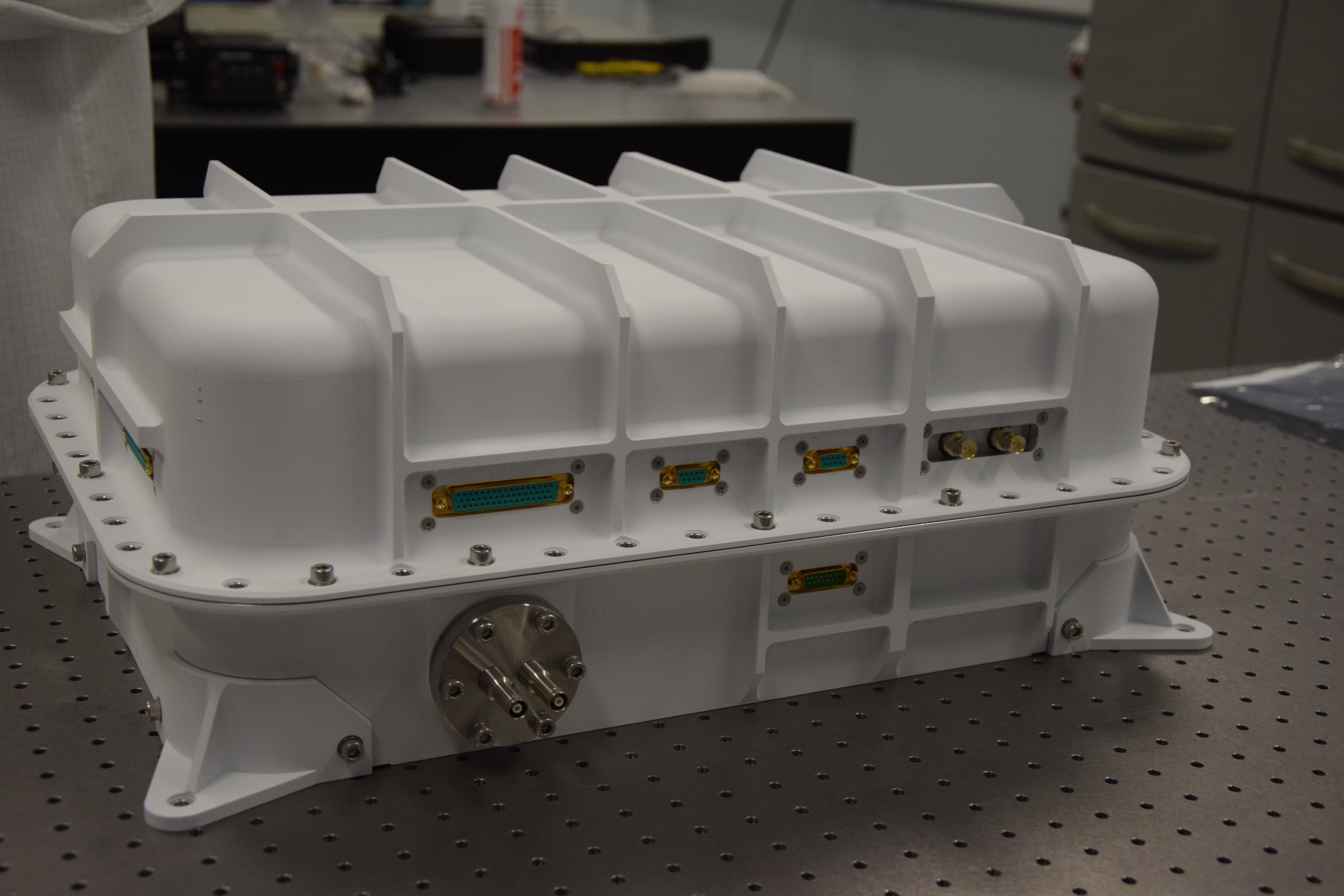}
	        \caption{TuMag's E-box. Connectors can be distinguished. In particular, the HVPS connectors can be seen on the circular flange at the bottom left of the box.}
	        \label{fig:Box}
        \end{figure}

        \begin{table}
        	\centering
        		\begin{tabular}{ l  r  r  r }
        			\hline
        			& Safe Mode & Operational Mode (max) & Ascent Mode (max) \\
        			\hline
        			Output & 30~W & 155~W & 126~W
        \\
        			PCM Losses & 20~W & 28~W & 25~W \\
        			\hline
        		\end{tabular}
        		\caption{TuMag approximate, mode dependent power consumption and PCM losses.}
        		\label{tab:PCM-allPower}
        \end{table}

   \subsection{The electronics pressurized box}
   \label{sec:ebox}
        The E-box (see Fig.~\ref{fig:Box}) is a pressurized housing that contains all the electronic boards (PCM, HVPS, DPU, and AMHD) that run  the necessary functions to meet the requirements of the instrument. It has been devised with a ``book-like'' design so that it is opened in two halves during AIV stages in order to ease the assembly and verification of the electronic components and their interconnections. Once all the components are attached, the two halves are closed and screwed together through a flange with an o-ring seal in order to prevent pressure losses.

        Stiffeners are included in the design to guarantee the required rigidity of the structure while maintaining a low wall thickness to reduce mass. The E-box envelope is 584~mm~$\times$~404~mm~$\times$~227~mm. It hosts electrical connectors and provides the interface attachments with the \sunriset electronic rack. Connectors are also included for data transmission and for power supplies (low voltages and high voltage). 

        The E-box surfaces are treated with a chromium-process coating in order to obtain a good thermal and electrical conductivity. The outer ones  are painted with a white painting (SG121-FD from MAP company).
        
    \begin{figure}
       \centering
       \includegraphics[width=1\textwidth]{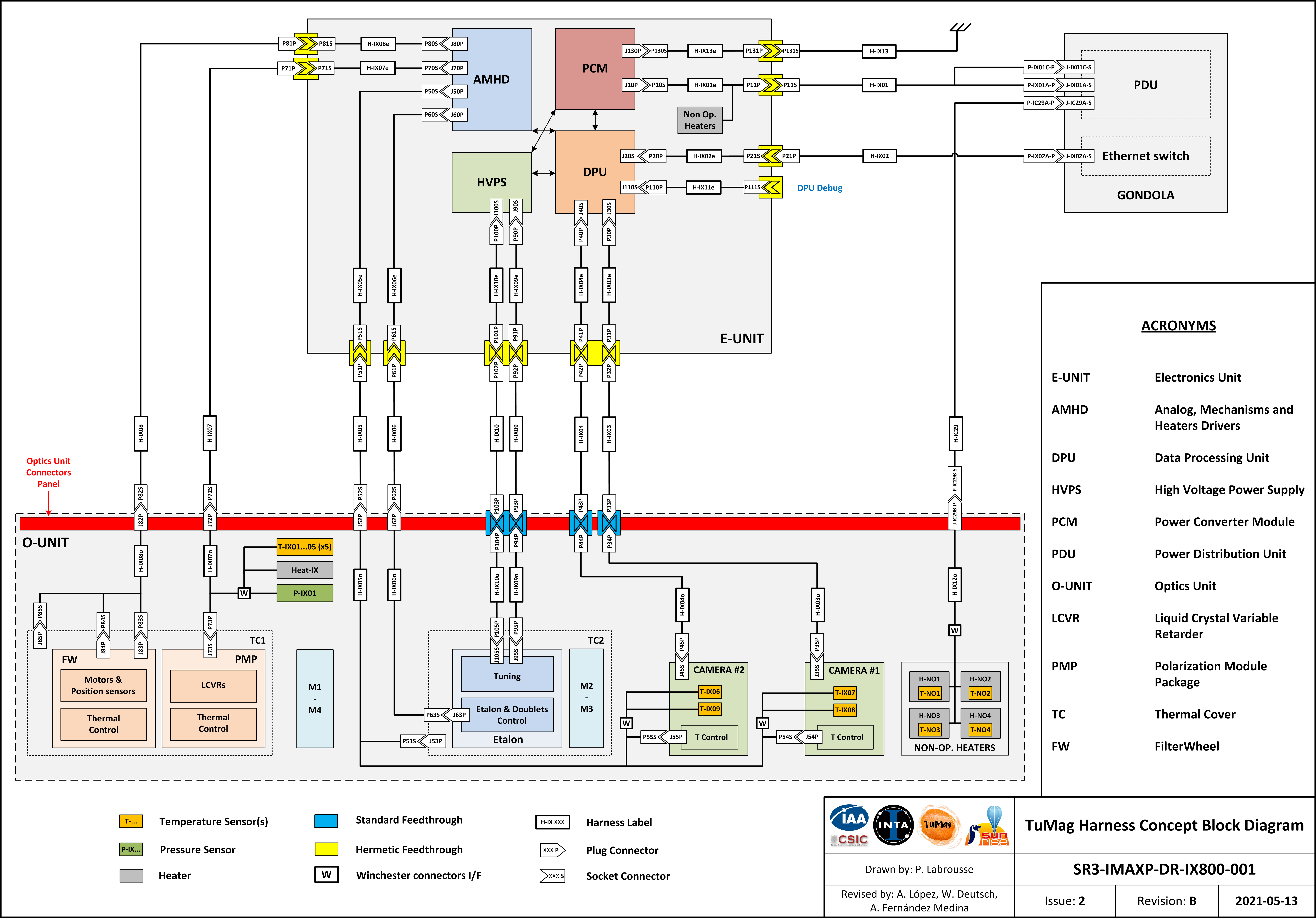}
       \caption{Detailed harness block diagram.}
        \label{fig:harness}
        \end{figure}

   \subsection{The harness}
   \label{sec:harness}
   
   The TuMag harness is a set of cable bundles that provide 1) control of the instrument, power supply, and debugging and reprogramming functions from outside and 2) an efficient interface between the E- and the O-Units. A detailed block diagram of the full set of harness is shown in Fig.~\ref{fig:harness}.

   From the gondola, three bundles provide the primary power (PPD in the diagram), the Ethernet connection with the ICS, and power to survival heaters inside the O-Unit.

   Inside the E-Unit, connections between the feedthroughs (yellow in the figure) and the boards are made with D-sub and Samtec high density connectors.

   The interconnection between the E-Unit and O-Unit uses six bundles for low-voltage and two for high-voltage signals. They are point-to-point labelled harnesses with D-sub, male-female connectors. They are made up of twisted, shielded pairs with extra external shields connected to the structure in order to minimize the electromagnetic interference of TuMag with itself and with the other instruments. Furthermore, they avoid emitted radiation and are manufactured using ethylene tetrafluoroethylene materials in order to fulfill the temperature requirements. The two cables for high voltage use especially potted plastic connectors to prevent arcs when the ambient pressure is lower than 10 mbar. 

   Inside the O-Unit, there also are the necessary bundles, linking the O-Unit connector panel and its corresponding subsystems. They use similar materials as those for the E-unit. Special care has been taken with the cable widths in order to avoid small voltage drops in the long harnesses connecting both units of the instrument.

   \section{Thermal design}
   \label{sec:thermal}

\begin{figure}
	        \centering
	        \includegraphics[width=\textwidth]{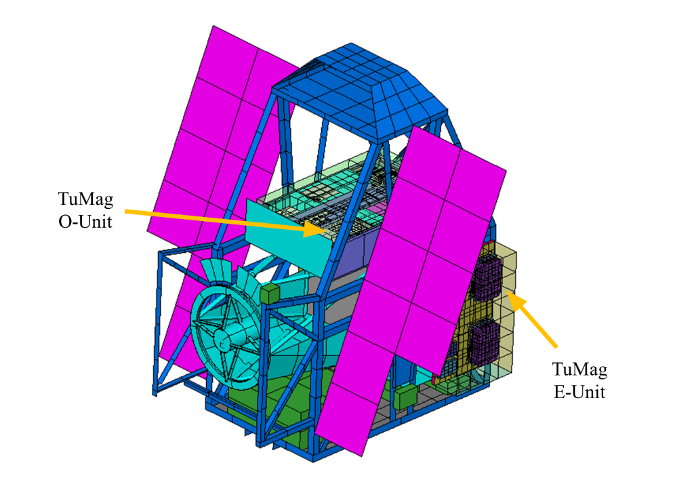}
	        \caption{\sunrise {\sc iii} geometrical mathematical model.}
	        \label{fig:sunrisethermal}
        \end{figure}

   The thermal environment encountered by the instruments onboard stratospheric balloons, similar in some aspects to the space environment, necessitates a thermal analysis to guarantee the correct performance of the systems during flight \citep{2011JAE...225...1037P}. Depending on the specific location on the balloon gondola, this environment may significantly differ from others. This is the case for the two units of TuMag: the O-Unit is located in the instrument platform (PFI) as a piggy-back of the {\sc Sunrise iii} telescope and the E-Unit is located in the electronics rack (see Fig.~\ref{fig:sunrisethermal}). Consequently, dedicated thermal analyses have been carried out that clearly differentiate the units. The thermal design of TuMag has been an iterative process, starting with a thorough study of the environmental parameters in order to identify the worst-case scenarios for both units.

   Science balloon instruments are typically designed to survive and properly work during the two main science phases of the mission: the ascent phase and the float phase. While the ascent could, in principle, be expected to be the coldest phase due to the chilly winds found in the tropopause, that is not the case for the \sunrise {\sc iii} instruments located in the PFI or in the E-racks. Both racks, are equipped with protective windshields that avoid the sub-cooling provoked by the relative wind speed. In addition, during the ascent phase, the gondola spins at approximately 1 rpm up to 32 km allowing uniform solar heating for all the instruments. A detailed study based on real data \citep{2020ActA...170...235G} was carried out to identify the worst-case scenarios for each instrument on board. As a result, we can conclude that TuMag finds the most extreme conditions during the float phase \citep{2022ActA...195...416G,2020ICES...105G}.

   For a float altitude of about 37 km, the air pressure is of the order of 500 Pa. At this pressure, heat transfer to the environment is dominated by thermal radiation. Although the low ambient pressure makes convection negligible, when dealing with small gaseous gaps between units, heat conduction has to be taken into account, as it is of the order of magnitude of thermal radiation \citep{2023AdSpR...04...10F}. The quantification of the environmental parameters for the hot and cold cases is not straightforward, as the values of albedo and Earth’s outgoing long-wave radiation are partially inversely correlated and the use of their extreme values would lead to an oversized system. A detailed description of the methodology used for the selection of the extreme parameters, based on the statistical analysis of local data retrieved from NASA's Clouds and Earth Radiant Energy System (CERES), can be found in \cite{2020ActA...170...235G} and \cite{2018AcA...148...276G}. The hottest case has been defined for the minimum solar zenith angle (SZA) whereas the coldest case corresponds to the Sun almost in the horizon (maximum SZA), because the flight is scheduled for the summer campaign, when there is no sunset at high latitudes of the Northern hemisphere. For both, the O-Unit and the E-Unit, the boundary conditions have been obtained from the \sunrise {\sc iii} system thermal model and they have been studied separately. The applied methodology is explained in \cite{2017EUCASS...G}. 

   The aim of the thermal control concept of TuMag is to achieve a robust and simple design, based mainly on passive elements (coatings, insulators and radiators), and on the use of heaters for temperature stabilization and for those elements that could reach excessively low temperatures. The following subsections describe the solution for each unit.

   \subsection{O-Unit thermal design}
   \label{sec:ounitthermal}

   The O-Unit thermal design is aimed at ensuring specific temperature set points and stability at various subsystems within the unit, while guaranteeing that the rest of internal parts and components are kept within their design temperature range at any mission scenario \citep{2023TuM_ICES139...G}. Values are specified in Table \ref{tab:temperatureounit}.

        \begin{table}
           \caption{O-Unit sybsystem's temperature set point and stability.}
           \label{tab:temperatureounit}
           \begin{tabular}{ccc}
        	  \hline
        	  Subsystem & Temperature set point & Temperature stability \\
                        & ($^{\circ}$C)          & ($^{\circ}$C/12\ {\rm h}) \\
        	  \hline
        	  NBFs & $+\, 27$  & $\pm\, 0.5$  \\
        	  LCVRs & $+\, 35$ & $\pm\, 0.5$ \\
        	  Etalon & $+\, 35$ & $\pm\, 0.05$ \\
        	  Detector & $+\, 20$ & $\pm\, 0.5$ \\
        	  \hline
          \end{tabular}
        \end{table}

\begin{figure}
	        \centering
	        \includegraphics[width=\textwidth]{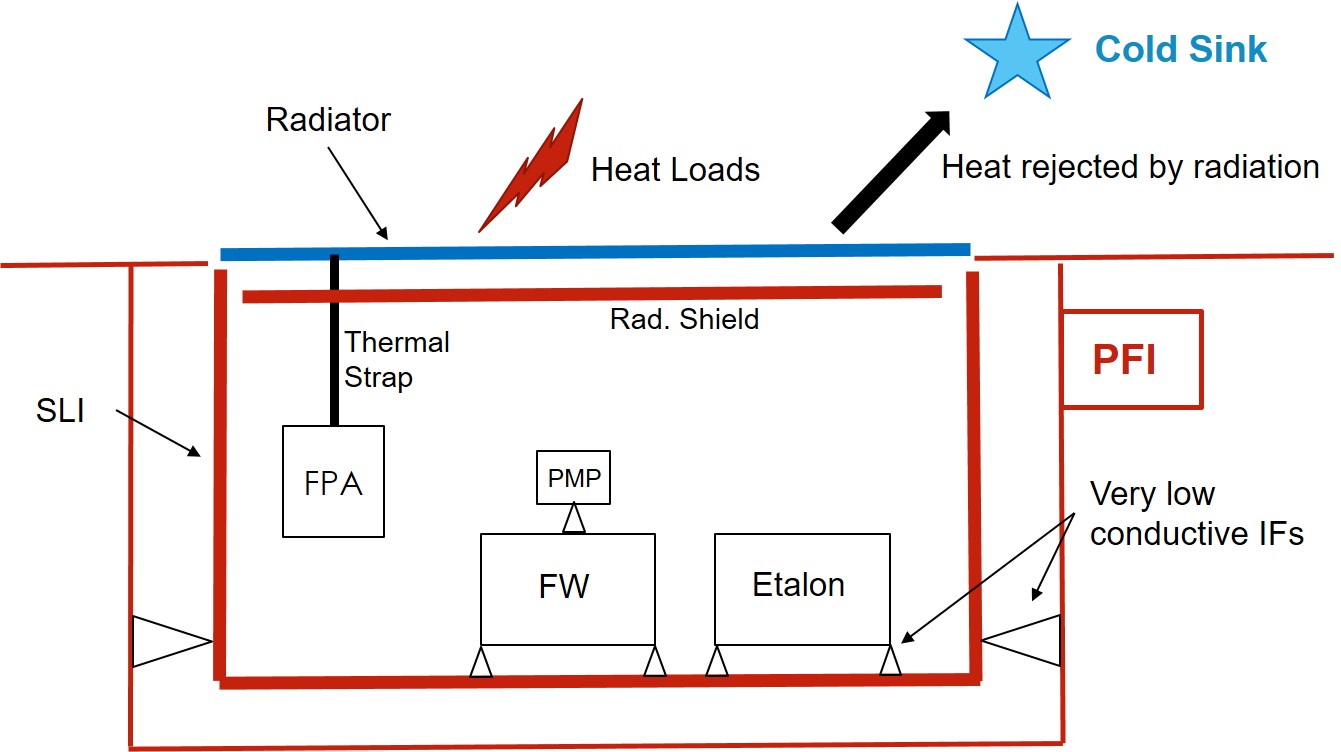}
	        \caption{O-Unit thermal architecture.}
	        \label{fig:othermalarc}
        \end{figure}

   The thermal control system (TCS) combines active and passive strategies in an architecture that is strongly determined by the O-unit environment. This environment consists of two boundaries: the PFI, which is a cavity surrounding the lateral and bottom walls of the Unit, and the open space, which represents the cold sink. The latter presents a view factor only to the top part of the unit. Thermal loads due to albedo, Earth outgoing long-wave radiation and direct solar flux, are thus impinging only on the top part of the unit. A diagram of the thermal architecture is shown in Fig.~\ref{fig:othermalarc}. A top cover acting as a thermal radiator has been set, which provides rejection of the waste heat generated by the internal subsystems. The radiator area is sized to achieve the required temperature when balancing the incoming heat loads with its heat rejection capability to the cold sink. It is coated with silver/teflon SSM tape (alp/eps = 0,066/0,78), which reflects most of the incoming solar light while allowing heat evacuation. The rest of the O-Unit is radiatively insulated from the PFI by means of single layer insulation, namely, a low-emissivity finish that minimizes the infrared heat flux exchange. It is made of an aluminum coated (two sides) polyethylene terephthalate layer, attached to the O-unit lateral walls and bottom optical bench by hook-and-pile fasteners.   

   The camera detectors are the only elements that require cooling. In order to maintain the adequate signal to noise ratio, a temperature set point below 20$^{\circ}$C has to be maintained. To do so, the waste heat from its proximity electronics is evacuated through the radiator via graphite fiber thermal straps (GFTS$^{\rm TM}$) provided by Technology Applications, Inc. For the rest of subsystems, the target set point is achieved by active heating via heaters, controlled in closed loop by a PID controller. To minimize the required power for these heaters, the various subsystems are conductively decoupled from the optical bench. To this end, they are mounted through small contact areas resulting in quite low conductive paths. Additionally, thermal spacers consisting of titanium bushings are interposed at the mounting points of the etalon and FW.

   It is worth noting that, the composite sandwich thermal conductivity plays a key role in the whole instrument thermal performance, since it constitutes the common conductive path for the whole set of internal assemblies. A dedicated thermal balance test was carried out to characterize the sandwich properties. A sandwich breadboard was used that includes embedded fiber Bragg gratings that allow measuring its thermo-elastic deformations.  

   Each of the critical TuMag O-Unit internal subsystems, namely the PMP, the cameras and the etalon, is provided with its own active control in closed loop. The thermal design of the filter wheel is explained in \cite{2022SPIE12188E..3AS} and that of the cameras is described in Sect.~\ref{sec:camerasthermal}.

        \begin{figure}
	        \centering
	        \includegraphics[width=0.45\textwidth]{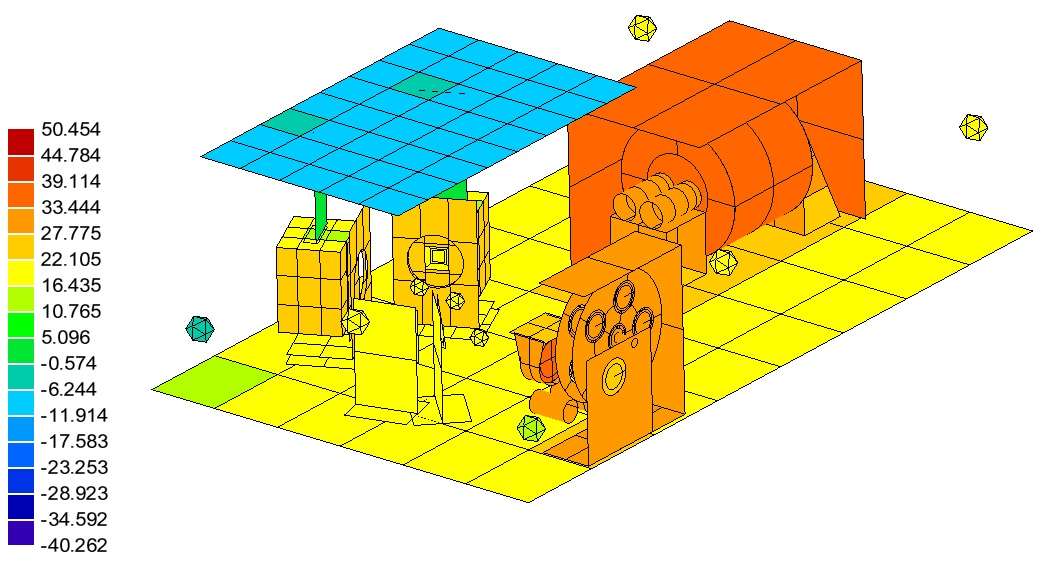} \includegraphics[width=0.45\textwidth]{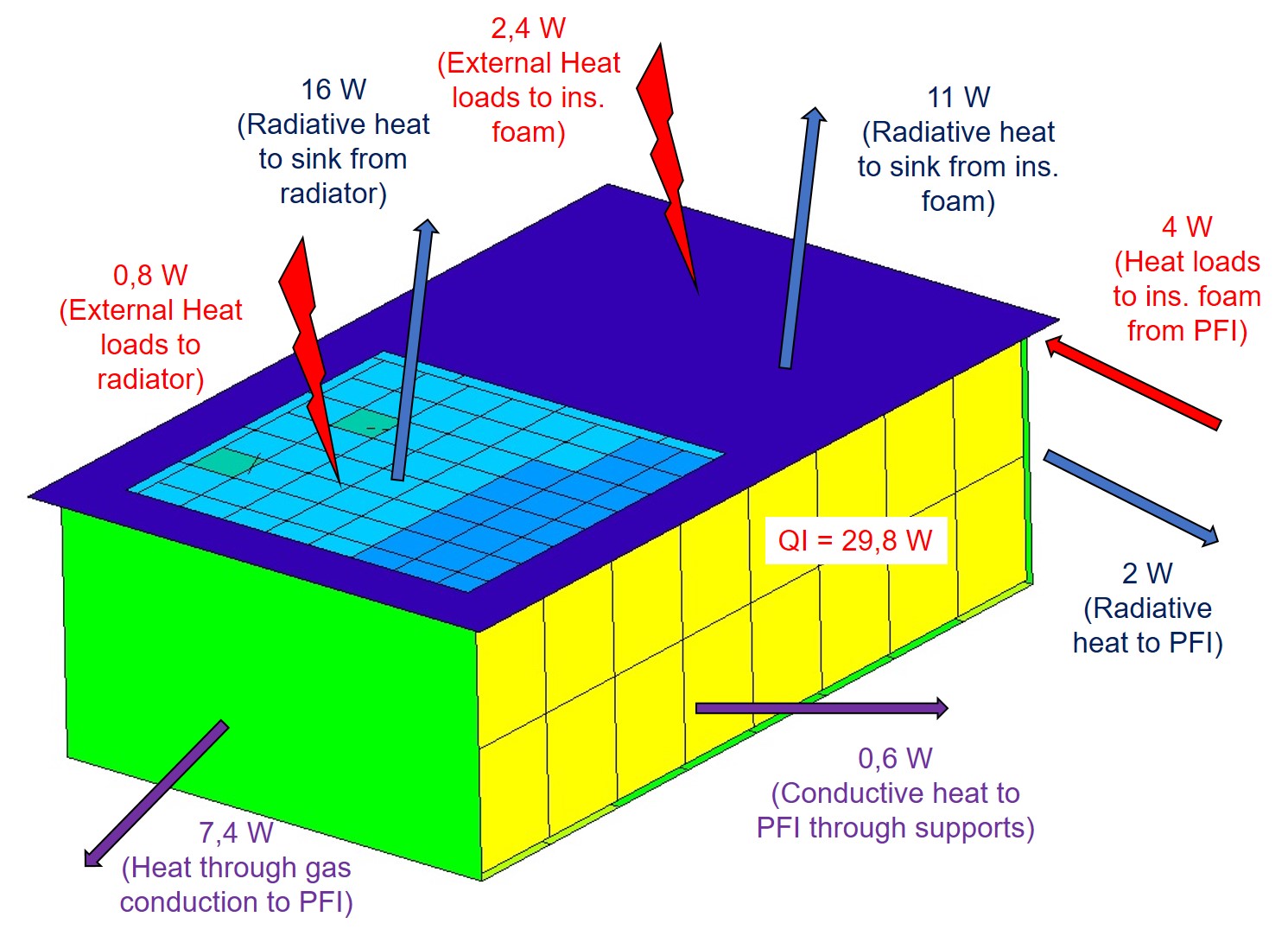}
	        \caption{Cold dimensional case analysis: temperature field (left), heat flow balance (right).}
	        \label{fig:thermmodel}
        \end{figure}

   ESATAN-TMS was used to build a thermal model (see Fig.~\ref{fig:thermmodel}) of the TuMag O-Unit and run simulations of its operation under the various expected thermal environments. In particular, the worst hot conditions determined the size of the radiator. The final combination of radiator patch area and external finish allow the rejection of the waste heat generated by the cameras plus the parasitic loads, while maintaining the detectors below its required temperature. The model also served to calculate the heater power for each of the internal sub-systems, to operate at their required operational temperature set-point according to Table~\ref{tab:temperatureounit}.

   As part of the O-Unit verification process, a thermal test was carried out upon the fully integrated unit. This test was performed in a thermal vacuum chamber with the capability of being filled with a pressure-controlled N2/air atmosphere. The part of the test at 5 mbar pressure was particularly relevant due to the uncertainties about the impact of the rarefied gas presence in the instrument performance. It pursued a triple goal: check the etalon high voltage connector reliability, check the instrument optical de-focus with regard to vacuum and laboratory conditions, and check the performance of the unit thermal control. Thermal-wise, the test was focused on two main aspects: on the one hand, assessment of the TCS capability to provide the correct operational set-points when subjected to boundary conditions replicating worst case mission scenarios and, on the other hand, tuning of the PID settings of the controllers to achieve the required stability. The ambient test revealed that heat losses in the FW and the etalon in the presence of the rarefied, 5 mbar atmosphere had been underestimated. This fact led to the redefinition of the heater power previously defined by analysis, with a significant increase in the case of the filter wheel.

        \begin{figure}
	        \centering
	        \includegraphics[width=\textwidth]{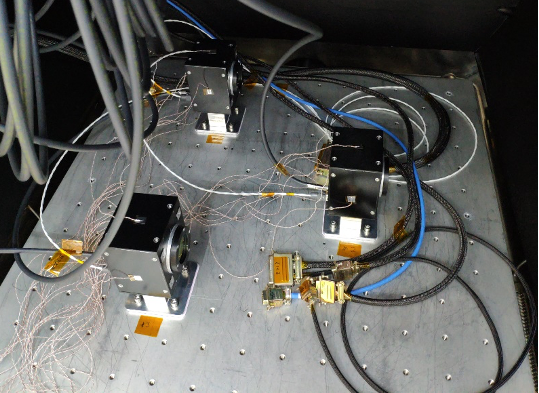}
	        \caption{TuMag cameras (two flight models and a spare model) inside the vacuum chamber and instrumented with 9 thermocouples each.}
	        \label{fig:camerathermal}
        \end{figure}

   \subsubsection{TuMag cameras thermal design}
   \label{sec:camerasthermal}

   From a thermal point of view, the two cameras (items 6 and 7 in Fig.~\ref{TuMagAIV}) are one of the critical elements in the TuMag O-Unit. The CMOS sensors of the cameras need to be maintained between 15 and 30$^{\circ}$C, with a temperature stability of $\pm\,  0.1^{\circ}$. The SPGcam cameras were designed not only to be used in TuMag but also in the SCIP instrument \citep{2023FrASS..1067540O}. For this reason, compatibility for both instruments was a design driver. Each camera dissipates a total of 3.6 W, distributed between the control printed circuit board (3 W) and in the sensor printed circuit board (0.6 W).

   To avoid overheating of the cameras during hot conditions, two graphite thermal straps, one per camera, are used to connect the top surface of the camera case to the top surface of the O-Unit housing that acts as a radiator. Internally, a copper cold finger, in direct contact with the sensor, is used to link the sensor to the thermal strap. These thermal links were sized to passively maintain the sensor at a temperature slightly below the operative one. Then, a 6 W dedicated heater, located on the cold finger and controlled by a proportional–integral–derivative controller, is used to reach the operative temperature of the sensor and to guarantee the stability requirements.

   The thermal analyses of the cameras were performed with ESTAN-TMS. Due to its criticality, each camera has been modeled with about 140 thermal nodes. Parametric and sensitivity analyses were performed to verify the robustness of the design, which was validated in a dedicated thermal test of an engineering model. Before the final integration inside the O-Unit, an acceptance test was carried out for the two flight cameras and the spare, as seen in Fig.~\ref{fig:camerathermal}. During the test, the performance of the cameras under nominal operation conditions was verified, as well as the capability to survive under non-operational conditions between $-25$ and 55$^{\circ}$C. The thermal models were correlated and accordingly updated with the results of the thermal vacuum tests measurements using a novel and efficient approach.

        \begin{figure}
	        \centering
	        \includegraphics[width=0.65\textwidth]{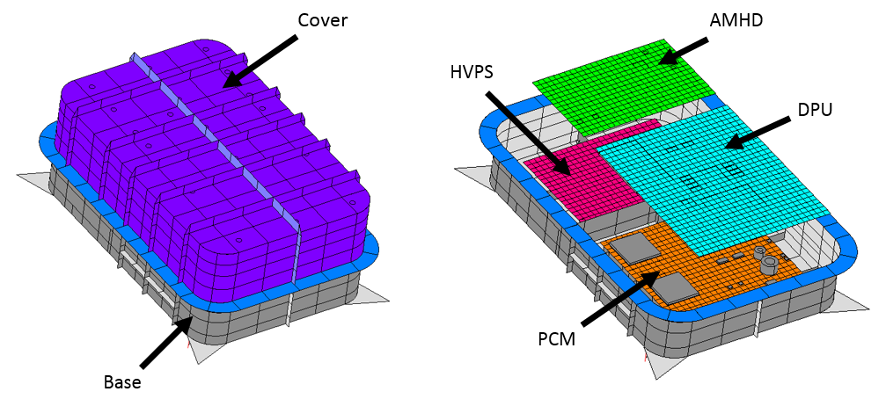}
	        \includegraphics[width=0.33\textwidth]{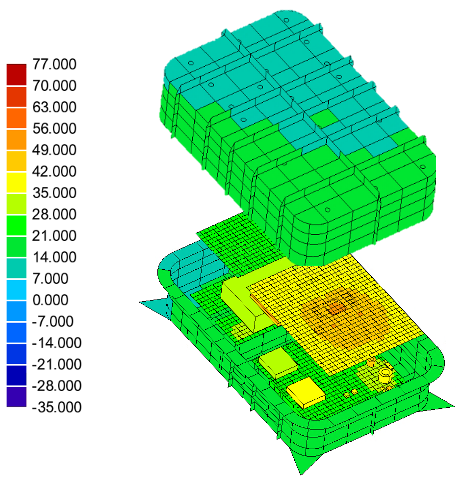}
	        \caption{Geometrical mathematical model (left and middle panels) and thermal model (right panel; temperatures are in $^{\circ}$C) of the E-Unit.}
	        \label{fig:eunitthermal}
        \end{figure}

   \subsection{E-unit thermal design}
   \label{sec:eunitthermal}

   The E-Unit is mounted on the E-rack, on one side of the gondola (see Fig.~\ref{fig:sunrisethermal}), in such a way that it is continuously shaded from direct solar radiation by the solar panels during the float phase. The absence of direct sunlight together with a direct view to the cold sky makes the E-rack suitable to place the highly dissipative electronic units. Specifically, the TuMag E-Unit dissipates 75 W during the worst, hot conditions.

   The cooling of the unit is mainly achieved through thermal radiation to the environment. The conduction to the rack is limited to avoid interference between units. Hence, the unit case acts a radiator and is painted white with SG121FD (solar absorptance $\alpha_{\rm s} = 0.18$ and infrared emissivity $\varepsilon_{\rm ir} = 0.88$). 

   A thermal model with 900 nodes was set up with ESATAN-TMS to predict the temperatures of the unit during worst-case conditions. The model contains the housing of the unit, all the four boards and the most dissipative or critical electronic components. A view of the geometrical mathematical model is shown in Fig.~\ref{fig:eunitthermal}. Thermal connections between components and boards, between boards and housing structure, and between all internal surfaces and the housing-filling nitrogen, were calculated using the typical procedure described in \cite{2020IEEE...56...186T}. Once designed and built, the E-Unit was tested in a thermal chamber filled with a nitrogen atmosphere at 300 Pa, near the expected flight pressure. The cases during the thermal testing campaign were selected such that they represent the most extreme possible environments (both in cold and hot conditions) for the mission. The design process was completed by correlating the model with the test data. After correlation, the model predictions showed a maximum discrepancy of less than 3$^{\circ}$C with respect to the test data. An example of the results for the hot operational case are shown in the right panel of Fig.~\ref{fig:eunitthermal}.

   \section{On-board software and data flow}
   \label{sec:software}
        The TuMag on-board software is divided into two main blocks: the control software (CSW) and the frame grabber software (FGSW). The CSW, which runs in the SyC, is mainly devoted to telecommand (TC) and telemetry (TM) management,\footnote{Telecommands are orders to the instrument through the {\sc Sunrise} ICS. Telemetries are data transferred from TuMag to the ICS.} image transfer from the frame grabber and to the {\sc Sunrise} instrument control system (ICS), to generating thumbnail from selected images, housekeeping acquisition, and contingency management. The FGSW, which runs within the FG's MicroBlaze processor, is in charge of commanding the different subsystems during the observation modes, taking images from the cameras, and handling the image transfer from the FG to the SyC.

        \begin{figure}
	        \centering
	        \includegraphics[width=0.3\textwidth]{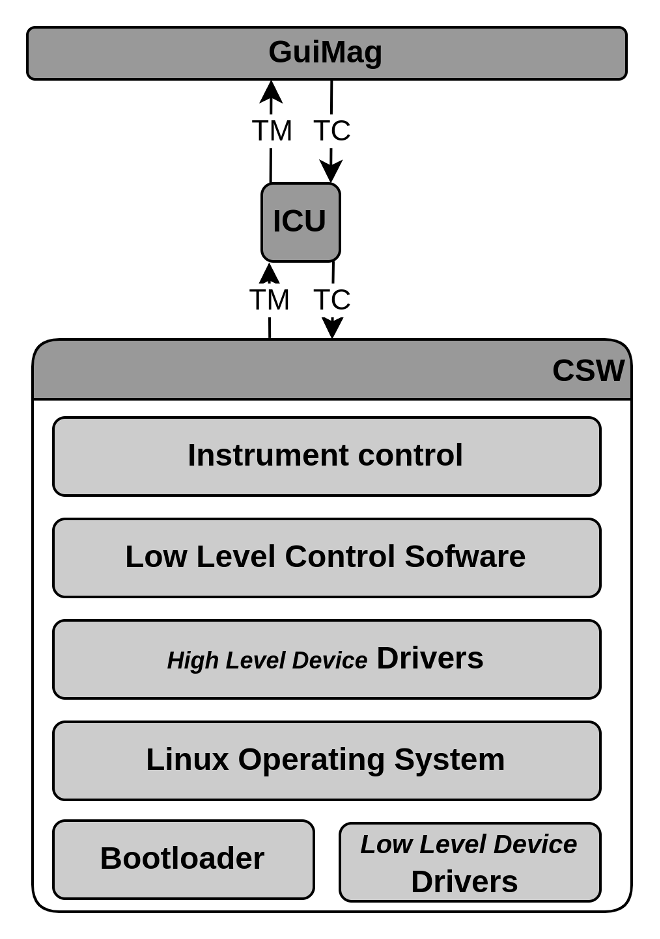}
	        \includegraphics[width=0.5\textwidth]{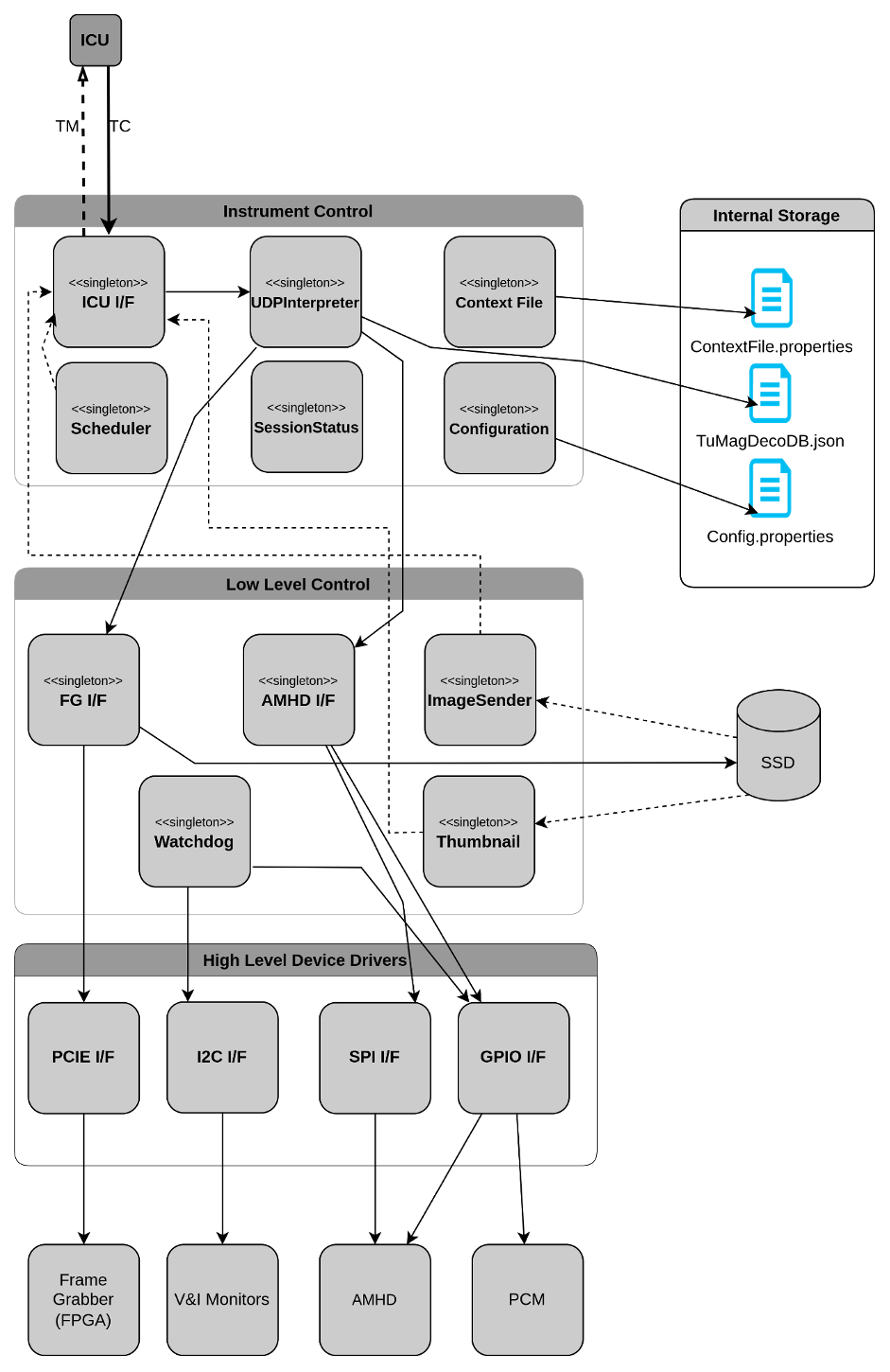}
	        \caption{TuMag control software layer (left panel) and functional (right panel) diagrams.}
	        \label{fig:CSW}
        \end{figure}
        
    \subsection{Control Software}
    \label{sec:control_software}
       CSW is developed using the C++11 language, which improves thread management and time library with respect to former versions. To achieve optimal memory management a \textit{singleton}\footnote{In software engineering, singleton is a creational design pattern that lets you ensure that a class has only one instance, while providing a global access point to this instance.} design pattern is used.
       
       A layer diagram of the CSW is shown in the left panel of Fig.~\ref{fig:CSW}. Each layer groups all the related functionalities, which are better described in the CSW functional diagram of the right panel (see paragraph below). Each layer has only one interface with the layer above and another with the layer below it. The first layer is composed of two blocks: the \textit{bootloader}, which is the software in charge of starting up the Jetson TX2 device after it is switched on, and the \textit{low level device drivers} that manage the necessary interfaces like PCIe, SPI, I2C and GPIO. The Linux Operating System layer refers to Linux for Tegra, which is a Linux-based system software distribution by Nvidia\textsuperscript{\textregistered} for Tegra devices, like the Jetson TX2 chip.  The \textit{high level device drivers} layer uses API’s of the Lower Level Device Drivers to implement the specific behavior for each subsystem interface. The \textit{low level control software} is the lowest level user layer, including the execution software and support to all actions carried out by each subsystem. The \textit{instrument control} is the highest level user layer, including the communication library (telecommands and telemetries), the scheduler, the housekeeping acquisition, its own status, and contingency management. GuiMag is the (ground equipment) graphical user interface (see Sect.~\ref{sec:gui}).
       
       A functional diagram of the CSW is shown in the right panel of Fig.~\ref{fig:CSW}. The ICS interface singleton manages the TCs (TMs) that are sent and received through it from (to) the {\sc Sunrise} ICS. The UDPInterpreter singleton manages the acceptance, validation, and execution of user defined programs (UDPs), which are sequences of DECOs (DEfined COmmands). A TC may (or not) contain one single UDP. This singleton is supported by a file called {\tt TuMagDecoDB.json} represented in the right-hand side, internal storage block of the figure. This file contains all the DECOs accepted by the instrument and is the same file as that used in GuiMag (see Sect.~\ref{sec:gui}). The SessionStatus singleton handles the current status of the instrument. The  \textit{context file} and \textit{configuration} singletons handle their own file each ({\tt ContextFile.prop\-erties} and {\tt Config.properties}, respectively). The {\tt ContextFile.properties} tell us the current status of TuMag (like the current observation mode, the status of each subsystem, the cameras' configuration, etc.); the {\tt Config.properties} file sets the internal configuration of the instrument (such as the IP address for the ICS, port number for TC/TMs, etc.). All these properties can be modified through appropriate TCs. The \textit{scheduler}  singleton handles the periodic event tasks like HK acquisition and housekeeping telemetry generation. The \textit{frame grabber interface} manages the observational modes and reads the images returned by the FG. These images are buffered in an external SSD. Each buffered image is sent to the ICS using the ImageSender singleton. The \textit{watchdog} singleton handles the contingency detection and recovery actions. The \textit{thumbnail} singleton determines whether the image to be sent to the ICS has to be converted to a thumbnail and sent to ground. The AMHD interface singleton manages the access to the AMHD board. It can read the housekeeping or command some action using the SPI and GPIO interfaces. The GPIO interface handles the access to the PCM, and some input lines to the AMHD such as the reset or force re-program functions. The SPI bus interface manages the communication with the AMHD. The I2C interface handles the reading of voltage and current sensors located in various places on the instrument. The PCIe interface manages access to the DMA in order to read images from the frame grabber.

    \subsection{Frame Grabber Software}
    \label{sec:fg_software}
       The FGSW is in charge of receiving and responding to general commands and those concerning the HK, sending accumulated images to the CSW, and informing on errors. The FGSW is developed in embedded C, provided by the Xilinx\textsuperscript{\textregistered} Software Development Kit 2017.4 version. The communication protocol consists in several internal messages to send/receive commands/answers to control all the various subsystems for all the observational modes. Internally, the message transfer is processed using a FIFO (first in-first out), through which they are paid attention to in order of arrival. Once the FGSW receives a command to execute an observational mode, it starts by setting the parameters of the different subsystems and by programming and enabling the trigger control unit that runs the cameras to grab the images. Every time that an image is ready for transfer, the FGSW sends a request command to the CSW. This command contains the memory address in which the image has been stored. The CSW can transfer the images from the FG’s DDR to the LPDDR4 by using DMA.
    
        \begin{figure}
	        \centering
	        \includegraphics[width=0.5\textwidth]{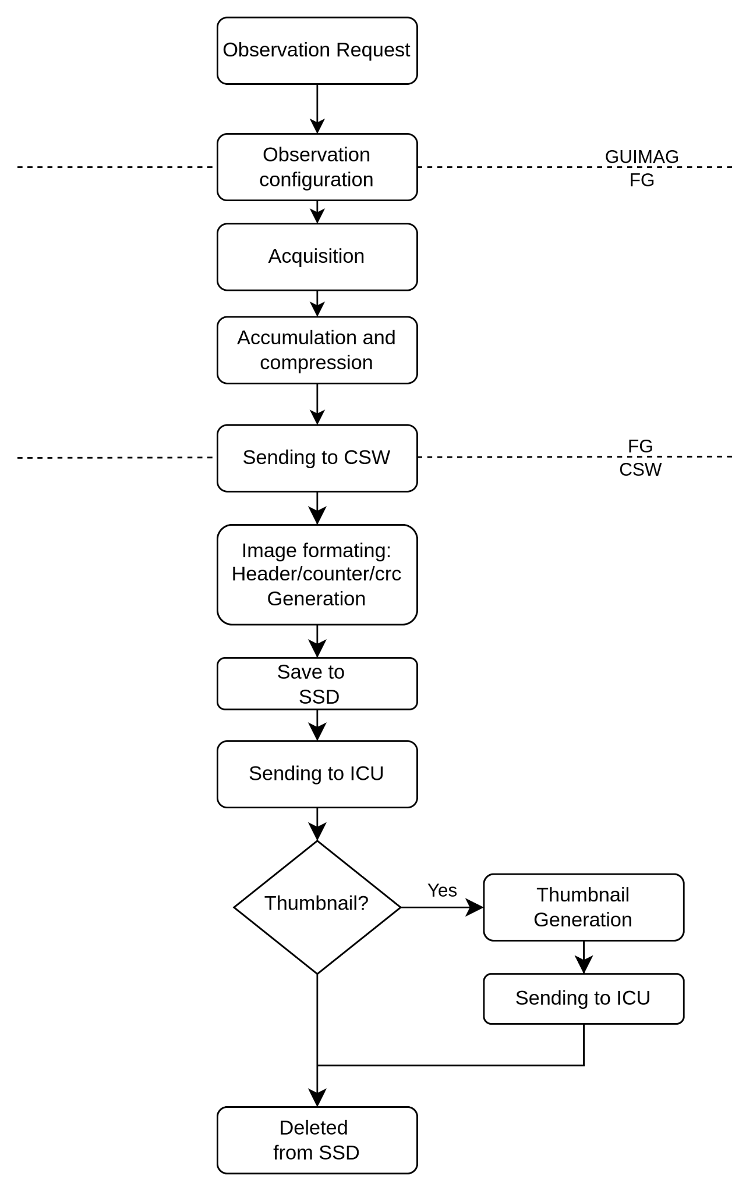}
	        \caption{Data flow diagram.}
	        \label{fig:dataflow}
        \end{figure}

    \subsection{Data Flow}
    \label{sec:data_flow}
       The data flow during an observation is depicted in Fig.~\ref{fig:dataflow}. An observation request is sent to the DPU through GuiMag. The FG decodes the command and configures its internal processing pipeline, the cameras, the filter wheel, the LCVRs, and the etalon before starting the image acquisition. In addition to the images, the FG collects metadata related to each image such as the time tag and the current O-Unit set-up. To reduce the final data rate, images are accumulated and compressed according to the configured observational mode. Finally, the sequence of images together with the metadata are transferred to the CSW.

       The first task of the CSW is to format the received data, create a header with the current status of the instrument, add the context file and the information reported by the platform. Finally, the CSW creates the telemetry packet containing the image data adding a counter and a cyclic redundancy check. Once created, the TM packet is stored into the external storage (SSD) because the platform bandwidth is not enough to send all the images to the ICS. Then, images are sent to the ICS. After the image has been sent, the thumbnail process checks whether this image has to become a thumbnail. If true, the CSW generates the thumbnail and sends it to the ICS. Finally, the original image is deleted from the SSD. 

   \section{Ground segment}
   \label{sec:gui}
       The ground control system is based on a desktop application running on a portable computer, which is in charge of sending TCs and receiving TMs to/from the CSW through the ICS. This application has been called GuiMag, which stands for GUI (Graphical User Interface) for the TuMag instrument, and whose main objective is to operate the instrument from the ground and to give the user a friendly interface to manage and visualize TuMag data. 

       GuiMag has been developed using JavaFX, which is a cross-platform GUI framework that has a high-quality, multiple threading support that allows  data to be received and displayed in real time. In order to make better use of the program memory and to allow future data analysis, all data received are saved in JSON format to disk, which is easily readable in any programming language.

       Apart from GuiMag, a database (DB) hosted in a Web service was set up. This DB stores the definition of certain information to be shared by GuiMag and the CSW:
    \begin{itemize} 
        \item DECO definitions.
        \item HK variables and values.
        \item Transfer function: Linear conversion from physical units to digital counts and vice versa. All HKs and DECOs are treated in digital counts in all the on-board software, but the user can work on physical units from GuiMag.
    \end{itemize}

       The communication between GuiMag and the ICS is done through transmission control protocol/internet protocol (TCP/IP) over Gigabit Ethernet. TCs are used to command TuMag actions from the ground. These TCs have been divided into groups and subgroups, depending on the type of action performed by each one:
      \begin{itemize}
         \item TC groups 0 to 3: standardized and common TCs used for all {\sc Sunrise iii} instruments. These are used to command general actions on the instrument, such as sending a ping, switching off the CSW, or aborting the execution of a previous command.
         \item TC groups from 4 onwards: user-defined commands. Each instrument uses these commands to perform specific actions. A special type of TCs are those carrying UDPs, which are defined as group 4 and subgroup 2. This TC type is composed of one or more DECOs, and can be created and stored in GuiMag for later use.
      \end{itemize}

      In addition to being able to turn off, turn on, and configure the different TuMag subsystems by sending TCs, the instrument has three operational  modes: {\tt safe}, {\tt debug}, and {\tt science}. In {\tt safe} mode all subsystems are off, and only certain DECOs can be sent. In {\tt debug} mode all subsystems are off, but all DECOs can be sent. In {\tt science} mode all subsystems required for observations are turned on and configured. We can switch between modes from GuiMag, but the instrument must pass through {\tt safe} mode before going to any other mode.

      The CSW sends information on the instrument status and observations to GuiMag (also through the ICS) using TM packages of two different types:
      \begin{itemize}
         \item HK: these packages are mainly dedicated to reporting on the instrument status. We receive information such as instrument sensor readings, subsystems status, gondola position, and the result of each observation. All this information is stored internally and presented graphically in GuiMag.
         \item Science: these packages contain thumbnails obtained in observing modes, which are composed of a header (containing information about the image and the instrument status) and the image itself. When a science package is received, both the header and the image are processed, saved in a file and displayed in GuiMag.
      \end{itemize}

      \section{Instrument assembly, integration, and verification}
      \label{sec:aiv}

      Since TuMag is one of the post-focus instruments of \sunriset\!\!, the assembly, integration and verification (AIV) activities were split into three phases: instrument stand-alone AIV, TuMag AIV in the post-focus instrument platform, and TuMag AIV in \sunriset. 
      
      \subsection{Instrument AIV phase}
      \label{sec:instrumentaiv}
      
      During the instrument AIV phase, the following subsystems were assembled and verified previously to their integration in the O-Unit due to their critical role in the instrument:

      \begin{itemize}
         \item The optical bench, whose inserts and pins (for the optical elements) had to be precisely positioned according to the restricted optical element tolerances.
         \item The PMP, to assure that the device provides the required polarization modulation efficiencies.
         \item The FW, whose elements, mainly the F4$^{\prime}$ frame and the NBFs, required high precision alignment since the wheel defines the optical axis for the rest of optical elements. The alignment of NBFs is also important because of their influence in wavelength tuning.
         \item The retro-projector mirror subsystem (M2 and M3 assembly) to guarantee an equal double pass through the etalon.
         \item The SPGCam cameras whose performances (i.e., S/N, dark current, full well, etc.) and its optical and mechanical interfaces had to be met to guarantee the TuMag performances.
      \end{itemize}

      A detailed description of the procedure for the assembly and alignment of the O-Unit can be found in \cite{2022SPIE12184E..2GA}. Once the O-unit, E-Unit, external harness, and software were ready, they were integrated and put in flight configuration. Then, a set of tests were carried out to verify that the instrument fulfils the requirements and to calibrate it: the so-called end-to-end (E2E) tests. Specifically, the goals of the E2E tests were to check the TuMag image quality as well as the polarimetric and spectroscopic performances. Each E2E test has a reduced version to be used as instrument health diagnosis after any critical event: e.g., environmental test campaign, transport, or integration in \sunriset. The results of the E2E tests and the setup description are detailed in \cite{2022SPIE12184E..2FA}. A summary of the results can be found below.

        \begin{figure}
	        \centering
	        \includegraphics[width= 0.32\textwidth]{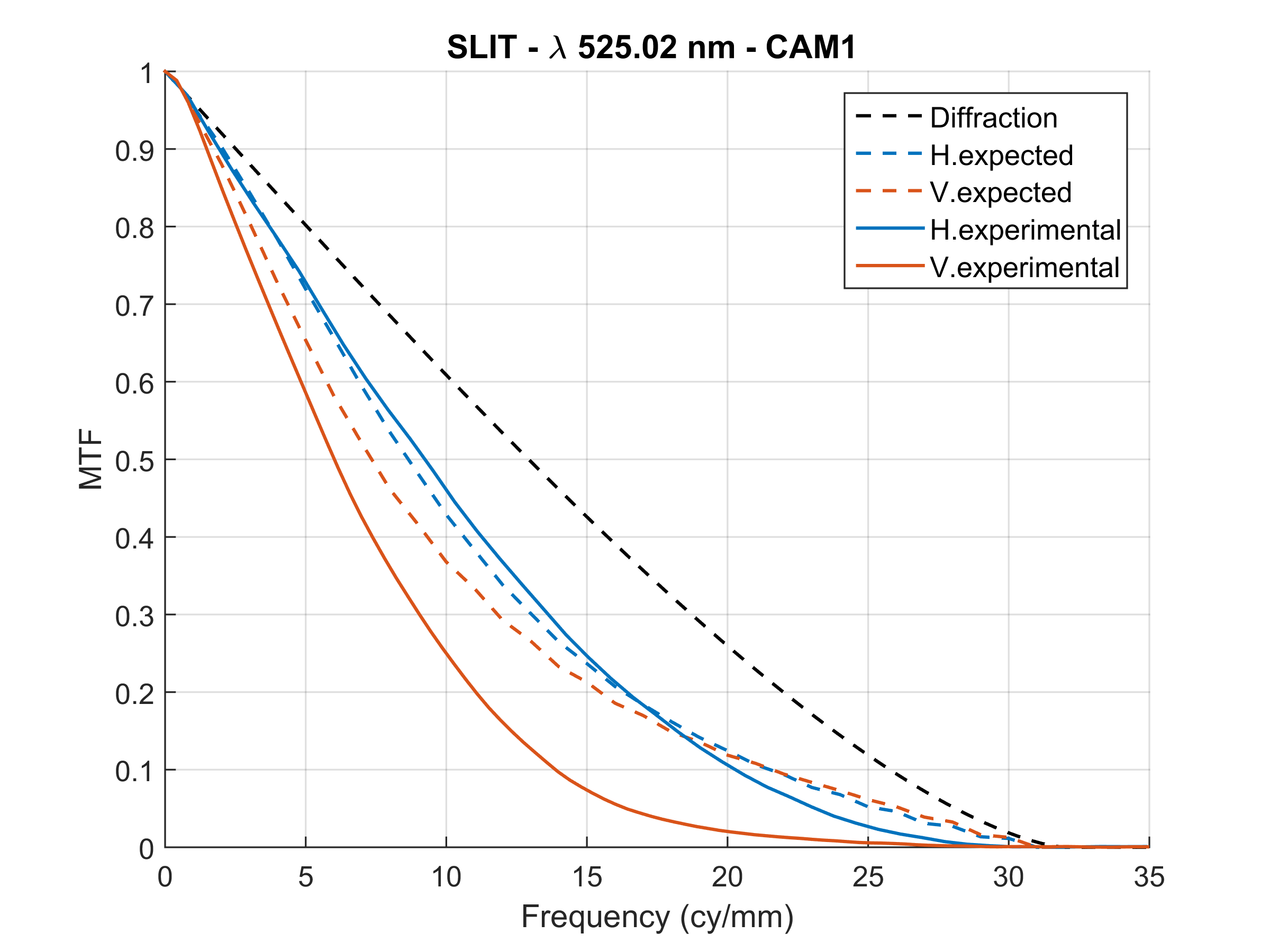} \includegraphics[width= 0.32\textwidth]{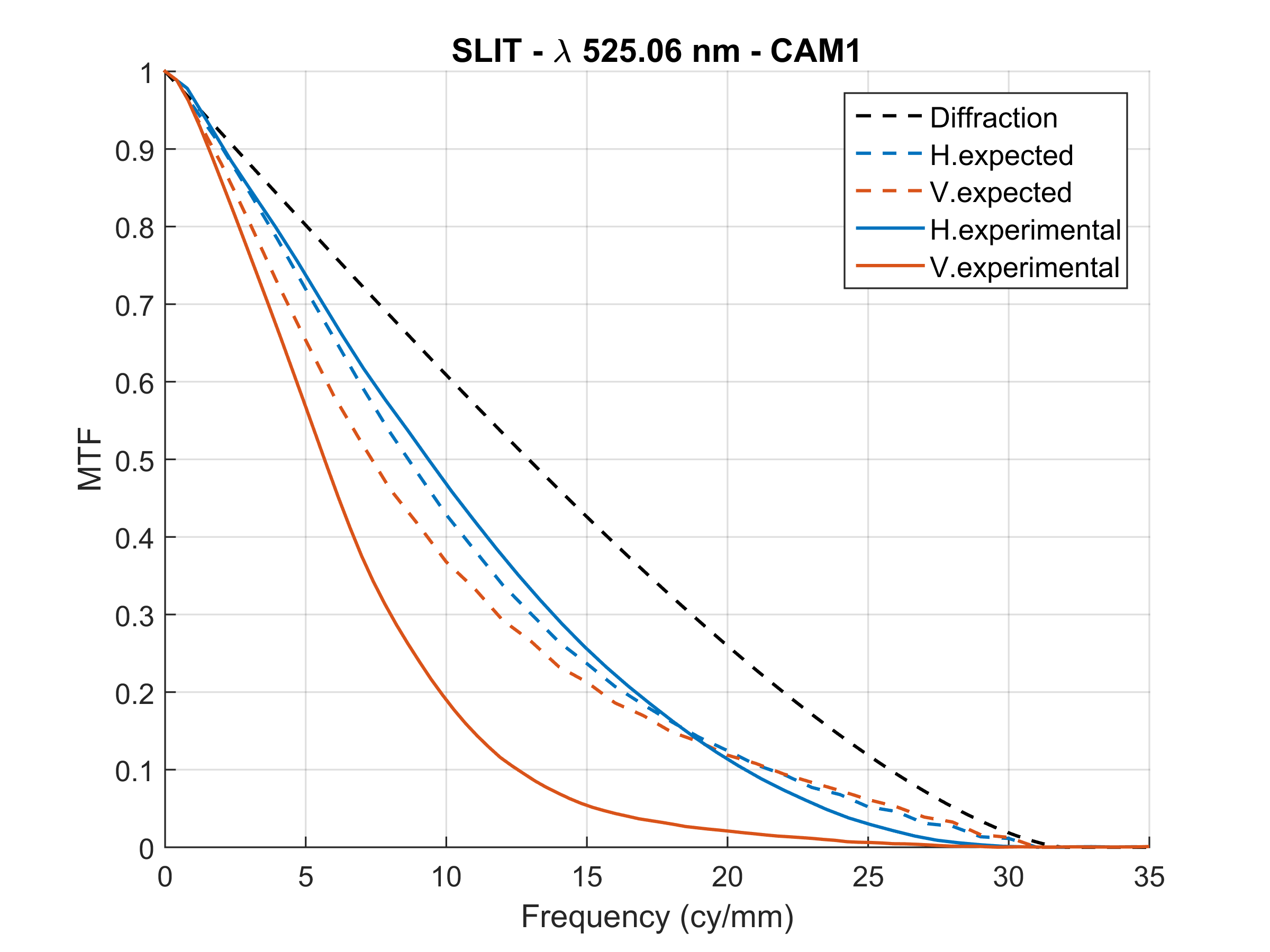} \includegraphics[width= 0.32\textwidth]{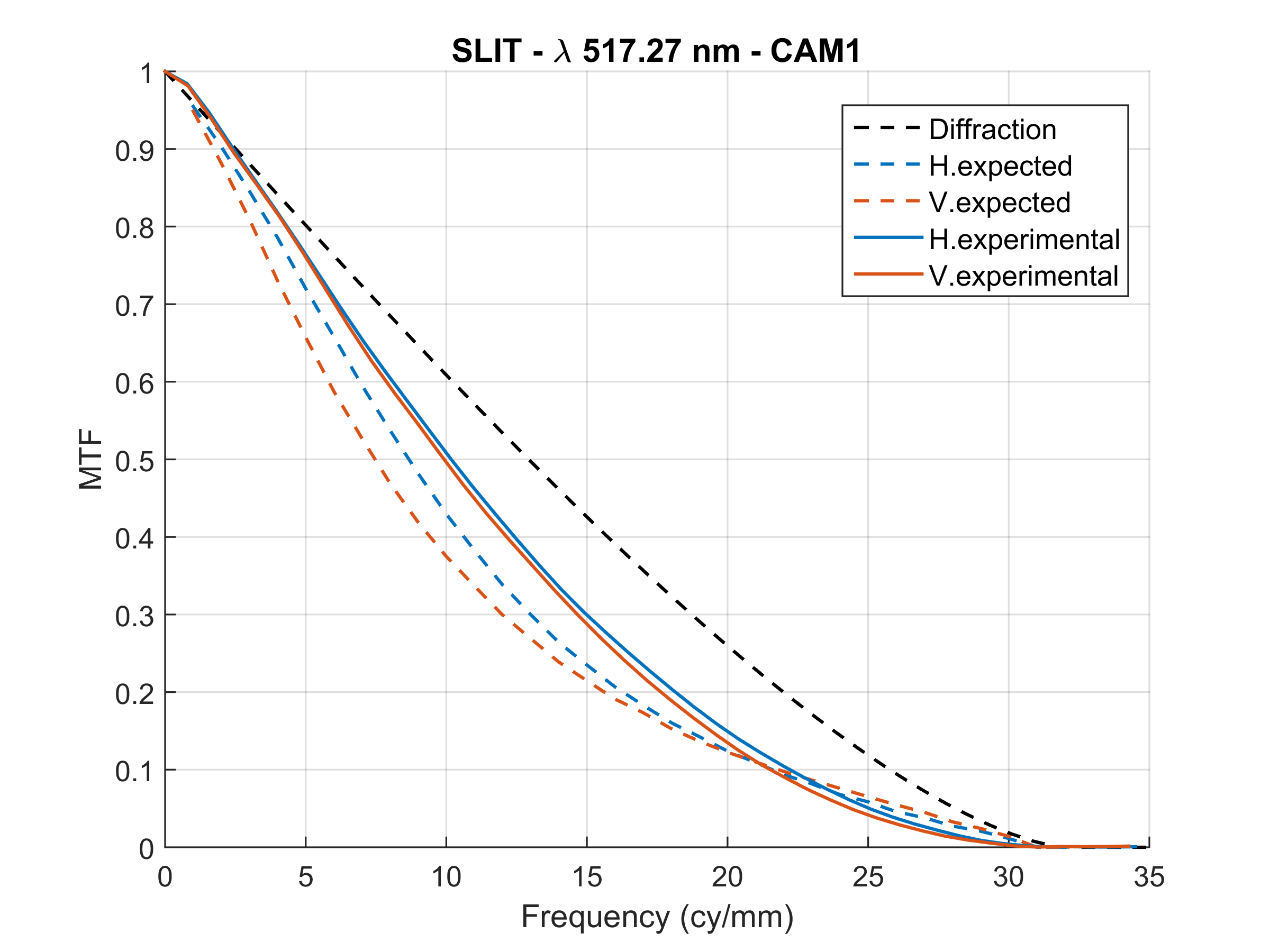} 
	        \caption{Camera \#1 MTFs with the three narrow band filters. The dashed black line represent the ideal curve only limited by diffraction. Blue lines are for the horizontal direction and orange lines are for the vertical direction. Solid lines are for experimentally measured MTFs and dashed lines are for theoretically expected ones.}
	        \label{fig:mtf}
        \end{figure}

      The image quality was checked by projecting several targets at F4$^{\prime}$, including a USAF test target, a star target, a grid, and a slit. The modulation transfer function (MTF) was calculated using the slanted-slit method \citep{2013SPIE.8788E..2JH}. The theoretically expected and measured MTFs of camera \#1 for the three wavelength ranges are plotted in Fig. \ref{fig:mtf} up to 32 mm$^{-1}$ (cutoff frequency). Results for camera \#2 are similar. Both the horizontal and vertical MTF curves for NBF3 (517.27 nm) agree very well with expectations. In the case of NBF1 (525.02 nm) and NBF2 (525.06 nm), only the horizontal behavior closely reaches theory. These results are very similar to those obtained with the USAF test. Some astigmatism was indeed expected due to the etalon, whose wavefront error was previously and separately measured. The difference between horizontal and vertical MTFs, however, cannot be explained solely by etalon effects. In addition, it was observed that the higher the tuning angle of incidence of the NBFs, the higher the MTF decay. This points to the mechanical method used to hold the NBFs in the FW as the cause of some deformations on them. In any case, the MTF degradation in the vertical direction for two of the pre-filters can be accepted as restorations through phase diversity techniques provide satisfactory corrections.

\begin{figure}
	        \centering
	        \includegraphics[width=\textwidth]{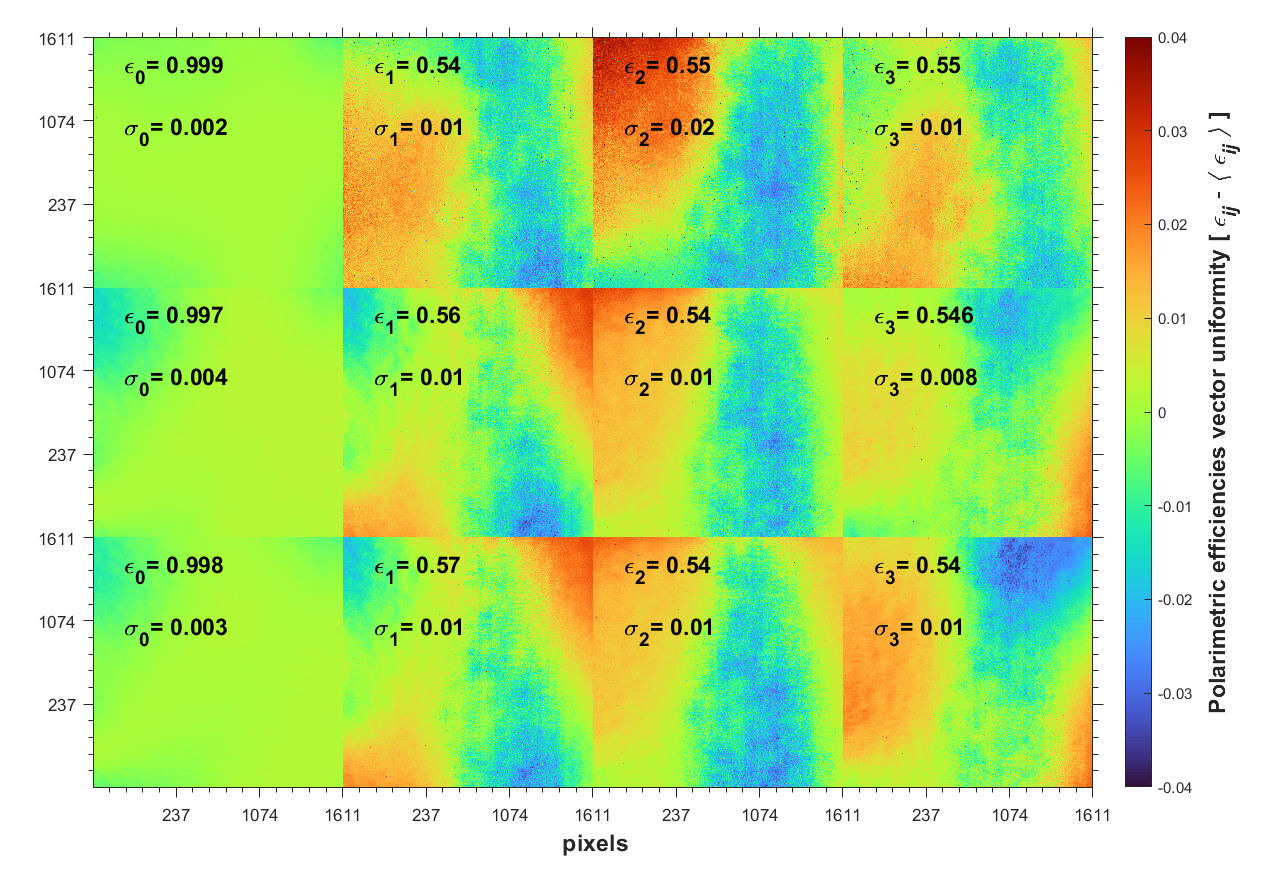}
	        \caption{Polarimetric efficiencies of the two cameras across the field of view.}
	        \label{fig:poleff}
        \end{figure}

      The polarimetric calibration was carried out at the three wavelengths of the instrument and the two cameras. A detailed description can be found in \cite{2024NOSE...C}. An optimum modulation scheme that maximizes the polarimetric efficiencies was selected \citep{2000ApOpt..39.1637D}. The retardances of the two liquid crystals at each of the four polarization modulation states are: [225, 225, 315, 315]\,deg for LCVR1 and [234.74, 125.26, 54.74, 305.26]\,deg for LCVR2. The initial voltages applied to the LCVR cells correspond to the calibration carried out with a variable angle spectroscopic ellipsometer from J. A. Woollam Co. Nevertheless, deviations from an ideal system can be produced for several reasons during the manufacturing and assembly process. A fine-tuning procedure \citep{2018OExpr..2612038A} was carried out in order to optimize the polarimetric efficiencies of the system. The polarimetric efficiencies achieved with this method, shown in Fig. \ref{fig:poleff} across the field of view, exceed the requirements: $\varepsilon_{\rm req} \geq [0.95, 0.45, 0.45, 0.45]$. 

      Finally, the spectroscopic E2E test was performed. Spectral measurements using sunlight from a coelostat for each target wavelength were made by scanning some 0.13 pm with the etalon with all the subsystem working temperatures secured in the laboratory. Additionally, an iodine cell illuminated with a diode (SOLIS-525C from Thorlabs Inc.) was used to verify the TuMag spectroscopic calibration. The expected TuMag spectral response was generated by convolution of the well-known iodine spectrum with the etalon transmission profile with a FWHM of 8.7 pm. The good match between measurements and simulations ensured the fulfillment of spectroscopic requirements. 

      After the E2E tests and calibrations, TuMag underwent a vacuum test for checking the F4$^{\prime}$ focus location at vacuum conditions; for checking the absence of electrical discharges when the etalon is performing a spectral scan; and for verifying the thermal control system. The test was passed successfully and TuMag was ready to be integrated in the \sunriset PFI.

      \subsection{PFI AIV phase}
      \label{sec:pfiaiv}

      TuMag was then integrated into the PFI together with the rest of the mission's instruments. To align the TuMag O-Unit inside the PFI, a reference cube on the TuMag optical bench was included and aligned with its optical axis. A double check was carried out by measuring the angle of incidence of the instrument's blocking filter with respect to the TuMag and PFI interface references. TuMag was located at the best focus position of the instrument in the PFI taking into account the vacuum defocus value. The image quality, polarimetric efficiencies and spectroscopic performances were checked while TuMag was in the PFI. A vacuum campaign was carried out, in order to verify the correct focus position of the instruments in the PFI, as well as the correct operation of the thermal and electrical control of the unit. Following the success of the campaign for all the instruments on board, the entire system was translated to the Esrange Space Center in Kiruna (Sweden). A characterization of TuMag inside the PFI was carried out upon arrival in order to check the instrument health and alignment after the transport.

      \subsection{System AIV phase}
      \label{sec:sunriseaiv}

      The PFI was installed on the telescope structure and E2E tests were performed to check the alignment and to calibrate the system. The last step was the integration of the telescope with the gondola. Later, pointing tests with solar light were done. The electromagnetic compatibility campaign and the communication tests were also carried out with the rest of the instruments. The result of the complete campaign was that TuMag fulfilled the science requirements for observations and was ready for flight.

      \section{Observing modes and data reduction}
      \label{sec:observinganddata}
        % All observational modes follow a similar scheme that will be detailed a little in Sect.~\ref{sec:obsfirmware}. The sequence is easy to follow. With the etalon tuned at a given wavelength, the LCVRs select a given polarization state. At this state, the instrument accumulates $N_{\rm acc}$ individual frames, needed to reach a given goal signal-to-noise ratio. Then the LCVRs change the polarization modulation state and the accumulations repeat as many times as necessary to get $N_{\rm p}$ polarization states. 
         %While LCVRs are tuning the new state, the frames coming out from the continuously running cameras are discarded. Only accumulated images are stored. Once the $N_{\rm p}$ polarization cycle is over, the etalon tunes the next wavelength sample. Again, no frames are stored whilst the etalon is tuning the next wavelength sample. All these processes, then, are repeated $N_\lambda$ times. For a single data set to be produced, the whole cycle is then repeated $N_{\rm c}$ times in order to reach the required polarimetric precision. As a result, $N_{\rm p} \times N_\lambda$ ($\times 2$ if two lines are observed) make such a single data set per camera.
         
   \begin{table}
      \caption[]{TuMag observing modes per single spectral line}
         \label{table:observing}
         \begin{minipage}{\textwidth}
            \begin{tabular}{lrrrrrrr}
      \hline
         {\rm Observing mode} & $N_{\lambda}$ & $N_{\rm p}$ & $N_{\rm a}$ & $N_{\rm c}$ & $t_{\rm eff}$ (s) & ${\rm (S/N)}_{1,2,3}$\footnote{For Obs. 0-s, ${\rm (S/N)}_0$.} & Data rate (Mb/s)\\
      \hline
         \noalign{\smallskip}
         Obs. 0-s & 12 & 1 &  2 &  1 &  6.30 & 500  & 11.22 \\
         Obs. 0-p & 12 & 4 & 16 &  1 & 37.62  & 1000 & 10.04 \\
         Obs. 1   & 10 & 4 & 16 &  1 & 31.81 & 1000 &  9.97 \\
         Obs. 2   &  8 & 4 & 16 &  1 & 23.40 & 1000 & 10.63 \\
         Obs. 3   &  5 & 2 & 20 &  1 & 10.04 & 1000 &  7.63 \\
         Obs. 4   &  3 & 4 & 10 & 10 & 54.01 & 2500 &  1.84 \\
         Obs. 5   &  3 & 4 & 10 & 10 & 53.60 & 2500 &  1.84 \\
         PD (100 frames)      &  1 & 4 &  1 &  1 & 10.67 & 1000 & 66.64 \\
         \noalign{\smallskip}
         \hline
   \end{tabular}
      \end{minipage}
   \end{table}         
   
      \subsection{Observing modes}
      \label{sec:observing}
         As explained in Sect.~\ref{sec:drivers}, TuMag aimed to increase the line sampling of single spectral lines, to tune among three possible ones, and to observe two out of those three lines in less than, say, 90 s. To allow for semi-automated operation, as necessary on an over-the-horizon stratospheric balloon-borne mission, a limited number of observing modes were implemented, which are summarized in Table~\ref{table:observing}. Modes are labelled in column 1 and characterized by the number of wavelength samples (second column), of polarization states (third column), of accumulations (fourth column), of cycles (fifth column), effective (total) time (sixth column), polarimetric precision (seventh column), and the  data rate (eighth column). The number of accumulations has been estimated with the expected photon budget in order to reach the polarimetric precision goal.
         
         Observing mode 0-s is a spectroscopic mode where no polarization modulation is carried out. For this mode ${\rm (S/N)}_{0}$ can be half the nominally needed for polarimetry. The line is sampled from $-40$\;pm through $60$\;pm in steps of 10\;pm plus another sample at 65\;pm.\footnote{Safe integrity of the etalon led us not to extend the samples symmetrically up to $-60$\;pm.} It is specifically thought to provide rapid scans of the Mg~{\sc i}{\small b}$_2$ line. Observing mode 0-p is also an \textit{extended} mode for Mg~{\sc i}{\small b}$_2$ line observations, but this time polarization modulation is included. Wavelength samples are the same as for Obs. 0-s. Observing mode 1 is also meant (in principle) for the Mg~{\sc i}{\small b}$_2$ line. Wavelength samples are at $[-30, -20, -10, -5, 0, 5, 10, 20, 30, 65]$ pm from line center. (The last sample corresponds to the nearby ``continuum''.) Combined with Obs. 2 for any of the Fe~{\sc i} lines, Obs. 1 allows recording two lines in less than 66\;s (including a worst-case time delay of 10\;s between the two lines). Wavelength samples for Obs. 2 are at $[-12, -8, -4, 0, 4, 8, 12, 22]$ pm from line center. Observing mode 3 is a longitudinal magnetic mode as only circular polarization analysis is carried out. Wavelength samples for this mode are at $[-8, -4, 4, 8, 22]$ pm. It is a fast mode and only Stokes $I$ and $V$ are measured. Observing mode 4 and 5 are very similar to each other. They are ``deep'' magnetic field modes because the number of accumulated frames is high enough to provide polarimetric precisions of 2500. The only difference between both is in the wavelength samples: Obs. 4 is meant for the Mg~{\sc i}{\small b}$_2$ line and samples it at $[-10, 0, 10]$ pm whilst Obs. 5 is meant for the Fe~{\sc i} lines and samples it at $[-8, 0, 8]$ pm. These observing modes have been appropriately used during the assembly, integration, and verification phase of the instrument (see Sect.~\ref{sec:aiv}). Total scan times correspond to single line observations. An extra 4 s overhead should be added for line selection. The PD mode consists in taking 100 (non-accumulated) frames at the continuum wavelength in all four polarization states.
         
        \begin{figure}
	        \centering
	        \includegraphics[width=\textwidth]{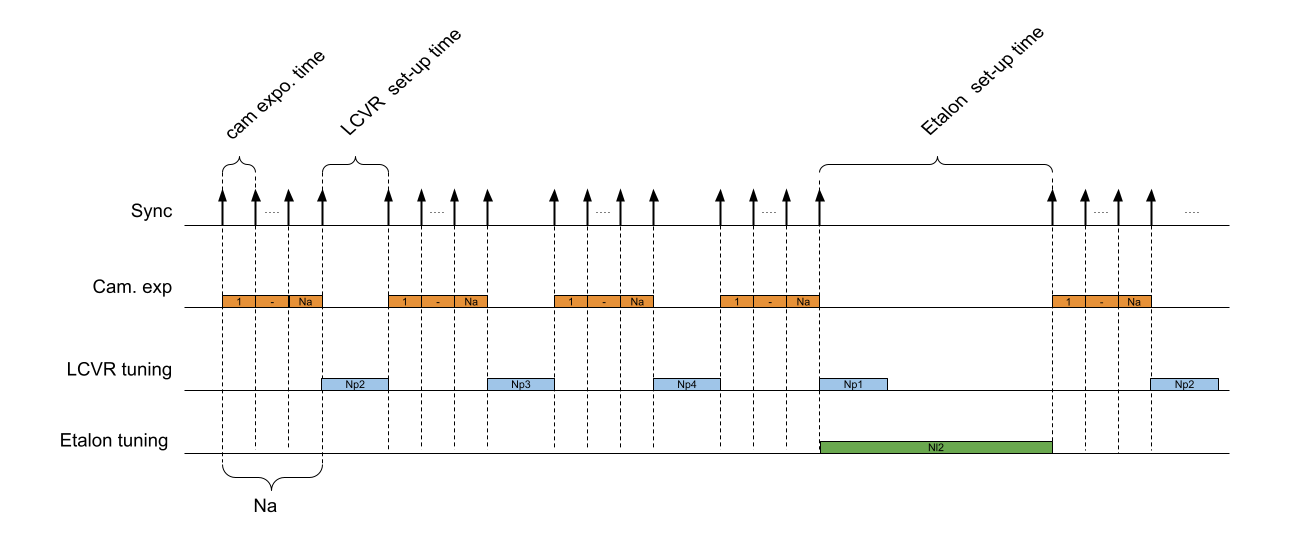}
	        \includegraphics[width=\textwidth]{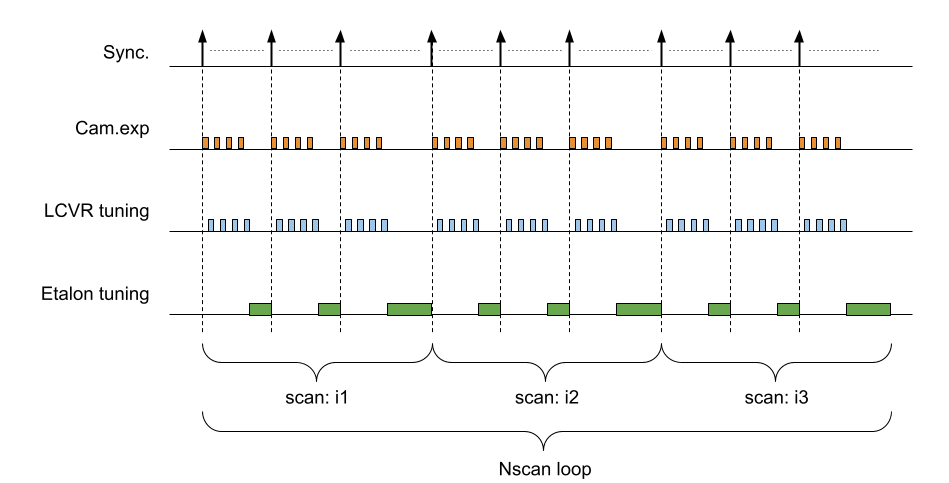}
	        \caption{Observing firmware timeline sequence diagram.}
	        \label{fig:timeline}
        \end{figure}

      \subsection{Observing modes: firmware implementation}
      \label{sec:obsfirmware}
         %\subsubsection{The observing modes in firmware}
         %\label{sec:obsfirmware}

      TuMAG's observation modes (see Sect.~\ref{sec:observing}) have been designed so that the instrument always works in the same way. The observational sequence starts sending initialization commands to all optical subsystems. Once the subsystems reach the target status, e.g., with the etalon tuned at a given wavelength and the LCVRs configured to measure a specific modulation state, the FGSW is set to that the cameras are triggered and grab $N_{\rm a}$ individual frames to generate a single, accumulated image with a target signal-to-noise ratio. Then the LCVRs are tuned to the next polarization modulation state and the instrument takes another set of $N_{\rm a}$ frames, resulting in an image. The operation is repeated $N_{\rm p}$ times to get the necessary polarization states, 4 in the case of vector modes and 2 in the case of longitudinal modes. This process can be repeated $N_{\rm c}$ times. Once the $N_{\rm p}N_{\rm c}$ polarization states have been recorded, the etalon is tuned (commanded) to the next wavelength sample. In total, $N_\lambda$ wavelength samples are recorded. In the end, a single dataset consists of $N_{\rm p}  N_{\rm c} N_\lambda$ images per camera. %During the whole process, both cameras take frames only when , but the FG discards those taken during etalon or LCVR tuning in such a way that $N_{\rm a}$ is kept constant over the whole dataset. 
                  
%The observational sequence starts sending initialization commands to all optical subsystems. Once these subsystems reach the target status, cameras are triggered to grab $N_{acc}$ frames. After that, LCVRs are tuned to the next polarization state and $N_{acc}$ frames are grabbed again. This sequence is repeated for a given number of polarization states, $N_p$, until the acquisition of the current wavelength sample is complete. Then, a new wavelength configuration is commanded to the etalon. The whole cycle is repeated $N_\lambda$ times corresponding to the wavelength samples to complete the scan of a given spectral line. 

     Figure~\ref{fig:timeline} shows two time diagrams of an observational sequence. On top, the processes corresponding to one wavelength sample (and the start of the next one) are displayed. On the bottom panel, the time axis has shrunk and three line scans are represented (specifically, three wavelength samples correspond to Obs. 4 and Obs. 5 modes). With a rolling-shutter camera continuously running at 24 fps, all commands in the frame grabber are synchronized by the {\tt sync} signal with an elementary time $T_{\rm sync} = 5\, \mu$s. Hence, all operations (top row) have a duration of a multiple of $T_{\rm sync}$.  Vertical dashed lines indicate some of those $n T_{\rm sync}$ moments. Orange rectangles (second row from top) mark the start and the end of a single-frame-acquisition exposure time, $t_{\rm exp}$. In the LCVR and etalon rows, rectangles  stand for the time it takes for these elements to reach the target state (polarization or wavelength, respectively) after having been commanded. Note that at the end of a line scan (bottom panel) the etalon tuning time is longer because the continuum sample is typically farther away than the in-line samples. Even longer rectangles had to be represented if a change in the line is needed because the filter wheel also takes time to move.
            
     Individual frame exposures are performed when the optical subsystems are ready. The set-up time of these subsystems is configurable in compilation time of the FPGA. In the case of LCVRs, the set-up time is fixed to the worst case of 100 ms. In contrast, the etalon set-up time depends on the wavelength separation between two consecutive steps. Hence, setting the etalon may take a different duration each time. These times are also configured in compilation time by a vector for each observational mode.

     Additionally, during the etalon setting time, the FG performs HK read-outs of the actual voltage of the etalon. In case the setting time ends and the etalon has not reached the target, an exception is thrown.

      \subsection{Internal image processing}
      \label{sec:preprocessing}

             TuMag generates an irregular data rate because of the various idle times during optical subsystems' reconfiguration, especially for the etalon setting time. The most demanding data rate situation takes place during PD observations, when the two cameras are operating at full-frame, $2048 \times 2048$ pixels, which can be kept for up to $\sim$ 5 s. In such a situation, the data rate is around 2416 Mbps. This figure is unmanageable by the DPU hardware, since it is more than twice the theoretical link capacity of 1 Gbps (Ethernet interface) between the instrument and the ICS. It is also greater than the theoretical throughput of the DPU's SSD, where images are temporally buffered.\footnote{A maximum of 2400 Mbps is supported by the SATA 2 standard interface.} Therefore, an image processing pipeline embedded in the FG is mandatory to reduce in real time the TuMag data rate while keeping the scientific quality of the observations. Such a processing pipeline performs image accumulation and lossless image compression. It is designed in an efficient architecture that processes one pixel per clock cycle at 150 MHz, with two independent pipeline instances working in parallel, one per camera. The data rate reduction factor after using the pipelines varies from 5 (i.e., 474 Mbps) for Obs. 0-s to 291 (i.e., 8.3 Mbps) for Obs. 4 and 5.

      \subsubsection{Camera frames, real-time accumulation}
      \label{sec:accumulator}

             The number $N_{\rm a}$ of accumulations must be programmable to set the different observational modes. To accomplish the requirements, we have developed a custom specific processing core (the \textit{accumulator}) which is instantiated inside the FPGA firmware. With it, full parallel processing is guaranteed in both data pipelines. The use of the FPGA capabilities to process the images allows the use of full parallel tasks, and this makes it possible to reduce the needed clock speed for processing. For that reason, the processing clock speed is 150MHz.. To save power and make the FPGA routing easier, the clock for commanding and programming is set to 50 MHz. As the entire FG block design is based on the Advanced eXtensible Interface (AXI) standard, the IP core has two AXI interfaces that connect to the MicroBlaze and Video Direct Memory Access (VDMAs), to be easily integrated in the design. Indeed, the accumulation core does not directly access the memory controller, but uses VDMAs to do this task. A region of interest can also be programmed to accept different image configurations. The number of accumulations programmed and the current number of accumulations are also accessible through the core configuration registers.
             
             The accumulator receives one input pixel, accumulates it and outputs the result through the AXI streaming interface. To fulfill AXI requirements and for a better flow control, it has an internal FIFO device to store the input pixels and another one to store the output pixels. To get the images it is necessary to access memory by using VDMAs, so the integrator also has the logic necessary to synchronize VDMAs to read and write to memory, and a finite state machine to correctly read and write the FIFOs. To synchronize with the VDMAs, it is necessary to send several pulses with every new image. This pulse generation is set by the synchronism generators. They generate synchronization pulses for the camera and the accumulator VDMAs. Each synchronism generator has a different finite state machine inside to correctly generate the pulses. To read the AXI stream interfaces and synchronize the addition of pixels, the core uses another finite state machine inside. This machine is designed to fit all the different possibilities in the process, from reading and writing FIFOs to accumulation. It is robust to prevent a possible failure in flow of input or previous pixels, which guarantees a perfect function of the system.

      \subsubsection{Image compression}
      \label{sec:compression}

             The last stage in the processing pipeline is an image compression block. The core of this block is the Sunrise Lossless Compression (SLOC) IP core which implements the standard CCSDS-LDC 121.0-B2, a low-complexity solution to compress any kind of data including 2D images. This core handles one pixel per clock cycle at a frequency of 150 MHz to achieve the real-time processing required in TuMag.
           
             To integrate the core in the FG architecture, a wrapper devised to standardize the interfaces to the AXI4 bus is included.  In this way, the core counts with AXI4-stream input/output interfaces for receiving the input image and returning the compressed one. A secondary output interface is also included for returning some status information, such as the input/output cyclic redundancy check  codes and the compressed image size. Additionally, the core counts with an AXI4-Lite interface, for handling the configuration register. 

      \subsection{Data reduction pipeline}
      \label{sec:pipeline}
        TuMag datasets need to be processed before being exploited for scientific purposes. A number of processes including various corrections are devised, such as dark current subtraction and flat-fielding. After this processing, demodulation of the polarized images provides the full Stokes vector. Further processing steps are necessary, e.g., blueshift and cross-talk correction, interference fringes cleansing (if needed) and, eventually, correction of the spatial PSF obtained through phase-diversity techniques.

        The pipeline for any observing mode consists of a series of well-defined consecutive steps:\footnote{The corresponding programs are ready for application and can be obtained upon request.}
        \begin{description}
            \item[Dark current calculation:] Identify the closest (in time) set of dark frame observations and generate a single dark frame for each camera by averaging all individual images. Both observing and calibration modes have the same number of accumulations and exposure times in order to ensure the consistency of the observations. 
            \item[Flat field calculation:] Identify the corresponding set of flat-field observations and find the closest (in time) one. Prior to any calculation, the dark current image must be subtracted from all frames. A single flat-field is then produced for each wavelength and modulation state by averaging and normalizing all frames.
            \item[Image correction and processing:]  Images must first be corrected from the dark current (subtraction) and flat field (division). At this point, the Stokes vector is computed by demodulating the data. The matrix used for the calculation depends on the type of the observing mode, which can either be vectorial or longitudinal. In the case of vectorial modes, the full Stokes vector is computed, whereas only Stokes $I$ and $V$ are provided in longitudinal modes. The demodulation process can either be done pixel by pixel (1D) or with a 2D matrix, depending on the performance for each observation mode. After the demodulation process, images from both cameras are combined. Prior to adding up the images, they should be aligned and their intensities equalized.
            
            With the resulting merged images, the cross-talk correction is computed. Only cross-talk from $I$ to $Q$, $U$ and $V$ and from $V$ to $Q$ and $U$ are corrected. The calculation is carried out by searching for correlations between the given Stokes parameters within a single set of four images. When a statistically significant correlation is found, a correction over the contaminated image is applied.
            \item[Blueshift correction:] As a consequence of using the Fabry--P\'erot etalon in collimated configuration, the transmission profile is shifted to the blue as one moves from the center of the image to the borders of the field of view \citep[see, e.g.,][]{2019ApJS..241....9B}. There is no need to correct for this blueshift, since it is considered during the inversion process: each pixel has a spectroscopic sampling different from the others by a small amount. To fully determine such a blueshift, the flat-field image is used. On it, the actual center (which might not coincide with the center of the detector) is determined, and the shifts are calculated according to the theoretical formula. Since the blueshift effect has a rotational symmetry, any possible linear illumination gradient which might appear in the image is also corrected in this step.
            \item[Interference fringes cleansing:] Any possible fringing that may remain in the images is removed in the Fourier space by conventional, i.e., manual, procedures.
            \item[Phase diversity restoration:] The closest (in time) set of phase diversity images is used to derive both the PSF and the incident wavefront aberration. Deconvolution from the PSF is carried out by (as a baseline) using the optimum-filter technique.
        \end{description}
        
        After all these steps are performed, the resulting images are ready for scientific exploitation. They are stored in FITS format with a header containing all needed metadata for a correct interpretation of the data, e.g., ephemerides, image size, wavelengths positions, or observing mode, among many other parameters.

\section{Conclusions}
\label{sec:conclu}

   We have presented the science drivers, the concept, design, integration, and characterization of the TuMag instrument and its subsystems. The Tunable Magnetograph is a tunable imaging spectropolarimeter for the \sunriset mission. It is conceived to explore the photosphere and low chromosphere of the Sun by sampling two out of three selected spectral lines. Combining the two most common observing modes, the total observation time does not exceed 66 s for obtaining a polarimetric precision of 10$^3$ in the worst case. Although TuMag inherits concepts and technologies from former instruments, it incorporates significant innovations that promise to increase the scientific return of the mission when combined with the other instruments on board, as compared with previous editions of the mission. 

%%%%%%%%%%%%%%%%%%%%%%%%%%%%%%%%%%%%%%%%%%%%%%%%%%%%%%%%%%%%%%%%%%%%%%%%%%%
%% Acknowledgements
%
\begin{acks}
      This work has been funded by the Spanish MCIN/AEI under projects RTI2018-096886-B-C5, and PID2021-125325OB-C5; and from the ``Center of Excellence Severo Ochoa'' awards to IAA-CSIC (SEV-2017-0709, CEX2021-001131-S), all co-funded by European REDEF funds, ``A way of making Europe''. D.O.S. acknowledges financial support from a {\em Ram\'on y Cajal} fellowship. We warmly thank the MPS personnel who manufactured the (non-high-voltage) external harness and who provided help during the various development phases and led the final integration in {\sc Sunrise}. This project has received funding from the European Research Council (ERC) under the European Union's Horizon 2020 research and innovation programme (grant agreement No. 101097844 — project WINSUN).
\end{acks}

\section{Additional statements}
 
\begin{authorcontribution}
J.C.T.I. was the principal investigator of the instrument, wrote most parts and assembled the rest of the manuscript. D.O.S. was the instrument co-PI, overviewed the overall development, the assembly, integration, and verification, and was responsible for the calibration. A.A.H. was responsible of the optical unit and the integration of the instrument. E.S.K. was responsible of power electronics and electronic box development. I.P.-G. was responsible of the thermal engineering of the development. B.R.C. participated in the scientific definition of the instrument. M.B.J. was the chief engineer of the project. A.C.L.J., D.A.G., J.L.R.M., P.L., A.J.M.M., I.B., and A.T. contributed to the hardware development of the cameras, the DPU, the AMHD, and HVPS. A.S.G. and E.B.M. contributed the control and ground segment software of the instrument. J.P.C.C., J.M.M.-F., B.A.M., D.H.E., E.P.M., E.M.C., and M.R.V. contributed to several subsystems of the instrument firmware. F.J.B. contributed to several aspects of the optical design and to the development of the phase diversity techniques for post-facto image reconstruction. H.S., A.L.S.-T., A.M.V., J.A.G., L.R.B.R, and A.J.D.M. contributed to the scientific preparation of observations and some aspects of calibration. P.S.G. was responsible for the data reduction pipeline. A.F.-M., A.N.P., M.C., D.G.-G., P.G.P., A.G.M., A.S.R., A.C.J., H.L., and M.S.-L. contributed to the optical unit design and development and to the integration of the instrument. J.B.R. contributed to the scientific definition of the instrument and its data reduction pipeline as well as to the preparation of observations and calibration. J.L.G.B., P.R.M., A.F., and D.G.P. contributed to the power electronics and the electronic box developments. I.T., J.P., D.G.-B., and A.J.F. contributed to the thermal engineering development and calibration. S.K.S. was the principal investigator of the mission. A.K.-L. was the project manager of the mission. A.G. was responsible of the overall optical design and integration of the mission. T.B., P.B., A.F. and Y.K. led the CWS, gondola, SUSI, and SCIP developments, respectively. H.N.S., M.K., and V.M.P. were other key people in the scientific working group of the mission. B.G. contributed to the optical design of the mission. A.B. contributed to the CWS development. M.C. contributed to the gondola development.
\end{authorcontribution}

\begin{fundinginformation}
This work has been funded by the Spanish MCIN/AEI under projects RTI2018-096886-B-C5, and PID2021-125325OB-C5; and from the ``Center of Excellence Severo Ochoa'' awards to IAA-CSIC (SEV-2017-0709, CEX2021-001131-S), all co-funded by European REDEF funds, ``A way of making Europe''. D.O.S. acknowledges financial support from a {\em Ram\'on y Cajal} fellowship. This project has received funding from the European Research Council (ERC) under the European Union's Horizon 2020 research and innovation programme (grant agreement No. 101097844 — project WINSUN).
\end{fundinginformation}

\begin{dataavailability}
Not applicable
\end{dataavailability}

\begin{ethics}
\begin{conflict}
The authors declare that they have no conflicts of interest.
\end{conflict}
\end{ethics}

%%% %%%%%%%%%%%%%%%%%%%%%%%%%%%%%%%%%%%%%%%%%%%%%%%%%%%%%%%%%%%
%% Bibliography
%
% Using BibTeX
%
\bibliographystyle{spr-mp-sola}
\bibliography{Biblioteca_de_citas}

\end{document}